\documentclass[runningheads]{llncs}

\usepackage{todonotes}
\usepackage{xr}
\usepackage[linesnumbered,lined,boxed,commentsnumbered]{algorithm2e}
\usepackage{indentfirst}
\usepackage{algorithm2e}
\RestyleAlgo{ruled}
\usepackage{array}
\usepackage{tabularx}
\usepackage{amsmath, amssymb, amsfonts}
\usepackage{tcolorbox}
\usepackage{xcolor}
\definecolor{grey}{RGB}{245,245,245}

\colorlet{mix}{red!50!black}

\usepackage{hyperref}
\hypersetup{colorlinks={true},linkcolor={blue},citecolor=blue}
\usepackage[noabbrev,capitalise,nameinlink]{cleveref}

\usepackage{hyperref}
\hypersetup{
	pdfnewwindow=true,      
	colorlinks=true,       
	linkcolor=mix,          
	citecolor=blue,        
	urlcolor=blue          
}

\usepackage{enumitem}

\SetKwComment{Comment}{/* }{ */}

\usepackage{array,tabularx}
\usepackage{colortbl}

\usepackage{tabularx}

\newtheorem{marking scheme}{Marking Scheme}
\newtheorem{observation}{Observation}

\newcolumntype{L}[1]{>{\raggedright\arraybackslash}p{#1}}
\newcolumntype{C}[1]{>{\centering\arraybackslash}p{#1}}

\newcommand{\No}{\texttt{No}\xspace} 
\newcommand{\Yes}{\texttt{Yes}\xspace}

\newcommand{\mr}{{\sc{Monopolar Recognition}}\xspace}

\newcommand{\me}{{\sc Monopolar Extension}\xspace}
\newcommand{\lmp}{{\sc List Monopolar Partition}\xspace}

\newcommand{\mvd}{{\sc Monopolar Vertex Deletion}\xspace}

\newcommand{\Oh}{{\mathcal{O}}}

\newtheorem{reduction rule}{Reduction Rule}

\newcommand{\tw}{\mathsf{tw}}

\usepackage{mdframed}
\usepackage{multicol}

\usepackage{booktabs}
\usepackage{xcolor}
\usepackage{graphicx}

\definecolor{grey}{RGB}{245,245,245}

\colorlet{mix}{red!50!black}

\usepackage{hyperref}
\hypersetup{
	pdfnewwindow=true,      
	colorlinks=true,       
	linkcolor=mix,          
	citecolor=blue,        
	urlcolor=blue          
}

\usepackage{tabularx,lipsum,environ}

\makeatletter
\newcommand{\problemtitle}[1]{\gdef\@problemtitle{#1}}
\newcommand{\probleminput}[1]{\gdef\@probleminput{#1}}
\newcommand{\problemquestion}[1]{\gdef\@problemquestion{#1}}

\begin{document}
\title{Improved Exact Algorithms and Parameterized Results for Monopolar Graph Problems} 

\titlerunning{Exact and Parameterized Algorithms for Monopolar Graphs} 
\author{Soumen Mandal \inst{1}\and Ashutosh Rai \inst{2}}

 \authorrunning{Soumen Mandal \and Ashutosh Rai}
 \institute{Tata Institute of Fundamental Research, Mumbai, India\and Indian Institute of Technology Delhi, New Delhi, India
\\ \email{soumen.mandal@tifr.res.in, ashutosh.rai@maths.iitd.ac.in}}

\maketitle

\begin{abstract}
A monopolar graph is a graph whose vertex set can be partitioned into a cluster
graph and an independent set. Deciding whether such a partition exists is
\textsc{Monopolar Recognition}. We also study three related problems:
\textsc{Monopolar Extension}, where a partial assignment of vertices to the two
sides is given and must be extended to a full monopolar partition;
\textsc{List Monopolar Partition}, where each vertex comes with a list of
allowed sides; and \textsc{Monopolar Vertex Deletion}, which asks for a
minimum-weight vertex set whose removal makes the graph monopolar. All four problems are NP-hard, which motivates the study of faster exact algorithms for these problems.

We design a new exact branching algorithm for \textsc{Monopolar Extension}
running in $\Oh^*(1.29803^n)$ time\footnote{The \(\mathcal{O}^*(\cdot)\)
notation suppresses factors that are bounded by a polynomial in the input
size.}, improving upon the previously best-known $\Oh^*(1.3734^n)$-time
algorithm of Philip and Sridhara [WG 2025]. The same bound applies to
\textsc{List Monopolar Partition}, which is an equivalent formulation, and to
\textsc{Monopolar Recognition}, which is a special case. It also gives an
$\Oh^*(2.29803^n)$-time algorithm for \textsc{Monopolar Vertex Deletion}.

We further present the first explicit fixed-parameter algorithms with concrete single-exponential dependence on treewidth for these problems. We show that \textsc{Monopolar Extension}, and
hence also \textsc{List Monopolar Partition} and \textsc{Monopolar
Recognition}, can be solved in $3^{\tw}\cdot n^{\Oh(1)}$ time. For
\textsc{Monopolar Vertex Deletion} we obtain a running time of
$5^{\tw}\cdot n^{\Oh(1)}$, which improves to $4^{\tw}\cdot n^{\Oh(1)}$ when the
weights are integers bounded by a polynomial in $n$.
\end{abstract}

\section{Introduction}
A graph is \emph{polar} if its vertex set admits a partition $(A,B)$ such that $G[A]$ is a complete
multipartite graph and $G[B]$ is a \emph{cluster graph}, that is, a disjoint union of complete graphs.
Polar graphs were introduced by Tyshkevich and Chernyak~\cite{tyshkevich1985decomposition} as a common
generalization of split and bipartite graphs, and their recognition is
\textsf{NP}-complete~\cite{chernyak1986recognizing}, though polynomial on several restricted
classes~\cite{DBLP:journals/siamdm/ChurchleyH11,DBLP:journals/ejc/EkimHM13,DBLP:journals/dam/EkimHSW08}.

A particularly well-studied subclass of polar graphs is the class of \emph{monopolar graphs}, where the vertex set admits a partition $(A,B)$ such that $A$ induces an independent set and $B$ induces a cluster graph. Monopolar graphs arise naturally in applications such as identifying core--periphery structures in protein interaction networks~\cite{DBLP:journals/almob/BrucknerHK15}. This class generalizes both split graphs and bipartite graphs: in split graphs, the cluster $B$ consists of a single clique, whereas in bipartite graphs, $B$ is a collection of singleton cliques. The problem of deciding whether a given graph is monopolar, formally defined below, is known to be \textsf{NP}-complete~\cite{DBLP:journals/combinatorics/Farrugia04}.

\noindent\fbox{
	\begin{minipage}{.95\textwidth}
		\begin{tabular*}{\textwidth}{@{\extracolsep{\fill}}lr}
			{{\sc Monopolar Recognition}} &
		\end{tabular*}
		{\bf Input:} A graph $G=(V,E)$. \\
		{\bf Question:} Is $G$ monopolar?
	\end{minipage}
}

The problem \mr remains \textsf{NP}-complete even when restricted to several graph classes, including triangle-free planar graphs of maximum degree~$3$ and $3$-colorable graphs~\cite{DBLP:conf/isaac/LeN11}. On the other hand, \mr is solvable in polynomial time for various well-known graph classes such as \emph{chordal graphs}~\cite{DBLP:journals/dam/EkimHSW08}, \emph{line graphs}~\cite{DBLP:journals/siamdm/ChurchleyH11}, \emph{claw-free graphs}, and \emph{chair-free graphs}~\cite{DBLP:journals/jgt/ChurchleyH14}. We refer the reader to~\cite{DBLP:journals/tcs/LeN14} for a comprehensive survey of graph classes where \mr is polynomial-time solvable or \textsf{NP}-complete.

Two closely related problems associated with the monopolar graph class are \me\ and \lmp. In \me, the
input is a graph $G$ together with two disjoint vertex subsets $I' \subseteq V$ and $C' \subseteq V$,
and the question is whether there exists a monopolar partition $(I,C)$ of $G$ with $I' \subseteq I$ and
$C' \subseteq C$. We do not require $I'$ to be an independent set or $G[C']$ to be a cluster graph; if
either fails, the instance is a \No-instance. In \lmp, the input is a graph $G$ together with a list
function $L$ defined on $V$ such that $\emptyset \neq L(v) \subseteq \{I,C\}$ for every $v \in V$, and
the question is whether $G$ admits a monopolar partition that respects $L$.

\me\ (setting $I^{\prime}=\emptyset$, $C^{\prime}=\emptyset$) and \lmp\ (assigning $L(v)=\{C,I\}$)
both generalize \mr, and were introduced in~\cite{DBLP:journals/tcs/LeN14}
and~\cite{DBLP:journals/dm/ChurchleyH12}, respectively. Since \mr\ is \textsf{NP}-complete, it follows
that both \me\ and \lmp\ are also \textsf{NP}-complete. Moreover, it is known that \me\ and \lmp\ are
computationally equivalent modulo polynomial time~\cite{DBLP:conf/wg/PhilipS25}.

Another natural problem related to monopolar graphs is \mvd. Here the input is a graph $G=(V,E)$
together with a weight function $w: V \rightarrow \mathbb{Q}_{\ge 0}$, and the task is to find a vertex
set $S$ of minimum weight such that $G-S$ is monopolar. Since \mr\ is \textsf{NP}-complete, so is the decision version of \mvd.

Recently, Philip and Sridhara~\cite{DBLP:conf/wg/PhilipS25} presented an exact
algorithm for \me\ running in $\Oh^*(1.3734^n)$ time, improving upon the
trivial $\Oh^*(2^n)$-time algorithm, where $n$ denotes the number of
vertices of the input graph. From the discussion above, it follows that
both \lmp\ and \mr\ also admit exact algorithms with running time
$\Oh^*(1.3734^n)$. A natural question is whether these problems admit
even faster exact algorithms.

Kanj \emph{et al.}~\cite{DBLP:journals/jcss/KanjKSL18} initiated the study of \mr\ in the parameterized
setting by considering the \emph{number of disjoint cliques} in the cluster part as a parameter, and
presented an $\Oh^*(2^k)$-time \textsf{FPT} algorithm. They also obtained a polynomial kernel of size
$\Oh(k^4)$ for \mr\ with respect to the same parameter~\cite{DBLP:journals/siamdm/KanjKSL20}. More recently,
Philip and Sridhara~\cite{DBLP:conf/wg/PhilipS25} studied \mr\ from a parameterized perspective by
considering the parameters \emph{vertex distance to claw-free graphs} ($k_v$) and
\emph{edge distance to claw-free graphs} ($k_e$), and presented \textsf{FPT} algorithms with running times
$\Oh^*(3.076^{k_v})$ and $\Oh^*(2.253^{k_e})$, respectively.

To the best of our knowledge, apart from these few results on \mr, no parameterized algorithms have been
studied for any of the other problems considered in this paper. Moreover, we are not aware of previous algorithmic work on \mvd.
Motivated by the recent progress on exact algorithms for monopolar graph problems and the lack of
parameterized results beyond \mr, we investigate the exact and parameterized complexity of these problems,
with a particular focus on algorithms parameterized by the well-known graph parameter \emph{treewidth}.

\paragraph{Our Contributions.} 
Our first contribution is an improved exact algorithm for \me, based on a branching strategy combined
with a collection of new reduction rules. The measure is the number of unassigned vertices, and at each
step we branch on a vertex with the maximum number of unassigned neighbors. Every branching vector has
branching number at most that of $(5,3,5)$, which gives the running time $\Oh^*(1.29803^n)$.

\me\ can also be solved by enumerating all maximal independent
sets~\cite{DBLP:journals/ipl/JohnsonP88,MoonMoser65}, which takes $\Oh^*(1.4423^n)$ time. The fastest
known algorithm is the $\Oh^*(1.3734^n)$-time algorithm of Philip and
Sridhara~\cite{DBLP:conf/wg/PhilipS25}, which our branching algorithm improves.

We also study these problems parameterized by \emph{treewidth}, a fundamental parameter introduced by
Robertson and Seymour~\cite{DBLP:journals/jal/RobertsonS86} that measures how close a graph is to being
a tree, and for which many otherwise intractable problems admit dynamic programming algorithms over
tree decompositions.

In this work, we present the first explicit fixed-parameter algorithms with concrete single-exponential dependence on treewidth for these problems. In particular, we show that \me\ can be solved in
$3^{\tw}\cdot n^{\Oh(1)}$ time and that \mvd\ can be solved in $5^{\tw}\cdot n^{\Oh(1)}$ time, the
latter improving to $4^{\tw}\cdot n^{\Oh(1)}$ when the weights are integers bounded by a polynomial
in $n$. Both algorithms are dynamic programs over a nice tree decomposition, where for each bag we record
which vertices lie in the independent set and which lie in cliques of the cluster part that can still
be extended. At a join node, a clique may descend into either of the two subtrees, but not into both,
and guessing which one costs an extra factor. We evaluate the join node using fast subset
convolution~\cite{DBLP:conf/stoc/BjorklundHKK07}, which is what keeps the bases at $3$ and $4$.

Our results for \me\ have immediate consequences for the related problems. As \mr\ is a special case
of \me\ and \lmp\ is an equivalent formulation~\cite{DBLP:conf/wg/PhilipS25}, both inherit the bounds
$\Oh^*(1.29803^n)$ and $3^{\tw}\cdot n^{\Oh(1)}$. Guessing the deletion set $S \subseteq V$ and then
testing whether $G-S$ is monopolar gives an $\Oh^*(2.29803^n)$-time algorithm for \mvd, which together
with our treewidth algorithm establishes the fixed-parameter tractability of all four problems with
respect to treewidth.

\paragraph{Organization.} \Cref{sec:preli} fixes notation. \Cref{sec:exactalgo} presents the exact
algorithm for \me, and \Cref{sec:fptmr} and \Cref{sec:fptmvd} the treewidth algorithms for \me\ and
\mvd. We conclude in \Cref{sec:con}. Proofs omitted from the main body are in the appendix.
 
\section{Preliminaries}\label{sec:preli}
Most of the preliminaries, including additional notation, background on tree decompositions, and the
analysis of branching algorithms, are deferred to Appendix~\ref{appensec:preliB}.

Throughout this paper, we only consider simple, undirected graphs. The subgraph obtained after
deleting $S$ and the edges incident to the vertices in $S$ is denoted by $G-S$. For a vertex
$u\in V(G)$ and a set $S\subseteq V(G)$, by $N_S(u)$ we denote $\{v\in S \mid vu\in E(G)\}$, and when
$S=V(G)$ we denote $N_S(u)$ by $N(u)$. We define $N_S[u]=N_S(u)\cup \{u\}$ and similarly
$N[u]=N(u)\cup \{u\}$. If a set of vertices $C$ induces a cluster graph, we sometimes refer to $C$ as
the cluster graph rather than $G[C]$. Whenever we refer to a clique in a cluster graph, we mean a
maximal clique, that is, a connected component of the cluster graph. For tree decompositions, nice
tree decompositions and treewidth, we refer to Appendix~\ref{appensec:preliB}; when the graph is clear from
the context we denote its treewidth by $\tw$. We use the standard measure-and-conquer framework for
analysing branching algorithms; the definitions of branching vector and branching number are in
Appendix~\ref{appensec:preliB}.

\section{Exact Algorithm}\label{sec:exactalgo}
We first give a set of reduction rules, applied exhaustively before each branching step. Rules~1, 3
and~5 are implicit in the structural observations of Philip and
Sridhara~\cite{DBLP:conf/wg/PhilipS25} on edges, induced paws and diamonds, and induced $P_3$s
respectively; the remaining rules are new.

 A reduction rule is said to be correct if, when applied to an instance $(G, I', C')$ of \me, it produces another instance $(G', I'', C'')$ of \me\ such that $(G, I', C')$ is a \Yes-instance if and only if $(G', I'', C'')$ is a \Yes-instance. For an instance $(G,I',C')$ of \me, we define $V'=V(G)\setminus (I'\cup C')$. We defer the correctness proofs of Reduction Rules~\ref{rule:3} to \ref{rule:11} to Appendix~\ref{appen:proofrr}.



\begin{reduction rule}\label{rule:1}
Given an instance $(G,I',C')$ of \me, if $I'$ is not an independent set or the subgraph $G[C']$ is not a cluster graph, then return a trivial \No-instance.
\end{reduction rule}

\begin{reduction rule}\label{rule:2}
Given an instance $(G,I',C')$ of \me, if $I'$ forms an independent set, $G[C']$ induces a cluster graph, and $V(G) = C' \cup I'$, then return a trivial \Yes-instance.
\end{reduction rule}

\begin{reduction rule}\label{rule:3}
Given an instance $(G,I',C')$ of \me, if there exists a vertex $v \in I'$ with
$N_{V'}(v) \neq \emptyset$, then move all vertices of $N_{V'}(v)$ to $C'$. The
resulting instance is $(G, I', C' \cup N_{V'}(v))$.
\end{reduction rule}


\begin{reduction rule}\label{rule:4}
Given an instance $(G,I',C')$ of \me, if there exists a vertex $v \in V'$ with
$N(v) \subseteq C'$, that is, both $N_{V'}(v)$ and $N_{I'}(v)$ are empty, then
include $v$ in $I'$. The new instance becomes $(G, I' \cup \{v\}, C')$.
\end{reduction rule}


\begin{reduction rule}\label{rule:6}
  For an instance $(G,I',C')$ of \me, if there exists a vertex $v \in V'$ such
that $N_{C'}(v)$ is non-empty and is not a clique of $C'$, then include $v$ in
$I'$. The new instance is $(G, I' \cup \{v\}, C')$.
\end{reduction rule}

\begin{reduction rule}\label{rule:5}
Given an instance $(G,I',C')$ of \me\ to which none of Reduction
Rules~\ref{rule:1} to \ref{rule:6} is applicable, suppose there exists a vertex
$v \in V'$ such that $B = N_{V'}[v]$ forms a clique, and there exists a
$B' \subseteq C'$ where $B'$ forms a clique and
$\bigcap_{u \in B} N_{C'}(u) = B'$. If there exists a vertex $w \in B'$ with
$N_{V'}(w) = B$, then add $v$ to $C'$. The resulting instance is
$(G, I', C' \cup \{v\})$.
\end{reduction rule}

\begin{reduction rule}\label{rule:7}
For an instance $(G,I',C')$ of \me, suppose there exists an edge $uv$ in $G[V']$ such that $N_{V'}(u)=\{v\}$, $N_{V'}(v)=\{u\}$ and $N_{I'}(u)$, $N_{I'}(v)$ both are empty. If $N_{C'}(u)$ is empty or a clique $B'$ in $G[C']$ and there exists a vertex $w\in B'$ (if $N_{C'}(u)$ is not empty) such that $w$ does not have neighbors in $V'$ other than $u$ and $v$, then add $u$ to $C'$. The new instance is $(G,I',C'\cup \{u\})$.
\end{reduction rule}

\begin{reduction rule}\label{rule:9}
Consider an instance $(G,I',C')$ of \me\ to which none of Reduction
Rules~\ref{rule:1} to \ref{rule:7} is applicable. If there exists a clique
$B \subseteq V'$ such that $N_{V'}[v] = B$ for every $v \in B$, and a set
$B' \subseteq C'$ such that $N_{C'}(v) = B'$ for every $v \in B$, then include
$B$ in $C'$. The resulting instance is $(G, I', C' \cup B)$.
\end{reduction rule}


\begin{reduction rule}\label{rule:10}
Consider an instance $(G,I',C')$ of \me. Suppose there exists an edge $uv$ in
$G[V']$ with $N_{C'}(u) \neq N_{C'}(v)$. If the set
$X = N_{V'}(u) \cap N_{V'}(v)$ is non-empty, then add $X$ to $C'$. The
resulting instance is $(G, I', C' \cup X)$.
\end{reduction rule}



\begin{reduction rule}\label{rule:11}
Consider an instance $(G,I',C')$ of \me\ to which none of Reduction
Rules~\ref{rule:1} to \ref{rule:10} is applicable. Suppose there exists a
connected component $M$ of $G[V']$ that induces a cycle of length at least $4$.
If $N(M) = \emptyset$, then select an arbitrary vertex $u \in M$ and add it to
$I'$; the resulting instance is $(G, I' \cup \{u\}, C')$. Otherwise, if
$N_{C'}(v)$ is the same clique in $C'$ for every $v \in M$, then return that it
is a trivial \No-instance.
\end{reduction rule}

\subsection{Algorithm}\label{subsec:algo}
In this subsection, we present the algorithm for \me. It is a branching algorithm that, for a given instance $(G,I',C')$ of \me, selects a vertex $v$ from $V' = V \setminus (I' \cup C')$ having the largest degree within $V'$, which we call as $V'$-degree and denote as $\deg_{V'}(v)$. The algorithm repeatedly applies the reduction rules whenever possible, and otherwise branches. The measure used for the analysis is $|V'|\leq n$.

We may assume that $G$ is connected, since otherwise the algorithm is applied to each component
separately, and that $G$ has at least $10$ vertices, since smaller instances are solved by brute force.
At the start of each case no reduction rule applies, so $I'$ is independent, $G[C']$ is a cluster
graph, no vertex of $V'$ has a neighbor in $I'$, and for every $u \in V'$ the set $N_{C'}(u)$ is either
empty or one clique of $C'$. We branch on a vertex of maximum $V'$-degree, distinguishing four cases according to that degree.




\textbf{Case 1: The maximum $V'$-degree is at least $5$.} Let $v \in V'$ be a vertex such that $|N_{V'}(v)| = x \ge 5$.
We branch on $v$ in two possible ways:

\begin{itemize}
    \item Add $v$ to $I'$. In this case, \cref{rule:3} would be applicable, leading to $N_{V'}(v)$ being added to $C'$. Consequently, at least $x + 1 \ge 6$ vertices are moved out of $V'$, and thus the measure decreases by at least $6$.

    \item Add $v$ to $C'$. In this case, the measure decreases by at least $1$.
\end{itemize}

Hence, the branching vector for this case is at least $(1, 6)$.

For Cases~2 to~4 we give only a high-level description here, highlighting the main branching ideas and
the resulting branching vectors; the detailed arguments are deferred to Appendix~\ref{appen:case34}.

\textbf{Case 2: The maximum $V'$-degree is exactly $4$.}
Let $v \in V'$ be a vertex with $\deg_{V'}(v) = 4$, and let
$N_{V'}(v) = \{u_1,u_2,u_3,u_4\}$. We branch on $v$ in two ways. If $v$ is added to $I'$, then by
\Cref{rule:3} all four vertices of $N_{V'}(v)$ are moved to $C'$, so the measure decreases by $5$. If
$v$ is added to $C'$, the measure decreases by at least $1$, and by at least $2$ if some reduction
rule becomes applicable afterwards, giving the branching vector $(5,2)$. If no reduction rule
applies, we branch further, according to the structure of $G[N_{V'}(v)]$.

If $N_{V'}(v)$ is an independent set, then each $u_i$ has a neighbor
$w_i \in V' \setminus N_{V'}[v]$, since otherwise $N(u_i) \subseteq C'$ after adding $v$ to $C'$ and
\Cref{rule:4} would apply to $u_i$. Branching on $u_1$ and applying \Cref{rule:3} and \Cref{rule:6}
gives the branching vector $(5,3,6)$.

If $N_{V'}(v)$ is a clique, then $\deg_{V'}(u_i) = 4$ for every $i$, so once $v$ is added to $C'$ we
have $N_{V'}[u_i] = N_{V'}(v)$. Let $D$ be the clique of $C'$ containing $v$. If $N_{C'}(u_i) \neq D$
for some $i$, then $N_{C'}(u_i)$ is not a clique of $C'$ and \Cref{rule:6} applies to $u_i$;
otherwise \Cref{rule:5} applies with $B = N_{V'}(v)$, $B' = D$ and witness $v$. In both cases a
reduction rule is applicable and the branching vector is $(5,2)$.

If the maximum clique size in $N_{V'}(v)$ is $3$, say $\{u_1,u_2,u_3\}$ is a triangle and
$u_1u_4 \notin E(G)$, then $u_4$ has a neighbor in $V' \setminus \{v\}$, again because
\Cref{rule:4} would apply otherwise. Branching on $u_4$ and applying \Cref{rule:3} and \Cref{rule:6}
gives the branching vector $(5,3,5)$.

If the maximum clique size in $N_{V'}(v)$ is $2$, say $u_1u_2 \in E(G)$ and $u_1u_3 \notin E(G)$,
then branching on $u_1$ and applying \Cref{rule:3}, \Cref{rule:4} and \Cref{rule:6} gives, depending
on the adjacencies of $u_2$ with $u_3$ and $u_4$, one of the branching vectors $(5,3,5)$, $(5,4,4)$
and $(5,4,5)$. In all subcases, the worst branching vector in Case~2 is $(5,3,5)$.

\textbf{Case 3: The maximum $V'$-degree is exactly $3$.}
Let $v \in V'$ be a vertex with $N_{V'}(v)=\{u_1,u_2,u_3\}$.
We branch on whether $v$ is added to $I'$ or to $C'$.
If $v$ is added to $I'$, then \Cref{rule:3} applies and forces all three neighbors
into $C'$, giving a decrease of $4$ in the measure.
If $v$ is added to $C'$, then either a reduction rule applies immediately,
leading to a decrease of at least $2$ (giving a branching vector $(4,2)$),
or no rule applies and we further branch based on the structure of $G[\{u_1,u_2,u_3\}]$.

When $\{u_1,u_2,u_3\}$ is an independent set, each $u_i$ must have a neighbor in
$V'$ outside $N_{V'}[v]$, otherwise \Cref{rule:4} would apply.
Branching on one of the $u_i$ and repeatedly applying \Cref{rule:3} and
\Cref{rule:6} guarantees branching vectors at least as good as $(4,4)$ or
$(4,3,7)$.

If $\{u_1,u_2,u_3\}$ forms a triangle, then after adding $v$ to $C'$ the set
$\{u_1,u_2,u_3\}$ is a connected component of $G[V']$, and either
\Cref{rule:9} or \Cref{rule:10} applies. The branching vector is $(4,2)$.

If $G[N_{V'}(v)]$ induces a $P_3$, or exactly one edge, then branching on an
appropriate vertex and applying \Cref{rule:3}, \Cref{rule:6},
and \Cref{rule:4} ensures that the measure decreases by at least $4$ in every
branch, giving branching vectors bounded by $(4,5,4)$ or $(4,4)$.
In all subcases, the worst branching vector in Case~3 is no worse than $(4,5,4)$.

\textbf{Case 4: The maximum $V'$-degree is at most $2$.}
In this case, every connected component of $G[V']$ is either a path or a cycle.
Let $A$ be any connected component of $G[V']$.

If $A$ is a singleton, then \Cref{rule:4} applies directly.
If $A$ consists of two vertices, then by \Cref{rule:7} both have non-empty
neighborhoods in $C'$, by \Cref{rule:6} these are cliques of $C'$, and by
\Cref{rule:9} they are distinct. Branching on one vertex and applying
\Cref{rule:3} and \Cref{rule:6} gives a branching vector of $(3,3)$.

If $A$ induces a path of exactly three vertices, branching on an internal vertex
and applying \Cref{rule:3} when it is added to $I'$ and \Cref{rule:4} when it is added to $C'$ again results in a
branching vector of $(3,3)$.

If $A$ induces a path $(u_1,u_2,u_3,u_4,...)$ of at least four vertices, we branch on an internal vertex $u_3$.
If $u_3$ is added to $I'$, then by \Cref{rule:3} and \Cref{rule:4}, $u_1,u_2, u_4$ are forced, giving a measure decrease of at least $4$.
If $u_3$ is added to $C'$, then depending on whether reduction rules apply to adjacent vertices, we obtain a decrease of at least $2$ or $3$.
In the remaining case, a further branching on $u_1$ ensures a stronger decrease, leading to vectors at least as good as $(4,5,4)$.
Overall, the branching vectors in this case are $(4,3)$, $(4,2)$, and $(5,4,4)$.

If $A$ induces a triangle, then either all three vertices have the same
neighborhood in $C'$, in which case \Cref{rule:9} applies, or two of these
neighborhoods differ, in which case \Cref{rule:10} applies to that edge.

Finally, if $A$ induces a cycle of length at least four, then either
\Cref{rule:11} applies directly, or we branch on a suitable vertex and apply
\Cref{rule:3} and \Cref{rule:6} in both branches, giving a branching vector of
$(3,3)$.

The branching vectors are summarized in \Cref{tab:branching-summary}.

\begin{theorem}\label{th:meexact}
    \me\ can be solved in $\Oh^*(1.29803^n)$ time.
\end{theorem}

\begin{corollary}\label{th:mrexact}
\mr\ and \lmp\ can be solved in $\Oh^*(1.29803^n)$ time.
\end{corollary}

\begin{proof}
\mr\ is the special case $I'=C'=\emptyset$ of \me, and \lmp\ is an equivalent
formulation~\cite{DBLP:conf/wg/PhilipS25}.\qed
\end{proof}

\begin{theorem}\label{th:mvdexact}
    \mvd\ can be solved exactly \\ in $\Oh^*(2.29803^n)$ time.
\end{theorem}

\begin{proof}
For each subset $S \subseteq V$ we test whether $G-S$ is monopolar using
\Cref{th:mrexact}, which takes $\Oh^*(1.29803^{n-|S|})$ time, and among all sets passing the test we
output one of minimum weight. The total running time is
\[
\sum_{i=0}^{n} \binom{n}{i} \cdot \Oh^*(1.29803^{\,n-i}) = \Oh^*(2.29803^n)
\]
by the binomial theorem.\qed
\end{proof}

\begin{table}[htbp]
\caption{Branching vectors of the four cases. In each row the worst vector, that is the one with the
largest branching number, is shown in bold, and the last column gives that number.}
\label{tab:branching-summary}
\centering
\renewcommand{\arraystretch}{1}
\setlength{\tabcolsep}{3pt}

\begin{tabular}{|
>{\raggedright\arraybackslash}p{3.4cm}|
>{\raggedright\arraybackslash}p{5.5cm}|
>{\centering\arraybackslash}p{2.4cm}|}
\hline
\textbf{Case} & \textbf{Branching Vectors} & \textbf{Worst Branching Number} \\ \hline

Case 1: Maximum $V'$-degree $\ge 5$ &
$\mathbf{(1,6)}$ &
$1.285199$ \\ \hline

Case 2: Maximum $V'$-degree $= 4$ &
$(5,2), (5,3,6), \mathbf{(5,3,5)}, (5,4,4), (5,4,5)$ &
$1.29803$ \\ \hline

Case 3: Maximum $V'$-degree $= 3$ &
$(4,2), (4,4), (4,3,7), \mathbf{(4,5,4)}$ &
$1.290649$ \\ \hline

Case 4: Maximum $V'$-degree $\le 2$ &
$(3,3), (4,3), (4,2), \mathbf{(4,5,4)}$ &
$1.290649$ \\ \hline
\end{tabular}

\end{table}

\section{FPT by Treewidth for Monopolar Extension}\label{sec:fptmr}
We present a dynamic programming algorithm on a nice tree decomposition
$\mathcal{T} = (T, \{{B}_t\}_{t \in V(T)})$ of $G$ of width $\tw$, which decides whether the given
pair $(\tilde{I},\tilde{C})$ extends to a monopolar partition $(I,C)$ of $G$. Processing $T$
bottom-up, we record for each bag which of its vertices lie in $I$, which lie in cliques of $C$ that
can still be extended by vertices of ancestor bags, and which lie in cliques of $C$ that cannot. This
three-way distinction gives the $3^{\tw}$ state space per bag.

For each node $t \in V(T)$ and every tripartition $(X_t, R_t, S_t)$ of the bag $B_t$, 
we compute a value $T[t, X_t, R_t, S_t] \in \{0,1\}$. We denote the graph induced by the vertices in the bags corresponding to the node $t$ and all of its descendants by $G_t$, and $V_t=V(G_t)$.
We set $T[t, X_t, R_t, S_t] = 1$ if and only if there exists a monopolar partition 
$(I_t, C_t)$ of the subgraph $G_t$ satisfying the following:

(1) $\tilde{I}\cap V_t\subseteq I_t$ and $\tilde{C}\cap V_t \subseteq C_t$; \quad
(2) $I_t \cap B_t = X_t$; \quad
(3) $C_t \cap B_t = R_t \cup S_t$;

(4) $R_t$ is the union of $C\cap B_t$ for all cliques $C$ in $C_t$ for which there exists a vertex
$u \in C \cap (V_t \setminus B_t)$.

\textbf{Recursion.} We now show how to compute $T[t, X_t, R_t, S_t]$ at each node $t$ in a nice tree decomposition. Throughout, we call a tripartition $(X_t,R_t,S_t)$ of $B_t$ \emph{valid} if $(X_t, R_t \cup S_t)$ is a
monopolar partition of $G[B_t]$, no connected component of $G[R_t \cup S_t]$ intersects both $R_t$ and
$S_t$, $\tilde{I}\cap B_t \subseteq X_t$ and $\tilde{C}\cap B_t \subseteq R_t \cup S_t$. At every node
we set $T[t,X_t,R_t,S_t]=0$ for every invalid tripartition, and the recursions below are stated for
valid ones.

\begin{itemize}\setlength\itemsep{0.5em}

    \item \textbf{Leaf Node.} We set $T[t,X_t,R_t,S_t] = 1$ if and only if $X_t = R_t = S_t = \emptyset$.

    \item \textbf{Introduce Node.} Let $v$ be the vertex introduced at node $t$, and let $t_1$ be its child.
    \[
    T[t,X_t,R_t,S_t] =
    \begin{cases}
        T[t_1,\, X_t \setminus \{v\},\, R_t,\, S_t], & \text{if } v \in X_t, \\[4pt]
        0, & \text{if } v \in R_t, \\[4pt]
        T[t_1,\, X_t,\, R_t,\, S_t \setminus \{v\}], & \text{if } v \in S_t.
    \end{cases}
    \]

    \item \textbf{Forget Node.} Let $v$ be the vertex forgotten at node $t$, and let $t_1$ be its child.

    \[
\begin{aligned}
T[t,X_t,R_t,S_t] 
&= T[t_1,\,X_t\cup\{v\},\,R_t,\,S_t] \\[3pt]
&\quad\lor\; [N_{Y_t}(v)\subseteq R_t]\; T[t_1,\,X_t,\,R_t\cup\{v\},\,S_t] \\[3pt]
&\quad\lor\; [N_{Y_t}(v)\subseteq R_t]\; T[t_1,\,X_t,\,R_t\setminus N_{Y_t}(v),\,S_t\cup N_{Y_t}[v]],
\end{aligned}
\]
 where $Y_t = R_t \cup S_t$ and by $[N_{Y_t}(v) \subseteq R_t]$ we mean we would consider the term only if $N_{Y_t}(v) \subseteq R_t$.

The three terms correspond to placing the forgotten vertex $v$ on the
independent side, or on the cluster side in a clique that already contains a
vertex of $V_{t_1} \setminus B_{t_1}$, or on the cluster side in a clique
contained in $B_{t_1}$; in the last case forgetting $v$ closes that clique, so
its bag part $N_{Y_t}(v)$ moves from the $S$-part to the $R$-part.

    \item \textbf{Join Node.} Let $t$ be a join node with children $t_1$ and $t_2$, and let
$(X_t, R_t, S_t)$ be a tripartition of $B_t$. Let $Q_1, \dots, Q_p$ be the
connected components of $G[R_t \cup S_t]$ that are contained in $R_t$, so that
$R_t = Q_1 \cup \cdots \cup Q_p$. We call a pair $(R^1_t, R^2_t)$ a \emph{split} of $R_t$ if $R^1_t$ and
$R^2_t$ are disjoint, $R^1_t \cup R^2_t = R_t$, and each of $R^1_t$ and
$R^2_t$ is a union of some of the sets $Q_1, \dots, Q_p$. Equivalently, a
split assigns each $Q_j$ to exactly one of $R^1_t$ and $R^2_t$, so there are
$2^p$ splits.
\[
  T[t, X_t, R_t, S_t] \;=\;
  \bigvee_{(R^1_t, R^2_t)}
  \Bigl( T[t_1, X_t, R^1_t, S_t \cup R^2_t]
   \;\wedge\;
   T[t_2, X_t, R^2_t, S_t \cup R^1_t] \Bigr),
\]
where the disjunction is over all splits $(R^1_t, R^2_t)$ of $R_t$.
A clique of a partial solution that meets $B_t$ may contain vertices of
$V_{t_1} \setminus B_t$ or of $V_{t_2} \setminus B_t$, but not of both, as
$B_t$ separates these two sets. The split records which of the two children
each such clique descends into; in the other child the same clique lies
entirely inside the bag, and therefore belongs to its $S$-part.

\end{itemize}

The proofs of correctness of the recursion for introduce, forget, and join nodes are deferred to Appendix~\ref{appen:fptmr}.

\begin{lemma}\label{lem:join-time}
Let $t$ be a join node with children $t_1$ and $t_2$. Given the tables of
$t_1$ and $t_2$, all entries $T[t, X_t, R_t, S_t]$ can be computed in time
$3^{|B_t|} \cdot n^{\mathcal{O}(1)}$.
\end{lemma}

\begin{proof}[Proof idea]
For a fixed $X_t$, the relevant tripartitions correspond to the subsets of the
$q \le |B_t \setminus X_t|$ connected components of $G[B_t \setminus X_t]$, and
the disjunction over splits is a subset convolution over these components,
computable for all $2^q$ subsets at once in time $2^q \cdot n^{\mathcal{O}(1)}$
by the algorithm of Björklund et al.~\cite{DBLP:conf/stoc/BjorklundHKK07}. Summing
$2^{|B_t \setminus X_t|}$ over all $X_t \subseteq B_t$ gives $3^{|B_t|}$. The
full proof is in Appendix~\ref{appen:join-time}.\qed
\end{proof}

\begin{theorem}\label{th:mrtw}
    For a given graph $G$, a pair of sets $(\tilde{I},\tilde{C})$, and a tree decomposition $\mathcal{T} = (T, \{{B}_t\}_{t \in V(T)})$ of width $\tw$, \me can be solved in time $3^{\tw}\cdot n^{\Oh(1)}$.
\end{theorem}

\begin{proof}[Proof idea]
We may assume that $\mathcal{T}$ is nice, as a nice tree decomposition of the same width can be
computed in polynomial time (see \Cref{lm:twnice}). We compute $T[t,X_t,R_t,S_t]$ for every node $t$
and every tripartition of $B_t$ using the recursions above, proceeding bottom-up along $T$. Since
$B_r=\emptyset$ and $G_r=G$ for the root $r$, the answer is \Yes\ if and only if
$T[r,\emptyset,\emptyset,\emptyset]=1$. Each bag admits at most $3^{\tw+1}$ tripartitions, entries at
leaf, introduce and forget nodes are computed in polynomial time from a constant number of entries of
the child, and join nodes are handled by \Cref{lem:join-time}. A full proof is in
Appendix~\ref{appenpf:mrtw}.\qed
\end{proof}

\begin{corollary}
Given a graph $G$ with a tree decomposition of width $\tw$, \mr\ and \lmp\ can be solved in
$3^{\tw}\cdot n^{\Oh(1)}$ time.
\end{corollary}

\section{FPT by Treewidth for Monopolar Vertex Deletion}\label{sec:fptmvd}
 We extend the algorithm of \Cref{sec:fptmr} by tracking, for each bag, which vertices belong to the
deletion set. This gives a fourth state per bag vertex and a $4^{\tw}$ state space. We use the
notation of \Cref{sec:fptmr} for $\mathcal{T}$, $G_t$ and $V_t$.

For every node $t \in V(T)$ and each partition $(A_t, X_t, R_t, S_t)$ of $B_t$,\\ let
$W[t, A_t, X_t, R_t, S_t]$ be the minimum weight of a monopolar vertex-deletion set $M_t$ for $G_t$
such that $G_t - M_t$ admits a monopolar partition $(I_t, C_t)$ with

(1) $M_t \cap B_t = A_t$; \quad (2) $I_t \cap B_t = X_t$; \quad (3) $C_t \cap B_t = R_t \cup S_t$; and
(4) $R_t$ is the union of $C \cap B_t$ over all cliques $C$ of $C_t$ containing a vertex of
$V_t \setminus B_t$. If no such set exists, we set $W[t, A_t, X_t, R_t, S_t] = \infty$.



\textbf{Recursion.} We now show how to compute $W[t, A_t, X_t, R_t, S_t]$ at each node $t$ in a nice tree decomposition. We call a partition $(A_t,X_t,R_t,S_t)$ of $B_t$ \emph{valid} if $(X_t, R_t \cup S_t)$ is a monopolar
partition of $G[B_t \setminus A_t]$ and no connected component of $G[R_t \cup S_t]$ intersects both
$R_t$ and $S_t$. At every node we set $W[t,A_t,X_t,R_t,S_t]=\infty$ for every invalid partition, and
the recurrences below are stated for valid ones.

\begin{itemize}\setlength\itemsep{0.5em}

    \item \textbf{Leaf Node.}
    We set $W[t,A_t,X_t,R_t,S_t]=0$ precisely when $A_t=X_t=R_t=S_t=\emptyset$; otherwise, $W[t,A_t,X_t,R_t,S_t]=\infty$.

    \item \textbf{Introduce Node.}
Let $v$ be the vertex introduced at $t$, and let $t_1$ be its child.
    \[
    W[t,A_t,X_t,R_t,S_t] =
    \begin{cases}
        W[t_1, A_t \setminus \{v\}, X_t, R_t, S_t] + w(v), & \text{if } v\in A_t, \\[4pt]
        W[t_1, A_t, X_t \setminus \{v\}, R_t, S_t], & \text{if } v \in X_t, \\[4pt]
        \infty, & \text{if } v \in R_t, \\[4pt]
        W[t_1, A_t, X_t, R_t, S_t \setminus \{v\}], & \text{if } v \in S_t.
    \end{cases}
    \]

    \item \textbf{Forget Node.} Let $v$ be the vertex forgotten at $t$, with child $t_1$.
    \[
    \begin{aligned}
    W[t,A_t,X_t,R_t,S_t]
    = \min \Big\{&
        W[t_1, A_t \cup \{v\}, X_t, R_t, S_t], W[t_1, A_t, X_t \cup \{v\}, R_t, S_t], \\[3pt]
        & [N_{Y_t}(v) \subseteq R_t]\; W[t_1, A_t, X_t, R_t \cup \{v\}, S_t], \\[3pt]
        & [N_{Y_t}(v) \subseteq R_t]\; W[t_1, A_t, X_t, R_t \setminus N_{Y_t}(v), S_t \cup N_{Y_t}[v]]
    \Big\},
    \end{aligned}
    \]
    where $Y_t = R_t \cup S_t$ and by $[N_{Y_t}(v) \subseteq R_t]$ we mean we would consider the term only if $N_{Y_t}(v) \subseteq R_t$.

The first term places the forgotten vertex $v$ in the deletion set; the remaining three are as in the forget node of \cref{sec:fptmr}.

\item \textbf{Join Node.} Let $t$ be a join node with children $t_1$ and $t_2$, and let
$(A_t, X_t, R_t, S_t)$ be a partition of $B_t$. As in \cref{sec:fptmr}, let
$Q_1, \dots, Q_p$ be the connected components of $G[R_t \cup S_t]$ contained
in $R_t$, and call $(R^1_t, R^2_t)$ a split of $R_t$ if $R^1_t$ and $R^2_t$
are disjoint, their union is $R_t$, and each is a union of some of
$Q_1, \dots, Q_p$.
\[
\begin{aligned}
W[t,A_t,X_t,R_t,S_t]
= \min_{(R^1_t, R^2_t)} \Bigl( & W[t_1, A_t, X_t, R^1_t, S_t \cup R^2_t] \\
& {} + W[t_2, A_t, X_t, R^2_t, S_t \cup R^1_t] \Bigr) - w(A_t),
\end{aligned}
\] 

where the minimum is over all splits $(R^1_t, R^2_t)$ of $R_t$. The term
$w(A_t)$ is subtracted because the deleted bag vertices are counted in both
children.

\end{itemize}

The proofs of correctness of the recursion for introduce, forget, and join nodes are deferred to Appendix~\ref{appen:fptmvd}.

\begin{lemma}\label{lem:join-time-mvd}
Let $t$ be a join node with children $t_1$ and $t_2$. Given the tables of
$t_1$ and $t_2$, all entries $W[t,A_t,X_t,R_t,S_t]$ can be computed in time
$5^{|B_t|}\cdot n^{\Oh(1)}$. If the weights are integers bounded by a
polynomial in $n$, they can be computed in time $4^{|B_t|}\cdot n^{\Oh(1)}$.
\end{lemma}

\begin{proof}[Proof idea]
For the first bound we evaluate the minimum in the recurrence directly: each
vertex of $B_t$ lies in one of $A_t$, $X_t$, $S_t$, $R^1_t$, $R^2_t$, so there
are at most $5^{|B_t|}$ pairs of a partition and a split. For the second bound
we proceed as in \cref{lem:join-time}, encoding an entry of value $a$ as the
monomial $x^a$, so that the minimum over splits becomes the lowest degree in a
subset convolution of polynomials. The full proof is in
Appendix~\ref{appen:join-time-mvd}.
\end{proof}

\begin{theorem}\label{th:mvdtw}
For a given graph $G$ with a weight function $w$ and a tree decomposition
$\mathcal{T} = (T, \{B_t\}_{t \in V(T)})$ of width $\tw$, \mvd can be solved in
time $5^{\tw}\cdot n^{\Oh(1)}$. If the weights are integers bounded by a
polynomial in $n$, then \mvd can be solved in time $4^{\tw}\cdot n^{\Oh(1)}$.
\end{theorem}

\begin{proof}[Proof idea] We use dynamic programming over a nice tree decomposition of width $\tw$,
where for each bag we store partial information describing a monopolar
partition together with a vertex-deletion set. Each vertex in a bag can be in
one of four states, giving at most $4^{\tw+1}$ states per bag. At leaf,
introduce and forget nodes each entry is computed in polynomial time from a
constant number of entries of the child; at join nodes we use
\cref{lem:join-time-mvd}. A full proof is in Appendix~\ref{appenpf:mvdtw}.\qed
\end{proof}

\section{Conclusion}\label{sec:con}
We presented an $\Oh^*(1.29803^n)$-time exact branching algorithm for \me, which also improves the
bounds for \mr, \lmp\ and \mvd, together with explicit
fixed-parameter algorithms with concrete single-exponential
dependence on treewidth for all four problems.

Several directions remain open. Since \mr\ is polynomial-time solvable on many graph classes, it would be natural to study parameterization by the distance to such classes. It also remains open whether the base of the exponent can be reduced further, or whether matching lower bounds can be established, for
instance under the Strong Exponential Time Hypothesis. Finally, the parameterized complexity of these
problems with respect to other structural parameters, such as clique-width or twin-width, is worth
investigating.

        









\bibliography{refer}

\newpage
\appendix
\section*{Appendix}

\section{Extended Preliminaries}\label{appensec:preliB}

We move some of the preliminaries to this part.
For a graph $G=(V,E)$, we denote the set of vertices by $V(G)$ and the set of edges by $E(G)$. 
We denote $|V (G)|$ by $n$, when the graph is clear from the context. For a set $S\subseteq V (G)$, the subgraph of $G$ {\em induced} by $S$ is denoted by $G[S]$ and defined as the subgraph of $G$ with the vertex set $S$ and the edge set $\{uv \in E(G) : u, v \in S\}.$ For any graph $G=(V,E)$ and $v\in V$, a vertex $u\in V$ is said to be a \textit{neighbor} of $v$ if $vu\in E(G)$. The \textit{open neighborhood} of $v$ is denoted by $N(v)$ and it is defined as $N(v)=\{u\in V : vu\in E(G)\}$. For a vertex $v\in V(G)$ and a set $S\subseteq V(G)$, by $N_S(u)$ we denote $\{v\in S \mid vu\in E(G)\}$. We denote the \textit{degree} of a vertex $v$ by $\deg(v)$ where $\deg(v)=|N(v)|$.

\begin{definition}[Parameterized problems and Parameterized algorithms]
    A \textit{parameterized problem} is a language $L \subseteq \Sigma^* \times \mathbb{N}$, where $\Sigma$ is a fixed, finite alphabet. For an instance
    $(x,k) \in \Sigma^* \times \mathbb{N}$, $k$ is called the \textit{parameter}.
    A parameterized problem $L \subseteq \Sigma^* \times \mathbb{N}$ is called \textit{fixed parameter tractable} (FPT) if there exists an algorithm $\mathcal{A}$ (called a \textit{fixed parameter algorithm}), a computable function $f: \mathbb{N} \rightarrow \mathbb{N}$, and a constant $c$ such that, given $(x, k) \in \Sigma^* \times \mathbb{N}$, the algorithm $\mathcal{A}$ correctly decides whether $(x, k) \in L$ in time bounded by $f(k) \cdot|(x, k)|^c$.
\end{definition}

 \begin{definition}[Tree decomposition and treewidth]\label{def:tdandtw}
    A \textit{tree decomposition} of a graph $G = (V,E)$ is a pair $(T, \mathcal{B})$ where $T$ is a tree and $\mathcal{B} = \{B_x \subseteq V : x \in T\}$ is a family of subsets of $V$ such that:
    \begin{enumerate}
        \item $\bigcup_{x \in T} B_x = V$.
        \item For every edge $e = \{u,v\} \in E$, there exists a node $x \in T$ such that $u,v \in B_x$.
        \item For every vertex $v \in V$, the set of nodes $\{x \in T : v \in B_x\}$ induces a connected subtree of $T$.
    \end{enumerate}
    Here $B_x \subseteq V$ is called a \textit{bag} of the node $x$ in $T$.
    The \textit{width} of a tree decomposition is $\max\limits_{x \in T} |B_x| - 1$.
    The \textit{treewidth} of a graph $G$ is the minimum width of a tree decomposition of $G$. We denote the \textit{treewidth} of a graph by $\tw$ only when the graph is clear from the context. 
\end{definition}

\begin{definition}[Nice tree decomposition]
    A tree decomposition $(T, \mathcal{B})$ of a graph $G$ is \textit{nice} if $T$ is rooted at an empty bag $r(T)$ and each of its bags is one of the following types:
    \begin{enumerate}
        \item \textit{Leaf node}: a leaf $x$ of $T$ with $B_x = \emptyset$ and no 
    children.
        \item \textit{Introduce node}: an inner node $x$ with one child $y$ such that $|B_x| = |B_y| + 1$ and $B_x \setminus B_y = \{u\}$ for some $u \in V$.
        \item \textit{Forget node}: an inner node $x$ with one child $y$ such that $|B_y| = |B_x| + 1$ and $B_y \setminus B_x = \{u\}$ for some $u \in V$.
        \item \textit{Join node}: an inner node $x$ with two children $y$ and $z$ such that $B_x = B_y = B_z$.
    \end{enumerate}
\end{definition}

The following lemma will be used. 
\begin{lemma}[Lemma 7.4 of \cite{DBLP:books/sp/CyganFKLMPPS15}]\label{lm:twnice}
    If a graph $G$ admits a tree decomposition of width at most $k$, then it also admits a nice tree decomposition of width at most $k$. Moreover, given a tree decomposition $\mathcal{T}=\left(T,\left\{X_t\right\}_{t \in V(T)}\right)$ of $G$ of width at most $k$, one can in time $\mathcal{O}\left(k^2 \cdot \max (|V(T)|,|V(G)|)\right)$ compute a nice tree decomposition of $G$ of width at most $k$ that has at most $\mathcal{O}(k|V(G)|)$ nodes.
\end{lemma}

\paragraph{Branching Algorithm Analysis.}
We need to define the notion of a branching tree for the analysis of our algorithms. 
An algorithm or a procedure $\mathcal{A}$ on an instance $I$ of a problem is called a branching algorithm if $\mathcal{A}$ can potentially recursively call itself on instances $I_1, I_2,\ldots, I_t$ for $t\geq 2$ such that output of $\mathcal{A}$ depends on the output of the recursive calls. A run of the algorithm on an instance $I$ is said to have executed a branching step, if it recursively calls itself on at least two instances. If the algorithm applies a reduction rule and makes a call to itself on only one instance, then we say that branching step is not executed. 

A branching tree associated with an algorithm $\mathcal{A}$ and an input instance $I$ is a tree where the root node of the tree is associated with the instance $I$. Further, if the algorithm on input $I$ makes a recursive call to instance $I'$ (which itself can be during a recursive call) without executing a branching step, then $I'$ is associated with the same node in the branching tree as the instance $I$ is. If on an instance $I$, branching step is executed and recursive call is made on the instances $I_1, I_2, \ldots, I_t$, then the node associated with $I$ in the branching tree would have $t$ children, each associated with one of the instances on which the recursive call was made. If there is no branching step executed on any of the instances associated with a node in the branching tree, then that node would be a leaf node in the branching tree. We also use the concept of {\em layers} in the branching tree, where layer $0$ consists of the root node of the branching tree and layer $i$ consists of the nodes which are at distance $i$ from the root in the branching tree.  

We define the branching tree in this way because application of a reduction rule will only take polynomial time in our case, and associating the instances obtained after applying a reduction rule with the same node in the branching tree will make the analysis easier. If there are multiple instances $I_1, I_2, \ldots, I_p$ associated with a node in a branching tree, then we use the term {\em first instance associated with that node} for the first instance that was associated with the node while executing the algorithm. In other words, first instance $I$ associated with a node in a branching tree is such that all the other instances are obtained by making successive recursive calls starting from $I$. We also define the {\em last instance associated with a node} in the branching tree to be an instance $I$ such that either no recursive call is made on $I$, or a branching step is executed when the algorithm is run on $I$. 

For the running-time analysis of a branching algorithm, we associate with each
instance $I$ a non-negative measure $\mu(I)$. Suppose that on an instance $I$ the
algorithm performs a branching step that results in $\ell$ recursive calls, and
that in the $i$-th recursive call the measure decreases by $a_i$. Then the vector
$(a_1,a_2,\ldots,a_\ell)$ is called the \emph{branching vector} of this branching
step. The \emph{branching number} of a branching vector $(a_1,a_2,\ldots,a_\ell)$
is defined as the unique positive real root of the equation
$\sum_{i=1}^{\ell} x^{-a_i} = 1$. Let $c$ denote the maximum branching number over all branching vectors that arise
in the algorithm. If the measure never becomes negative, then the total number of
nodes in the resulting branching tree is bounded by $\Oh(c^k)$, where $k$ is the
initial value of the measure. Furthermore, if each node of the branching tree can
be processed in polynomial time, then the overall running time of the algorithm
is $\Oh^*(c^k)$.
For a more in-depth introduction to branching algorithms, one can see chapter 3 of \cite{DBLP:books/sp/CyganFKLMPPS15}. 

Next we state three problems formally here.

\vspace{0.2cm}
\noindent\fbox{
	\begin{minipage}{.95\textwidth}
		\begin{tabular*}{\textwidth}{@{\extracolsep{\fill}}lr}
			{{\sc Monopolar Extension}} &
		\end{tabular*}
		{\bf Input:} A graph $G$ and two disjoint vertex subsets $I' \subseteq V$ and $C' \subseteq V$.\\
		{\bf Question:} Does there exist a monopolar partition $(I,C)$ such that $I'\subseteq I$ and $C'\subseteq C$?
	\end{minipage}
}

\vspace{0.2cm}
\noindent\fbox{
	\begin{minipage}{.95\textwidth}
		\begin{tabular*}{\textwidth}{@{\extracolsep{\fill}}lr}
			{{\sc List Monopolar Partition}} &
		\end{tabular*}
		{\bf Input:} An undirected graph $G=(V,E)$ on $n$ vertices, together with a list function $L$ defined on $V$ such that $\emptyset \neq L(v) \subseteq \{{I},{C}\}$ for every $v \in V$. \\
		{\bf Question:} Does there exist a monopolar partition of $G$ that respects the list function $L$?
	\end{minipage}
}
\vspace{0.2cm}

\vspace{0.2cm}
\noindent\fbox{
	\begin{minipage}{.95\textwidth}
		\begin{tabular*}{\textwidth}{@{\extracolsep{\fill}}lr}
			{{\sc Monopolar Vertex Deletion}} &
		\end{tabular*}
		{\bf Input:} A graph $G=(V,E)$, a weight function $w: V\rightarrow {\mathbb{Q}_{\ge 0}}$ and a positive rational $W$. \\
		{\bf Question:} Does there exist a vertex set $S$ of weight at most $W$ such that $G-S$ is monopolar?
	\end{minipage}
}
 \vspace{0.2cm}

\section{Correctness Proofs for the Reduction Rules of \Cref{sec:exactalgo}}\label{appen:proofrr}

We begin with a simple observation that will be repeatedly used in the
correctness proofs of the reduction rules.

\begin{observation}\label{obs:1}
If either instance $(G,I', C' \cup S)$ or $(G, I' \cup S, C')$ is a
\Yes-instance for some subset $S \subseteq V'$, where
$V' = V \setminus (I' \cup C')$, then the instance $(G, I', C')$ is also
a \Yes-instance.
\end{observation}

Throughout the proofs of the reduction rules, we implicitly use this
observation and assume that $V' = V \setminus (C' \cup I')$.

\begin{lemma}
    \cref{rule:3} is correct.
\end{lemma}
\begin{proof}
        By \cref{obs:1}, if $(G, I', C' \cup N_{V'}(v))$ is a \Yes-instance, then
    $(G, I', C')$ is also a \Yes-instance. For the converse, note that if the instance $(G, I', C')$ is a \Yes-instance, then there is a monopolar partition $(I,C)$ such that $C'\subseteq C$ and $I'\subseteq I$. Let $w$ be a vertex in $N_{V'}(v)$ for some $v\in I'$. Then, we cannot have $w\in I$, as in that case $I$ would no longer be an independent set. Thus, if $(G, I', C')$ is a \Yes-instance, then $N_{V'}(v)\subseteq C$ and hence our reduction rule is correct. 
\end{proof}

\begin{lemma}
    \cref{rule:4} is correct.
\end{lemma}
\begin{proof}
   By \cref{obs:1}, if $(G, I' \cup \{v\}, C')$ is a \Yes-instance,
then so is $(G,I',C')$. For the converse, let $(I,C)$ be a monopolar partition
extending $(I',C')$. If $v \in I$, the claim is immediate, so assume $v \in C$.
Since $N_{V'}(v)$ and $N_{I'}(v)$ are empty, we have $N(v) \subseteq C'
\subseteq C$, and $I \cap C = \emptyset$ gives $N_I(v) = \emptyset$. Hence
$I \cup \{v\}$ is an independent set, and $C \setminus \{v\}$ induces a cluster
graph because the class of cluster graphs is hereditary. Therefore
$(I \cup \{v\}, C \setminus \{v\})$ is a monopolar partition of $G$ extending
$(I' \cup \{v\}, C')$.
\end{proof}

\begin{lemma}
    \cref{rule:6} is correct.
\end{lemma}

\begin{proof}
    By \cref{obs:1}, if $(G, I'\cup\{v\}, C')$ is a \Yes-instance, then so is
$(G,I',C')$. For the converse, it suffices to show that $v\in I$ for every
monopolar partition $(I,C)$ extending $(I',C')$. Suppose $v \in C$, and let $K$
be the clique of $C$ containing $v$. Fix $y \in N_{C'}(v)$, which exists as
$N_{C'}(v) \neq \emptyset$, and let $B'$ be the clique of $C'$ containing $y$.
As $B'$ is a clique of $G$ and $B' \subseteq C' \subseteq C$, the set $B'$ is
contained in the clique of $C$ containing $y$, which is $K$ since $y$ and $v$
are adjacent. Every vertex of $B'$ therefore lies in $K$ and is adjacent to
$v$, so $B' \subseteq N_{C'}(v)$.
Since $B'$ is a clique of $C'$ while $N_{C'}(v)$ is not, we have
$N_{C'}(v) \neq B'$, and hence there is a vertex $y' \in N_{C'}(v) \setminus B'$.
Then $y' \in K$ as well, but $y$ and $y'$ lie in distinct cliques of the
cluster graph $G[C']$ and are therefore non-adjacent, contradicting that $K$ is
a clique. Hence $v \in I$, and adding $v$ to $I'$ is safe.
\end{proof}

\begin{lemma}
    \cref{rule:5} is correct.
\end{lemma}
\begin{proof}
  By \cref{obs:1}, if $(G,I',C' \cup \{v\})$ is a \Yes-instance,
then so is $(G,I',C')$. For the converse, let $(I,C)$ be a monopolar partition
extending $(I',C')$. If $v \in C$ the claim is immediate, so assume $v \in I$.
As $v \in I$ and $I$ is an independent set, $v$ has no neighbor in $I' \subseteq I$,
that is, $N_{I'}(v) = \emptyset$, and $N(v) \subseteq C$.

Since $w \in B' = \bigcap_{u \in B} N_{C'}(u)$, every $u \in B$ has a neighbor
in $C'$, since \cref{rule:6} is not applicable to $u$ and $N_{C'}(u)\neq \emptyset$, the set $N_{C'}(u)$ is a clique of $C'$; as it contains $w$, it is the clique of $C'$ containing $w$. Hence
$N_{C'}(u) = B'$ for all $u \in B$, and therefore $B \cup B'$ is a clique of
$G$ with $N[v] \subseteq B \cup B'$.
The vertex $w$ may have neighbors in $I'$, but these vertices do not
belong to $C$. Since $N_{V'}(w)=B$ and $N_{C'}(w)\subseteq B'$,
we therefore have $N[w]\cap C\subseteq B\cup B'$.

Let $K$ be the clique of $C$ containing $w$. Then
$K\subseteq N[w]\cap C\subseteq B\cup B'\subseteq N[v]$,
so $K\cup\{v\}$ is a clique.
Moreover $N(v) \subseteq B \cup B'$ meets no clique of $C$ other than $K$,
since $B' \subseteq K$ and every vertex of $B \setminus \{v\}$ is a neighbor
of $v \in I$, hence lies in $C$, and is adjacent to $w \in K$. Therefore
$(I \setminus \{v\},\, C \cup \{v\})$ is a monopolar partition extending
$(I', C' \cup \{v\})$.
\end{proof}



\begin{lemma}
    \cref{rule:7} is correct.
\end{lemma}
\begin{proof}
From \cref{obs:1}, if $(G, I', C' \cup \{u\})$ is a \Yes-instance, then
$(G, I', C')$ must also be a \Yes-instance. For the converse, assume that
$(G, I', C')$ is a \Yes-instance, and let $(I,C)$ be a monopolar partition
extending $(I', C')$. If $u \in C$, the claim holds trivially. Otherwise,
suppose $u \in I$; then $v \in C$, as $u$ and $v$ are adjacent.

First, we consider the case $N_{C'}(u)$ is empty. Then $N(u) = \{v\}$. Since the
only neighbor of $v$ in $V'$ is $u$, and $N_{I'}(v)$ is empty, $v$ has no
neighbor in $I \setminus \{u\}$; hence $(I \setminus \{u\}) \cup \{v\}$ is
still an independent set and it extends $I'$. On the other hand, in
$(C \setminus \{v\}) \cup \{u\}$ the vertex $u$ has no neighbor at all, so it
forms a clique of its own, and every other clique of $C$ is unchanged or loses
$v$. Thus $(C \setminus \{v\}) \cup \{u\}$ is a cluster extending
$C' \cup \{u\}$, and the two sets form a monopolar partition of $G$.

Next, we consider the case $N_{C'}(u) = B'$ is non-empty. Let $K$ be the clique
of $G[C]$ containing $w$. The vertices of $B'$ are pairwise adjacent and lie in
$C$, so $B' \subseteq K$. Conversely, every vertex of $K$ equals $w$ or is
adjacent to $w$; hence a vertex of $K \cap C'$ lies in the clique of $C'$
containing $w$, that is in $B'$, and a vertex of $K \cap V'$ lies in
$N_{V'}(w) \subseteq \{u,v\}$ and therefore equals $v$, as $u \in I$. Thus
$K$ is either $B'$ or $B' \cup \{v\}$.

If $K = B' \cup \{v\}$, then $N(u) = B' \cup \{v\} = K$, so $K \cup \{u\}$ is a
clique and $u$ has no neighbor in any other clique of $C$. Hence
$(I \setminus \{u\}, C \cup \{u\})$ is a monopolar partition extending
$(I', C' \cup \{u\})$.

If $K = B'$, then $v$ belongs to a different clique $C_1$ of $G[C]$. Consider
the partition $((I \setminus \{u\}) \cup \{v\}, (C \setminus \{v\}) \cup \{u\})$.
As in the first case, $v$ has no neighbor in $I \setminus \{u\}$, so the first
set is independent. In the second set, $u$ is adjacent exactly to the vertices
of $B' = K$, so $K \cup \{u\}$ is a clique, while $C_1 \setminus \{v\}$ remains
a clique and all other cliques of $C$ are unchanged. Hence this is a monopolar
partition extending $(I', C' \cup \{u\})$.

In all cases $(G, I', C' \cup \{u\})$ is a \Yes-instance, and the reduction
rule is correct.
\end{proof}


\begin{lemma}
    \Cref{rule:9} is correct.
\end{lemma}
\begin{proof}
         By \cref{obs:1}, it suffices to show that if $(G,I',C')$ is a
\Yes-instance, then so is $(G,I',C' \cup B)$. Let $(I,C)$ be a monopolar
partition extending $(I',C')$.

First, $|B| \ge 2$: if $B = \{u\}$, then $N_{V'}(u) = \emptyset$, and
$N_{I'}(u) = \emptyset$ as well, since otherwise \cref{rule:3}
would be applicable; but then $N(u) \subseteq C'$ and \cref{rule:4} would be applicable, contrary to the guard.

Since $B$ is a clique, at most one vertex of $B$ lies in $I$. If
$B \cap I = \emptyset$, then $B \subseteq C$ and we are done, so let
$B \cap I = \{u\}$ and fix $u' \in B \setminus \{u\}$. As $u \in I$ we have
$N_{I'}(u) = \emptyset$, hence
$N(u) = (B \setminus \{u\}) \cup B' \subseteq C$.

The set $(B \setminus \{u\}) \cup B'$ is a clique of $G$: the vertices of
$B \setminus \{u\}$ are pairwise adjacent, those of $B'$ are pairwise adjacent,
and every vertex of $B \setminus \{u\}$ is adjacent to all of $B'$ because
$N_{C'}(u'') = B'$ for every $u'' \in B$. Hence it lies in a single clique $K$
of $C$, and $N(u) \subseteq K$.

Conversely, let $z \in K$. Then $z$ is adjacent to $u' \in K$ or $z = u'$. If
$z \in V'$, then $z \in N_{V'}[u'] = B$; if $z \in C'$, then
$z \in N_{C'}(u') = B'$; and $z \notin I'$ since $K \subseteq C$. Hence
$K \subseteq B \cup B' \subseteq N[u]$.

Therefore $K \cup \{u\}$ is a clique and $N(u)$ meets no clique of $C$ other
than $K$, so $(I \setminus \{u\}, C \cup \{u\})$ is a monopolar partition
extending $(I', C' \cup B)$.
\end{proof}

\begin{lemma}
    \Cref{rule:10} is correct.
\end{lemma}

\begin{proof}
   By \cref{obs:1}, it is enough to show that if $(G,I',C')$ is a
\Yes-instance, then so is $(G,I',C' \cup X)$. Let $(I,C)$ be a monopolar
partition extending $(I',C')$.

We first show that $u$ and $v$ cannot both lie in $C$. Suppose they do. Since
$uv$ is an edge, they lie in the same clique $K$ of $C$. Every neighbor of $u$
in $C'$ lies in $C$ and is adjacent to $u$, so it lies in $K$; being in $K$, it
is also adjacent to $v$. This gives $N_{C'}(u) \subseteq N_{C'}(v)$, and by the
same argument with the roles of $u$ and $v$ exchanged,
$N_{C'}(v) \subseteq N_{C'}(u)$. So $N_{C'}(u) = N_{C'}(v)$, contradicting the
hypothesis.

Hence at least one of $u$ and $v$ lies in $I$. All neighbors of a vertex of
$I$ lie in $C$, and every vertex of $X$ is a neighbor of both $u$ and $v$, so
$X \subseteq C$. Therefore $(I,C)$ is a monopolar partition extending
$(I', C' \cup X)$. 
\end{proof}

\begin{lemma}
    \Cref{rule:11} is correct.
\end{lemma}

\begin{proof}
    We consider the two cases of the rule separately.

\medskip
\noindent\emph{Case 1: $N(M) = \emptyset$.} Here $M$ is a connected component
of $G$, so no vertex of $M$ has a neighbor outside $M$. By
Observation~\ref{obs:1}, it is enough to show that if $(G,I',C')$ is a
\Yes-instance, then so is $(G, I' \cup \{u\}, C')$.

We first partition $M$ on its own, starting at $u$. Walk along the cycle from
$u$ and cut it into consecutive blocks, each of size $2$ or $3$; this is
possible because every integer at least $4$ is a sum of $2$'s and $3$'s. Put
the first vertex of every block into $I_M$ and the remaining one or two
vertices of that block into $C_M$. Then $u \in I_M$, any two vertices of $I_M$
are at distance at least $2$ on the cycle and hence non-adjacent, so $I_M$ is
an independent set. Each block leaves a single vertex or a single edge in
$C_M$, and the vertex right after a block belongs to $I_M$, so these pieces are
pairwise non-adjacent and $C_M$ induces a cluster graph.

Now take any monopolar partition $(I,C)$ extending $(I',C')$ and replace its
restriction to $M$ by $(I_M, C_M)$. As $M$ has no neighbors outside itself,
this does not affect the rest of the partition, so the result is again a
monopolar partition of $G$, and it extends $(I' \cup \{u\}, C')$.

\medskip
\noindent\emph{Case 2: $N_{C'}(v)$ is the same clique $B'$ for every
$v \in M$.} Since Reduction Rule~\ref{rule:3} is not applicable, no vertex of
$V'$ has a neighbor in $I'$, so all neighbors of $M$ lie in $C'$; as
$N(M) \neq \emptyset$, the clique $B'$ is non-empty. Fix a vertex $y \in B'$,
which is adjacent to every vertex of $M$.

Suppose $(I,C)$ is a monopolar partition extending $(I',C')$. Every vertex of
$M$ that lies in $C$ is adjacent to $y$, so it lies in the same clique of $C$
as $y$. Hence any two vertices of $M$ lying in $C$ are adjacent. A cycle of
length at least $4$ has no triangle, so at most two vertices of $M$ lie in $C$,
and if there are two, they are consecutive on the cycle. Removing at most two
consecutive vertices from a cycle of length at least $4$ leaves at least two
consecutive vertices, which are adjacent and both lie in $I$. This contradicts
$I$ being an independent set, so no such partition exists and $(G,I',C')$ is a
\No-instance.
\end{proof}

\section{Detailed Branching Arguments for Case~2, Case~3, and Case~4 of \Cref{subsec:algo}}\label{appen:case34}

\noindent\textbf{Case 2: The maximum $V'$-degree is exactly $4$.}
Let $v \in V'$ be a vertex with $\deg_{V'}(v) = 4$, and let $N_{V'}(v) = \{u_1, u_2, u_3, u_4\}$. We branch as follows:

\begin{itemize}
\item Add $v$ to $I'$. Then, by \Cref{rule:3}, all vertices in $N_{V'}(v)$ are moved to $C'$. Consequently, the measure decreases by $5$.

\item Add $v$ to $C'$.  
If any reduction rule becomes applicable afterwards, the measure decreases by at least $2$, giving a branching vector of $(5,2)$.  
If no reduction rule applies after moving $v$ to $C'$, the measure still decreases by at least $1$. In this case, we further analyze the structure of $N_{V'}(v)$ to obtain a stronger decrease.

\begin{enumerate}
    \item \textbf{$\{u_1,u_2,u_3,u_4\}$ forms an independent set.}
    Each $u_i$ must have a neighbor $w_i \in V' \setminus \{v,u_1,u_2,u_3,u_4\}$.
    Indeed, since \Cref{rule:3} is not applicable, no vertex of $V'$ has a
    neighbor in $I'$; so if $u_i$ had no such neighbor, then after adding $v$
    to $C'$ we would have $N(u_i) \subseteq C'$ and \Cref{rule:4} would apply
    to $u_i$. After adding $v$ to $C'$, we branch as follows:
    \begin{itemize}
        \item Add $u_1$ to $I'$. Then, by \Cref{rule:3}, $w_1$ moves to $C'$,
        and the measure decreases by at least $3$.
        \item Add $u_1$ to $C'$. Then $u_1$ and $v$ are adjacent and both lie
        in $C'$, so they belong to the same clique of $C'$, while $u_2,u_3,u_4$
        are adjacent to $v$ but not to $u_1$; hence by \Cref{rule:6} all three
        move to $I'$. Also $w_1$ is adjacent to $u_1$ but not to $v$, so by the
        same argument $w_1$ moves to $I'$. The measure decreases by $6$.
    \end{itemize}
    Therefore, when $\{u_1,u_2,u_3,u_4\}$ is independent, the branching vector is $(5,3,6)$.

    \item \textbf{$\{u_1,u_2,u_3,u_4\}$ forms a clique.}
Each $u_i$ is adjacent to $v$ and to the three other vertices of the clique,
so $\deg_{V'}(u_i) = 4$ before $v$ is added to $C'$. As the maximum
$V'$-degree is $4$ in this case, no $u_i$ has a neighbor in $V'$ outside
$N_{V'}[v]$, and hence $N_{V'}[u_i] = \{u_1,u_2,u_3,u_4\}$ for every $i$ once
$v$ is added to $C'$.

Let $D$ be the clique of $C'$ containing $v$. If some $u_i$ has a neighbor in
a clique of $C'$ other than $D$, or $N_{C'}(u_i)$ is a non-empty proper subset
of $D$, then $N_{C'}(u_i)$ is not a clique of $C'$ and \Cref{rule:6} applies to
$u_i$. Otherwise $N_{C'}(u_i) = D$ for every $i$, and \Cref{rule:5} applies
with $B = \{u_1,u_2,u_3,u_4\}$, $B' = \bigcap_{i} N_{C'}(u_i) = D$ and witness
$v$, since $v \in D$ and $N_{V'}(v) = B$. In both cases a reduction rule is
applicable after $v$ is added to $C'$, so the measure decreases by at least
$2$. Hence, when $\{u_1,u_2,u_3,u_4\}$ is a clique, the branching vector is at
least $(5,2)$.

    \item \textbf{The maximum clique size in $\{u_1,u_2,u_3,u_4\}$ is $3$.}
Let $\{u_1, u_2, u_3\}$ form a triangle. Since $\{u_1,u_2,u_3,u_4\}$ is not a clique, there must exist at least one non-adjacent pair of vertices. Without loss of generality, assume that $u_1$ and $u_4$ are non-adjacent.
Vertex $u_4$ must have a neighbor $w_4 \in V' \setminus \{v\}$. Indeed, since
\Cref{rule:3} is not applicable, no vertex of $V'$ has a neighbor in $I'$; so
if $u_4$ had no neighbor in $V' \setminus \{v\}$, then after adding $v$ to
$C'$ we would have $N(u_4) \subseteq C'$ and \Cref{rule:4} would apply to
$u_4$. We then proceed as follows:
\begin{itemize}
\item Add $u_4$ to $I'$. Then, by \Cref{rule:3}, $w_4$ is added to $C'$, leading to a decrease of at least $3$ in the measure.
\item Add $u_4$ to $C'$. Since $v$ and $u_4$ are adjacent and both lie in
$C'$, they belong to the same clique of $C'$. The vertex $u_1$ is adjacent to
$v$ but not to $u_4$, so $N_{C'}(u_1)$ is a non-empty proper subset of that
clique, and by \Cref{rule:6} the vertex $u_1$ moves to $I'$. Then $u_2$ and
$u_3$ are neighbors of $u_1 \in I'$, so by \Cref{rule:3} they are added to
$C'$. The overall decrease in the measure is at least $5$.
\end{itemize}
Hence, in this case, the branching vector is $(5,3,5)$.

\item \textbf{The maximum clique size in $\{u_1,u_2,u_3,u_4\}$ is $2$.}
Here, there must exist an edge $u_iu_j$ among $\{u_1,u_2,u_3,u_4\}$ such that $u_i$ is non-adjacent to some other vertex $u_k$, for distinct $i,j,k \in \{1,2,3,4\}$.
Without loss of generality, assume $u_1u_2\in E(G)$ but $u_1$ and $u_3$ are not adjacent. We branch as follows:
\begin{itemize}
\item Add $u_1$ to $I'$. Then, by \Cref{rule:3}, $u_2$ is added to $C'$.
If there exists a vertex $u_i$ (for some $i \in \{3,4\}$) that is not adjacent to $u_2$, then by \Cref{rule:6}, $u_i$ is added to $I'$.
Therefore, if such a $u_i$ exists, the measure decreases by at least $4$; otherwise, the decrease is at least $3$. So note that the decrease is $3$ only if both $u_3$ and $u_4$ are adjacent to
$u_2$, which is the situation of subcase~(a) below; in the other three
subcases the decrease is at least $4$.
\item Add $u_1$ to $C'$. Since $v$ and $u_1$ are adjacent and both lie in $C'$,
they belong to the same clique of $C'$. The vertex $u_3$ is adjacent to $v$ but
not to $u_1$, so $N_{C'}(u_3)$ is a non-empty proper subset of that clique, and
by \Cref{rule:6} the vertex $u_3$ moves to $I'$. We now consider the following
subcases, according to the adjacencies of $u_2$ with $u_3$ and $u_4$.
\begin{enumerate}
\item If $u_2u_3 \in E(G)$ and $u_2u_4 \in E(G)$, then $u_2$ is a neighbor of
$u_3 \in I'$, so by \Cref{rule:3} it is added to $C'$. Since $u_1u_4 \notin
E(G)$, the same argument as for $u_3$ shows that \Cref{rule:6} moves $u_4$ to
$I'$. Together with $v$, the vertices $u_1, u_2, u_3, u_4$ leave $V'$, so the
measure decreases by $5$, and the branching vector is $(5,3,5)$.

\item If $u_2u_3 \in E(G)$ but $u_2u_4 \notin E(G)$, then again $u_2$ is a
neighbor of $u_3 \in I'$, so by \Cref{rule:3} it is added to $C'$. The measure
decreases by $4$, and the branching vector is $(5,4,4)$.

\item If $u_2u_3 \notin E(G)$ and $u_2u_4 \in E(G)$, then $u_1u_4 \notin E(G)$:
otherwise $\{u_1,u_2,u_4\}$ would be a clique of size $3$ in
$\{u_1,u_2,u_3,u_4\}$, contradicting that the maximum clique size here is $2$.
As for $u_3$, the vertex $u_4$ is adjacent to $v$ but not to $u_1$, so
\Cref{rule:6} moves $u_4$ to $I'$. Then $u_2$ is a neighbor of $u_4 \in I'$,
so by \Cref{rule:3} it is added to $C'$. The measure decreases by $5$, and the
branching vector is $(5,4,5)$.

\item If $u_2u_3 \notin E(G)$ and $u_2u_4 \notin E(G)$, then $u_3$ must have a
neighbor $w_3 \in V' \setminus \{v\}$. Indeed, since \Cref{rule:3} is not
applicable, no vertex of $V'$ has a neighbor in $I'$; so if $u_3$ had no
neighbor in $V' \setminus \{v\}$, then after adding $v$ to $C'$ we would have
$N(u_3) \subseteq C'$ and \Cref{rule:4} would apply to $u_3$. By \Cref{rule:3},
the vertex $w_3$ is added to $C'$, and the measure decreases by $4$. Hence the
branching vector in this subcase is $(5,4,4)$.
\end{enumerate}
\end{itemize}
Therefore, for this case, the possible branching vectors are $(5,3,5)$,
$(5,4,4)$ and $(5,4,5)$.
\end{enumerate}
\end{itemize}

\paragraph{\bf Case 3: The maximum $V'$-degree is exactly $3$.}
Let $v \in V'$ such that $|N_{V'}(v)| = 3$, and let $N_{V'}(v) = \{u_1, u_2, u_3\}$. We branch as follows:

\begin{itemize}
    \item Add $v$ to $I'$. Now, \cref{rule:3} is applicable and it adds $N_{V'}(v)$ to $C'$, resulting in a decrease of $4$ in the measure.

    \item Add $v$ to $C'$. After this, if any of the reduction rules are applicable, then the measure would decrease by at least two, and the branching vector would be $(4,2)$.

    In the case where none of the reduction rules applies after moving $v$ to $C'$, the measure decreases by at least one. However, we aim to further decrease the measure based on the following structural cases.

    \begin{enumerate}
                \item \textbf{$\{u_1, u_2, u_3\}$ is independent.}
        For each $1 \le i \le 3$, the vertex $u_i$ must have a neighbor
        $w_i \in V'$ outside $N_{V'}[v]$. Indeed, since \Cref{rule:3} is not
        applicable, no vertex of $V'$ has a neighbor in $I'$; so if $u_i$ had
        no such neighbor, then after adding $v$ to $C'$ we would have
        $N(u_i) \subseteq C'$, as $u_1,u_2,u_3$ are pairwise non-adjacent, and
        \Cref{rule:4} would apply to $u_i$.

        Note also that at most one of $u_1,u_2,u_3$ can be added to $C'$ once
        $v$ is in $C'$: two of them would be adjacent to $v \in C'$ and hence
        lie in the same clique of $C'$, while $u_1,u_2,u_3$ are pairwise
        non-adjacent.

        \noindent
        After adding $v$ to $C'$, we branch according to the following cases.

        \begin{enumerate}
            \item {\bf Any two of $w_1,w_2,w_3$ are same.} Without loss of
            generality, let us assume $w_1=w_2=w$. Then branch as follows:
            \begin{itemize}
                \item Add $u_1$ to $I'$. Then, by \Cref{rule:3}, $w$ is added
                to $C'$. Now $v$ and $w$ both lie in $C'$ and are
                non-adjacent, so they belong to distinct cliques of $C'$,
                while $u_2$ is adjacent to both; hence by \Cref{rule:6} the
                vertex $u_2$ moves to $I'$. Thus, assuming $v$ is already in
                $C'$, the measure decreases by at least $4$.
                \item Add $u_1$ to $C'$. Then $u_1$ and $v$ are adjacent and
                both lie in $C'$, so they belong to the same clique of $C'$,
                while $u_2$ is adjacent to $v$ but not to $u_1$; hence by
                \Cref{rule:6} the vertex $u_2$ moves to $I'$. As $w$ is
                adjacent to $u_1 \in C'$ but not to $v$, the same argument puts
                $w$ into $I'$. Since $u_2$ and $w$ are adjacent, the set $I'$
                is no longer independent, so \Cref{rule:1} returns a trivial
                \No-instance and this branch is discarded.
            \end{itemize}
            Thus in this case, the branching vector is $(4,4)$.

            \item {\bf All of $w_1,w_2,w_3$ are different.} We branch as
            follows:
            \begin{itemize}
            \item Add $u_1$ to $I'$. Then, by \Cref{rule:3}, $w_1$ is added to
            $C'$. Thus, assuming $v$ is already in $C'$, the measure decreases
            by at least $3$.
            \item Add $u_1$ to $C'$. Then $u_1$ and $v$ belong to the same
            clique of $C'$, while $u_2$ and $u_3$ are adjacent to $v$ but not
            to $u_1$; hence by \Cref{rule:6} both move to $I'$. Also $w_1$ is
            adjacent to $u_1$ but not to $v$, so by the same argument $w_1$
            moves to $I'$. Then $w_2$ and $w_3$ are neighbors of $u_2$ and
            $u_3$ in $I'$, so by \Cref{rule:3} they are added to $C'$.
            Counting $v$, the vertices $u_1,u_2,u_3,w_1,w_2,w_3$ leave $V'$
            and the measure decreases by $7$.
        \end{itemize}
        Thus, the branching vector in this case is $(4,3,7)$.
        \end{enumerate}
        Hence, when $\{u_1, u_2, u_3\}$ is an independent set, the branching
        vector is either $(4,4)$ or $(4, 3, 7)$.

        \item \textbf{$\{u_1, u_2, u_3\}$ forms a triangle.}
Each $u_i$ is adjacent to $v$ and to the two other vertices of the triangle,
so $\deg_{V'}(u_i) = 3$ before $v$ is added to $C'$. As the maximum
$V'$-degree is $3$ in this case, no $u_i$ has a neighbor in $V'$ outside
$N_{V'}[v]$. Hence, after adding $v$ to $C'$, we have
$N_{V'}[u_i] = \{u_1,u_2,u_3\}$ for every $i$, so $\{u_1,u_2,u_3\}$ is a
connected component of $G[V']$.

If $N_{C'}(u_1) = N_{C'}(u_2) = N_{C'}(u_3)$, then \Cref{rule:9} applies with
$B = \{u_1,u_2,u_3\}$. Otherwise two of these sets differ, say
$N_{C'}(u_1) \neq N_{C'}(u_2)$, and \Cref{rule:10} applies to the edge
$u_1u_2$, whose common $V'$-neighborhood contains $u_3$. In both cases a
reduction rule is applicable, so the measure decreases by at least $2$ in this
branch, and the branching vector is $(4,2)$.

        \item \textbf{$\{u_1, u_2, u_3\}$ induces a $P_3$.}
        Without loss of generality, assume that\\ $(u_1, u_2, u_3)$ forms the $P_3$, i.e., $u_1u_3 \notin E(G)$.  
        We analyze this case in two subcases depending on whether $u_1$ or $u_3$ has neighbors in $V'$ outside $N_{V'}[v]$.

        \begin{enumerate}
            \item Suppose at least one of $u_1$ or $u_3$ has a neighbor in $V'$ outside $N_{V'}[v]$.  
            Without loss of generality, let $u_1$ have a neighbor $w_1$ in $V'$ outside $N_{V'}[v]$.

                        \begin{itemize}
                \item Add $u_1$ to $C'$. Since $v$ and $u_1$ are adjacent and
                both lie in $C'$, they belong to the same clique of $C'$. The
                vertex $u_3$ is adjacent to $v$ but not to $u_1$, so
                $N_{C'}(u_3)$ is a non-empty proper subset of that clique, and
                by \Cref{rule:6} the vertex $u_3$ moves to $I'$. Then $u_2$ is
                a neighbor of $u_3 \in I'$, so by \Cref{rule:3} it is added to
                $C'$. Finally, $w_1$ is adjacent to $u_1$ but not to $v$, so by
                the same argument \Cref{rule:6} moves $w_1$ to $I'$. Thus the
                measure decreases by at least $5$.
                \item Add $u_1$ to $I'$. Then, by \Cref{rule:3}, both $u_2$ and
                $w_1$ are added to $C'$. The measure decreases by at least $4$.
            \end{itemize}
            Hence, when at least one of $u_1$ or $u_3$ has a neighbor in $V'$ outside $N_{V'}[v]$, the branching vector is $(4, 5, 4)$.

            \item Now consider the case where both $u_1$ and $u_3$ have no neighbors in $V'$ outside $N_{V'}[v]$.  
            We branch as follows:
            \begin{itemize}
                \item Add $u_2$ to $C'$. Then \cref{rule:4} applies to both $u_1$ and $u_3$, resulting in a measure drop of $4$.
                \item Add $u_2$ to $I'$. Then, by \cref{rule:3}, both $u_1$ and
                $u_3$ are added to $C'$. Now $u_1$ and $u_3$ are both adjacent
                to $v \in C'$ but are not adjacent to each other, so $G[C']$ is
                not a cluster graph and \Cref{rule:1} returns a trivial
                \No-instance. This branch is therefore discarded.
            \end{itemize}
            Thus, the branching vector in this case is $(4, 4)$.
        \end{enumerate}
        Hence, when $\{u_1,u_2,u_3\}$ induces a $P_3$, the branching vector is either $(4,5,4)$ or $(4,4)$.

        \item \textbf{Exactly two of $\{u_1, u_2, u_3\}$ are adjacent.}
        Without loss of generality, assume that $u_2$ and $u_3$ are adjacent.  
        Then there must exist a vertex $w_1 \in V'$ which is a neighbor of $u_1$ in $V'$ outside $N_{V'}[v]$, otherwise \cref{rule:4} would be applicable.

                The branching is as follows:
        \begin{itemize}
            \item Add $u_2$ to $C'$. Since $v$ and $u_2$ are adjacent and both
            lie in $C'$, they belong to the same clique of $C'$, while $u_1$ is
            adjacent to $v$ but not to $u_2$; hence by \Cref{rule:6} the vertex
            $u_1$ moves to $I'$, and by \Cref{rule:3} its neighbor $w_1$ is
            added to $C'$. The measure decreases by $4$.
            \item Add $u_2$ to $I'$. Then, by \Cref{rule:3}, $u_3$ is added to
            $C'$. Now $v$ and $u_3$ belong to the same clique of $C'$, while
            $u_1$ is adjacent to $v$ but not to $u_3$; hence by \Cref{rule:6}
            the vertex $u_1$ moves to $I'$, and by \Cref{rule:3} its neighbor
            $w_1$ is added to $C'$. The measure decreases by $5$.
        \end{itemize}
        Hence, the branching vector in this case is $(4,4,5)$.
    \end{enumerate}
\end{itemize}

\noindent
\paragraph{\bf Case 4: The maximum $V'$-degree is at most $2$.}
In this scenario, each connected component of $G[V']$ forms either a path or a cycle. Let $A \subseteq V'$ denote an arbitrary connected component of $G[V']$. We analyze the following subcases.

\begin{enumerate}
\item \textbf{$A$ is a singleton.}
In this case, \cref{rule:4} applies directly.

\item \label{caseref}\textbf{$A$ contains exactly two vertices.}
Let $A = \{u,v\}$, so $N_{V'}(u) = \{v\}$ and $N_{V'}(v) = \{u\}$. Since
\Cref{rule:3} is not applicable, neither $u$ nor $v$ has a neighbor in $I'$.

We first record what the non-applicability of the reduction rules gives us.
If $N_{C'}(u)$ were empty, then \Cref{rule:7} would apply to $u$; hence
$N_{C'}(u) \neq \emptyset$, and by \Cref{rule:6} it is a clique $C_1$ of $C'$.
Symmetrically, $N_{C'}(v)$ is a clique $C_2$ of $C'$. If $C_1 = C_2$, then
\Cref{rule:9} applies with $B = \{u,v\}$, so we have $C_1 \neq C_2$. Finally,
since \Cref{rule:7} is not applicable to $u$, every vertex of $C_1$ has a
neighbor in $V'$ other than $u$ and $v$; fix such a neighbor $w_1$. In the same
way, $C_2$ contains a vertex with a neighbor $w_2 \in V' \setminus \{u,v\}$.

Our branching strategy is as follows:
\begin{itemize}
    \item Add $u$ to $I'$. Then by \Cref{rule:3}, $v$ moves to $C'$, and the
    clique of $C'$ containing $v$ contains $C_2$. As $w_2$ is adjacent to a
    vertex of $C_2$ but not to $v$, its neighborhood in $C'$ is a non-empty
    proper subset of that clique, so by \Cref{rule:6} the vertex $w_2$ moves
    to $I'$. Thus the measure decreases by $3$.
    \item Add $u$ to $C'$. Then the clique of $C'$ containing $u$ contains
    $C_1$. As $v$ is adjacent to $u$ but $N_{C'}(v) = C_2 \neq C_1$, by
    \Cref{rule:6} the vertex $v$ moves to $I'$, and by the same argument as
    above \Cref{rule:6} also moves $w_1$ to $I'$. Again the measure decreases
    by $3$.
\end{itemize}
Therefore, for a two-vertex component $A$, the branching vector is $(3,3)$.

\item \textbf{$A$ induces a path on exactly three vertices.}  
Let $\{u_1,u_2,u_3\}= A$ where $u_1$ is an endpoint of the path, and $u_1u_2,u_2u_3\in E(G)$. We branch as follows:
\begin{itemize}
    \item Add $u_2$ to $I'$. Then by \Cref{rule:3}, both $u_1$ and $u_3$ are added to $C'$, decreasing the measure by $3$.
    \item Add $u_2$ to $C'$. By \Cref{rule:4}, both $u_1$ and $u_3$ are added to $I'$. Thus, total measure decreases by $3$.
\end{itemize}
Hence, for a path of exactly three vertices, the branching vector is $(3,3)$.

\item \textbf{$A$ induces a path on at least four vertices.}  
Let $\{u_1,u_2,u_3,u_4\}\subseteq A$ where $u_1$ is an endpoint of the path, and $u_1u_2,u_2u_3, u_3u_4\in E(G)$. We branch as follows:
\begin{itemize}
    \item Add $u_3$ to $I'$. Then by \Cref{rule:3}, both $u_2$ and $u_4$ are added to $C'$. Also by \cref{rule:4}, $u_1$ is added to $I'$. Thus the measure decreases by $4$.
    \item Add $u_3$ to $C'$. Now $u_1u_2$ is an edge and a component in $V'$. 
    
    \begin{enumerate}

     \item Suppose some reduction rule applies to $u_1$ or $u_2$. By \cref{rule:3} and \cref{rule:4}, this places one of them in $I'$ and the other one in $C'$. Either way the measure decreases by $3$ and hence, the branching vector would be $(4,3)$.
\item If \cref{rule:6} or \cref{rule:4} is applicable to $u_4$, then we add $u_4$ to $I'$. Thus, decrease in measure would be $2$ and hence, the branching vector is $(4,2)$.
                \item If none of the reduction rules is applicable to $u_1$,
        $u_2$ or $u_4$, then, as in \cref{caseref}, the set $N_{C'}(u_1)$ is a
        clique $C_1$ of $C'$. Let $D$ be the clique of $C'$ containing $u_3$.
        Since \cref{rule:6} is not applicable to $u_2$, its neighborhood
        $N_{C'}(u_2)$ is a clique of $C'$; as $u_2$ is adjacent to $u_3 \in D$,
        this clique is $D$. In particular $C_1 \neq D$, since otherwise
        \cref{rule:9} would apply with $B = \{u_1,u_2\}$. Similarly, as
        \cref{rule:6} is not applicable to $u_4$, we have $N_{C'}(u_4) = D$,
        and as \cref{rule:4} is not applicable to $u_4$, it has a neighbor
        $u_5 \in V'$. Finally, since \cref{rule:7} is not applicable to the
        edge $u_1u_2$, the clique $C_1$ contains a vertex with a neighbor
        $w_1 \in V' \setminus \{u_1,u_2\}$. Our branching strategy is as
        follows:
\begin{itemize}
    \item Add $u_1$ to $I'$. Then by \cref{rule:3}, $u_2$ moves to $C'$, and
    it joins the clique $D$, since $N_{C'}(u_2) = D$. Now $u_4$ is adjacent to
    $u_3 \in D$ but not to $u_2 \in D$, so $N_{C'}(u_4)$ is a non-empty proper
    subset of $D \cup \{u_2\}$, and by \cref{rule:6} the vertex $u_4$ moves to
    $I'$. Then by \cref{rule:3}, $u_5$ moves to $C'$. Counting $u_3$, which
    was already added to $C'$, the measure decreases by $5$.
    \item Add $u_1$ to $C'$. Then $u_1$ joins the clique $C_1$, while $u_3$
    lies in $D$, and $C_1 \neq D$. As $u_2$ is adjacent to both $u_1$ and
    $u_3$, its neighborhood in $C'$ meets two distinct cliques, so by
    \cref{rule:6} the vertex $u_2$ moves to $I'$. Also $w_1$ is adjacent to a
    vertex of $C_1$ but not to $u_1$, so $N_{C'}(w_1)$ is a non-empty proper
    subset of $C_1 \cup \{u_1\}$ and by \cref{rule:6} the vertex $w_1$ moves
    to $I'$ as well. Again counting $u_3$, the measure decreases by $4$.
\end{itemize}
Hence the branching vector in this subcase is $(4,5,4)$.
\end{enumerate}
    
\end{itemize}
Hence, for a path of length at least four, the branching vectors can be $(4,3)$, $(4,2)$ or $(4,5,4)$.

\item \textbf{$A$ induces a triangle.}
Let $A=\{u_1,u_2,u_3\}$. We claim that at least one reduction rule applies in
this case. Since $A$ is a connected component of $G[V']$, we have
$N_{V'}[u_i] = A$ for every $i$.

If $N_{C'}(u_1) = N_{C'}(u_2) = N_{C'}(u_3)$, then \Cref{rule:9} applies with
$B = A$; note that this includes the case where all three sets are empty.
Otherwise two of these sets differ, say $N_{C'}(u_1) \neq N_{C'}(u_2)$, and
\Cref{rule:10} applies to the edge $u_1u_2$, whose common $V'$-neighborhood
contains $u_3$.

\item \textbf{$A$ induces a cycle of length at least four.}
Since \Cref{rule:3} is not applicable, no vertex of $A$ has a neighbor in
$I'$, so $N(A) \subseteq C'$. For $u \in A$ write $C_u = N_{C'}(u)$; since
\Cref{rule:6} is not applicable to $u$, the set $C_u$ is either empty or a
clique of $C'$.

If $C_u = \emptyset$ for every $u \in A$, then $N(A) = \emptyset$ and
\Cref{rule:11} applies, adding an arbitrary vertex of $A$ to $I'$. If $C_u$ is
the same non-empty clique for every $u \in A$, then \Cref{rule:11} again
applies and returns a trivial \No-instance, so this branch is discarded.
In the remaining cases we branch as follows.

\begin{enumerate}
    \item If some $u \in A$ has $C_u = \emptyset$ and some other vertex of $A$
    has a non-empty $C$-neighborhood, then, as $A$ is connected, we can find
    an edge $w_1w_2$ in $G[A]$ with $C_{w_1} \neq \emptyset$ and
    $C_{w_2} = \emptyset$.
    \begin{itemize}
        \item Add $w_1$ to $I'$. Then by \Cref{rule:3}, both neighbors of
        $w_1$ in $V'$ are added to $C'$, so the measure decreases by $3$.
        \item Add $w_1$ to $C'$. Then $w_2$ is adjacent to $w_1 \in C'$ but
        has no other neighbor in $C'$, while the clique of $C'$ containing
        $w_1$ also contains $C_{w_1} \neq \emptyset$; hence $N_{C'}(w_2)$ is a
        non-empty proper subset of that clique and by \Cref{rule:6} the vertex
        $w_2$ moves to $I'$. By \Cref{rule:3}, the other neighbor of $w_2$
        along the cycle is then added to $C'$, and the measure decreases by
        $3$.
    \end{itemize}
    Thus the branching vector in this case is $(3,3)$.

    \item If $C_u \neq \emptyset$ for every $u \in A$ but the cliques $C_u$ are
    not all equal, then, as $A$ is connected, there are two adjacent vertices
    $w_1, w_2 \in A$ with $C_{w_1} \neq C_{w_2}$. Note that \Cref{rule:10} is
    not applicable to the edge $w_1w_2$, since two adjacent vertices of a cycle
    of length at least four have no common neighbor in $V'$. We branch as
    follows.
    \begin{itemize}
        \item Add $w_1$ to $I'$. Then by \Cref{rule:3}, both neighbors of
        $w_1$ in $V'$ are added to $C'$, so the measure decreases by $3$.
        \item Add $w_1$ to $C'$. Then the clique of $C'$ containing $w_1$ also
        contains $C_{w_1}$, while $w_2$ is adjacent to $w_1$ and satisfies
        $N_{C'}(w_2) = C_{w_2} \cup \{w_1\}$ with $C_{w_2} \neq C_{w_1}$; hence
        $N_{C'}(w_2)$ meets two distinct cliques of $C'$ and by \Cref{rule:6}
        the vertex $w_2$ moves to $I'$. By \Cref{rule:3}, the next neighbor
        $w_3$ of $w_2$ along the cycle is added to $C'$, and the measure
        decreases by $3$.
    \end{itemize}
    Thus, the branching vector in this case is also $(3,3)$.
\end{enumerate}
Hence, when $A$ induces a cycle of length at least four, and no reduction rule
applies directly, every branching step decreases the measure by at least $3$
in each branch, giving a worst-case branching vector of $(3,3)$.
\end{enumerate}

\section{Missing Proofs of \Cref{sec:fptmr}}\label{appen:fptmr}

\begin{lemma}\label{lm:mrintroduce}
The recursion for the introduce node is correct.
\end{lemma}

\begin{proof}
Let $t$ be an introduce node with child $t_1$, and let $v$ be the vertex introduced at node $t$.
We know that $B_t = B_{t_1} \cup \{v\}$, and that $v$ has no neighbors in $V_t \setminus B_t$.

For any tripartition $(X_t, R_t, S_t)$ of $B_t$, the table entry
$T[t, X_t, R_t, S_t]$ is set to zero if $(X_t, R_t \cup S_t)$ is not a monopolar partition of $G[B_t]$, or if some connected component of $G[R_t \cup S_t]$ intersects both $R_t$ and $S_t$, or if $\tilde{I}\cap B_t \nsubseteq X_t$, or if $\tilde{C}\cap B_t\nsubseteq R_t\cup S_t$.
Otherwise, it is computed using the stated recursion.
We prove correctness by showing both directions.

\paragraph{Forward Direction.}
Let us assume that $T[t, X_t, R_t, S_t] = 1$. We need to show that there exists a monopolar partition $(I_t, C_t)$ of the subgraph $G_t$ satisfying $\tilde{I}\cap V_t\subseteq I_t$ and $\tilde{C}\cap V_t \subseteq C_t$; $I_t \cap B_t = X_t$; $C_t \cap B_t = R_t \cup S_t$; and $R_t$ is the union of $C\cap B_t$ for all cliques $C$ in $C_t$ for which there exists a vertex $u \in C \cap (V_t \setminus B_t)$.
By the recursion, $v \notin R_t$, since in that case the entry is set to zero.
Hence one of the following must hold:
\begin{itemize}
    \item $v \in X_t$ and $T[t_1, X_t \setminus \{v\}, R_t, S_t] = 1$, or
    \item $v \in S_t$ and $T[t_1, X_t, R_t, S_t \setminus \{v\}] = 1$.
\end{itemize}

\medskip
\noindent\textbf{Case 1: $v \in X_t$.} In this case, it follows that $v \notin \tilde{C}$. Since
$T[t_1, X_t \setminus \{v\}, R_t, S_t] = 1$, there exists a monopolar partition
$(I_{t_1}, C_{t_1})$ of $G_{t_1}$ satisfying
\[
\tilde{I} \cap V_{t_1} \subseteq I_{t_1}, \qquad
\tilde{C} \cap V_{t_1} \subseteq C_{t_1}, \qquad
I_{t_1} \cap B_{t_1} = X_t \setminus \{v\}, \qquad \text{$R_t$ and $S_t$ are disjoint},
\]

$C_{t_1} \cap B_{t_1} = R_t \cup S_t,$
and $R_t$ is the union of $C_1 \cap B_{t_1}$ over all cliques $C_1$ of $C_{t_1}$ that contain a vertex from
$V_{t_1} \setminus B_{t_1}$.

Let us define
\[
I_t = I_{t_1} \cup \{v\}
\quad \text{and} \quad
C_t = C_{t_1}.
\]
Since $(X_t, R_t \cup S_t)$ is a monopolar partition of $G[B_t]$ and $v$ has no neighbors in
$V_t \setminus B_t = V_{t_1} \setminus B_{t_1}$, the set $I_t$ is independent in $G_t$.
Moreover, since $C_t = C_{t_1}$ and $C_{t_1}$ is a cluster graph, $C_t$ is also a cluster graph. Again since $(I_{t_1},C_{t_1})$ is a partition for $V_{t_1}$, $(I_t,C_t)$ is a partition for $V_{t_1}\cup \{v\}=V_t$ and thus, $(I_t, C_t)$ is a monopolar partition of $G_t$.

Furthermore,

\begin{itemize}
    \item $\tilde{I}\cap V_t=\tilde{I}\cap (V_{t_1}\cup \{v\})=(\tilde{I}\cap V_{t_1})\cup (\tilde{I}\cap \{v\})\subseteq I_{t_1}\cup \{v\}= I_t$;

    \item $\tilde{C}\cap V_t=\tilde{C}\cap (V_{t_1}\cup \{v\})=(\tilde{C}\cap V_{t_1})\cup (\tilde{C}\cap \{v\})\subseteq C_{t_1}= C_t$, as $v\notin \tilde{C}$;

    \item $I_t \cap B_t
= (I_{t_1} \cup \{v\}) \cap (B_{t_1} \cup \{v\})=(I_{t_1}\cap B_{t_1})\cup \{v\}
= X_t$;

\item $R_t$ and $S_t$ are disjoint;

\item $C_t \cap B_t = C_{t_1} \cap (B_{t_1}\cup \{v\})=C_{t_1}\cap B_{t_1} = R_t \cup S_t$, as $v\notin C_{t_1}$;

\item Since $V_t \setminus B_t = V_{t_1} \setminus B_{t_1}$ and $v\notin C_t$, the definition of $R_t$ is preserved.
\end{itemize}

Hence, we are done in this case.

\medskip
\noindent\textbf{Case 2: $v \in S_t$.} In this case, it follows that $v \notin \tilde{I}$.
Since $T[t_1, X_t, R_t, S_t \setminus \{v\}] = 1$, there exists a monopolar partition
$(I_{t_1}, C_{t_1})$ of $G_{t_1}$ such that
\[
\tilde{I} \cap V_{t_1} \subseteq I_{t_1}, \quad
\tilde{C} \cap V_{t_1} \subseteq C_{t_1}, \quad
I_{t_1} \cap B_{t_1} = X_t, \quad \text{$R_t$ and $S_t\setminus \{v\}$ are disjoint},
\]

$C_{t_1} \cap B_{t_1} = R_t \cup (S_t\setminus \{v\}),$
and $R_t$ is the union of $C_1 \cap B_{t_1}$ over all cliques $C_1$ of $C_{t_1}$ that contain a vertex from
$V_{t_1} \setminus B_{t_1}$.

Again we define
\[
I_t = I_{t_1}
\quad \text{and} \quad
C_t = C_{t_1} \cup \{v\}.
\]
By the validity condition, the connected component of $v$ in
$G[R_t \cup S_t]$ is contained in $S_t$, so $v$ has no neighbor in $R_t$.
Since $v$ has no neighbor in $V_t \setminus B_t$, the neighbors of $v$ in
$C_{t_1}$ are exactly the vertices of $N_{S_t}(v)$.

If $N_{S_t}(v) = \emptyset$, then $v$ has no neighbor in $C_{t_1}$, so
$C_t = C_{t_1} \cup \{v\}$ is a cluster graph in which $v$ forms a clique of
its own, and this clique contains no vertex of $V_t \setminus B_t$.

Otherwise, fix $u \in N_{S_t}(v)$ and let $K$ be the clique of $C_{t_1}$
containing $u$. The vertex $u$ lies in $S_t \setminus \{v\}$, so the clique $K$ of $C_{t_1}$
containing $u$ is not one of the cliques contributing to $R_t$. By the
definition of $R_t$ at node $t_1$, such a clique contains no vertex of
$V_{t_1} \setminus B_{t_1}$. Hence $K \subseteq B_{t_1}$. Hence $K \cup \{v\} \subseteq R_t \cup S_t$. Now
$G[R_t \cup S_t]$ is a cluster graph, since $(X_t, R_t \cup S_t)$ is a
monopolar partition of $G[B_t]$; as $v$ is adjacent to $u$, the vertex $v$
lies in the same clique of $G[R_t \cup S_t]$ as $u$, and so does every vertex
of $K$, because $K$ is a clique containing $u$. Therefore $v$ is adjacent to
every vertex of $K$, that is, $N_{S_t}(v) = K$. Consequently
$C_t = C_{t_1} \cup \{v\}$ is a cluster graph, obtained from $C_{t_1}$ by
replacing $K$ with the clique $K \cup \{v\}$, and this clique still contains no
vertex of $V_t \setminus B_t$. Also the set $I_t$ remains independent in $G_t$ and since $(I_{t_1},C_{t_1})$ is a partition for $V_{t_1}$, $(I_t,C_t)$ is a partition for $V_{t_1}\cup \{v\}=V_t$. Hence, $(I_t, C_t)$ is a monopolar partition of $G_t$.

Moreover,

\begin{itemize}
    \item $\tilde{I}\cap V_t=\tilde{I}\cap (V_{t_1}\cup \{v\})=(\tilde{I}\cap V_{t_1})\cup (\tilde{I}\cap \{v\})\subseteq I_{t_1}= I_t$, as $v\notin \tilde{I}$;

    \item $\tilde{C}\cap V_t=\tilde{C}\cap (V_{t_1}\cup \{v\})=(\tilde{C}\cap V_{t_1})\cup (\tilde{C}\cap \{v\})\subseteq C_{t_1}\cup \{v\}= C_t$;

    \item $I_t \cap B_t
= I_{t_1} \cap (B_{t_1} \cup \{v\})
= I_{t_1}\cap B_{t_1}=X_t$, as $v\notin I_{t_1}$;

\item $R_t$ and $S_t$ are disjoint as $R_t$ and $S_t\setminus \{v\}$ are disjoint and $v\notin R_t$;

\item $C_t \cap B_t = (C_{t_1}\cap B_{t_1})\cup \{v\}= R_t \cup S_t, \text{ as } v\in C_t\cap B_t$;

\item Since $V_t \setminus B_t = V_{t_1} \setminus B_{t_1}$ and $v\notin R_t$, the definition of $R_t$ is preserved.
\end{itemize}
 Hence, we are done with this case and the forward direction.

\paragraph{Converse Direction.}
Let us assume that there exists a monopolar partition $(I_t, C_t)$ of $G_t$ satisfying
\[
\tilde{I} \cap V_{t} \subseteq I_{t}, \quad
\tilde{C} \cap V_{t} \subseteq C_{t}, \quad I_t \cap B_t = X_t,
\quad
C_t \cap B_t = R_t \cup S_t,
\quad
R_t \cap S_t = \emptyset,
\]
and $R_t$ is the union of $C\cap B_t$ for all cliques $C$ in $C_t$ containing a vertex from
$V_t \setminus B_t$. We need to show that $T[t,X_t,R_t,S_t]=1$.

Let us define
\[
I_{t_1} = I_t \cap V_{t_1}
\quad \text{and} \quad
C_{t_1} = C_t \cap V_{t_1},
\]
where $V_{t_1}=V_t\setminus \{v\}$.
Since $I_t$ is independent in $G_t$ and $C_t$ is a cluster graph, $I_{t_1}$ is an independent set and $C_{t_1}$ is a cluster in $G_{t_1}$. Again, $(I_{t_1}, C_{t_1})$ would be a partition for $V_{t_1}$ and thus,
a monopolar partition of $G_{t_1}$.

\medskip
\noindent\textbf{Case 1: $v \in I_t$.} Thus, in this case, $I_{t_1}=I_t\setminus \{v\}$ and $C_{t_1}=C_t$.
Since $v \in B_t$ and $I_t \cap B_t = X_t$, we have $v \in X_t$. Also, we must have $v\notin \tilde{C}$.

Moreover,
\begin{itemize}
    \item $\tilde{I}\cap V_{t_1}=\tilde{I}\cap (V_t\setminus \{v\})=(\tilde{I}\cap V_t)\setminus \{v\}\subseteq I_t\setminus \{v\}=I_{t_1}$;

    \item $\tilde{C}\cap V_{t_1}=\tilde{C}\cap (V_t\setminus \{v\})=(\tilde{C}\cap V_t)\setminus \{v\}\subseteq C_t\setminus \{v\}=C_{t_1}$, as $v\notin C_t$;

    \item $I_{t_1} \cap B_{t_1} = (I_t\setminus \{v\})\cap (B_{t}\setminus \{v\})=(I_t\cap B_t)\setminus \{v\} = X_t\setminus \{v\}$;
    \item $R_t$ and $S_t$ are disjoint;
    \item $C_{t_1} \cap B_{t_1}=C_t\cap (B_{t_1}\cup \{v\})=C_t\cap B_t = R_t \cup S_t$, as $v\notin C_t$;
    \item Note that for any clique $C$ in $C_t=C_{t_1}$, $C\cap B_{t_1}=C\cap B_t$, as $v\notin C_t$ and $V_t\setminus B_t=V_{t_1}\setminus B_{t_1}$. Thus, $R_t$ is the union of $C\cap B_{t_1}$ for all cliques $C$ of $C_t=C_{t_1}$ for which 
    $C$ contains a vertex in $V_{t_1} \setminus B_{t_1}$.
\end{itemize}

Hence, all the conditions for
$T[t_1, X_t \setminus \{v\}, R_t, S_t] = 1$ are satisfied, which implies $T[t, X_t, R_t, S_t] = 1$ from the recursion.

\medskip
\noindent\textbf{Case 2: $v \in C_t$.} Thus, in this case, $I_{t_1}=I_t$ and $C_{t_1}=C_t\setminus \{v\}$.
Since $C_t \cap B_t = R_t \cup S_t$, we have $v \in R_t \cup S_t$. Also, we must have $v\notin \tilde{I}$.

We now show that $v \notin R_t$. By definition, $R_t$ consists of the bag vertices belonging to cliques in $C_t$ that contain a vertex
from $V_t \setminus B_t$.
Since $v$ is introduced at node $t$, it has no neighbors in $V_t \setminus B_t$.
Therefore, the clique of $C_t$ containing $v$ does not contain any vertex from
$V_t \setminus B_t$.
Hence, $v \notin R_t$, and thus $v \in S_t$.

Now,

\begin{itemize}
    \item $\tilde{I}\cap V_{t_1}=\tilde{I}\cap (V_t\setminus \{v\})=(\tilde{I}\cap V_t)\setminus \{v\}\subseteq I_t\setminus \{v\}=I_{t_1}$, as $v\notin I_t$;

    \item $\tilde{C}\cap V_{t_1}=\tilde{C}\cap (V_t\setminus \{v\})=(\tilde{C}\cap V_t)\setminus \{v\}\subseteq C_t\setminus \{v\}=C_{t_1}$;

    \item $I_{t_1} \cap B_{t_1} = I_t\cap (B_{t_1}\cup \{v\})=I_t\cap B_t = X_t$, as $v\notin I_t$;
    \item $R_t$ and $S_t\setminus \{v\}$ are disjoint as $R_t$ and $S_t$ are disjoint;
    \item $C_{t_1} \cap B_{t_1}=(C_t\setminus \{v\})\cap (B_{t}\setminus \{v\})=(C_t\cap B_t)\setminus \{v\} = R_t \cup (S_t\setminus \{v\})$;
    \item Note that $v\notin R_t$ and $V_t\setminus B_t=V_{t_1}\setminus B_{t_1}$. Thus, $R_t$ is the union of $C\cap B_{t_1}$ for all cliques $C$ of $C_{t_1}=C_t\setminus \{v\}$ for which $C$ contains a vertex in $V_{t_1} \setminus B_{t_1}$.

\end{itemize}

Therefore, all the conditions for $T[t_1, X_t, R_t, S_t \setminus \{v\}] = 1$ are satisfied, which implies $T[t, X_t, R_t, S_t] = 1$ from the recursion of the introduce node.

This completes the proof.
\end{proof}

\begin{lemma}\label{lm:mrforget}
The recursion for the forget node is correct.
\end{lemma}
\begin{proof}
Let $t$ be a forget node with child $t_1$, and let $v$ be the vertex forgotten at node $t$.
Thus,
\[
B_{t_1} = B_t \cup \{v\}
\quad\text{and}\quad
V_t = V_{t_1}.
\]

Throughout, we use that $C_t$ is a cluster graph, so for $v \in C_t$ the
vertices of $B_{t_1}$ lying in the clique of $v$ are exactly $N_{Y_t}[v]$,
where $Y_t = R_t \cup S_t$. This is why the condition
$N_{Y_t}(v) \subseteq R_t$ appears in the second and third terms: in both, the
clique of $v$ meets $V_t \setminus B_t$ once $v$ is forgotten, and since $R_t$ is the union of the bag parts of exactly those cliques that contain a vertex outside the bag, its bag part $N_{Y_t}(v)$ must lie in $R_t$.

The table entry $T[t, X_t, R_t, S_t]$ is set to zero if $(X_t, R_t \cup S_t)$
is not a monopolar partition of $G[B_t]$, or if some connected component of
$G[R_t \cup S_t]$ intersects both $R_t$ and $S_t$, or if
$\tilde{I} \cap B_t \nsubseteq X_t$, or if
$\tilde{C} \cap B_t \nsubseteq R_t \cup S_t$.
Otherwise, it is defined by the stated recursion.
We prove correctness by showing both directions.

\paragraph{Forward Direction.}
Let us assume that $T[t, X_t, R_t, S_t] = 1$.
We need to show that there exists a monopolar partition $(I_t, C_t)$ of the subgraph $G_t$
satisfying:
$\tilde{I}\cap V_t \subseteq I_t$; $\tilde{C}\cap V_t \subseteq C_t$;  $I_t \cap B_t = X_t$; $R_t$ and $S_t$ are disjoint;
$C_t \cap B_t = R_t \cup S_t$; and
$R_t$ is the union of $C\cap B_t$ for all cliques $C$ in $C_t$ for which there exists a vertex
$u \in C \cap (V_t \setminus B_t)$.

By the recursion, one of the following holds:
\begin{itemize}
    \item $T[t_1, X_t \cup \{v\}, R_t, S_t] = 1$, or
    \item $N_{Y_t}(v) \subseteq R_t$ and $T[t_1, X_t, R_t \cup \{v\}, S_t] = 1$, or
    \item $N_{Y_t}(v) \subseteq R_t$ and
    $T[t_1, X_t, R_t \setminus N_{Y_t}(v), S_t \cup N_{Y_t}[v]] = 1$.
\end{itemize}

\medskip
\noindent\textbf{Case 1: $T[t_1, X_t \cup \{v\}, R_t, S_t] = 1$.}

Since $T[t_1, X_t \cup \{v\}, R_t, S_t] = 1$, there exists a monopolar partition
$(I_{t_1}, C_{t_1})$ of $G_{t_1}$ such that
\[
\tilde{I} \cap V_{t_1} \subseteq I_{t_1}, \qquad
\tilde{C} \cap V_{t_1} \subseteq C_{t_1}, \qquad
I_{t_1} \cap B_{t_1} = X_t \cup \{v\}, \qquad \text{$R_t$ and $S_t$ are disjoint},
\]

$C_{t_1} \cap B_{t_1} = R_t \cup S_t,$
and $R_t$ is the union of $C_1 \cap B_{t_1}$ over all cliques $C_1$ of $C_{t_1}$ that contain a vertex from
$V_{t_1} \setminus B_{t_1}$.

Clearly, $v \in I_{t_1}$, and thus $v\notin \tilde{C}$.
Since $V_t = V_{t_1}$, we have $G_t=G_{t_1}$. Hence, $(I_{t_1},C_{t_1})$ is a monopolar partition for $G_t$. 

Moreover,
\begin{itemize}
    \item $\tilde{I}\cap V_t=\tilde{I}\cap V_{t_1}\subseteq I_{t_1}$;

    \item $\tilde{C}\cap V_t=\tilde{C}\cap V_{t_1}\subseteq C_{t_1}$;

    \item $I_{t_1} \cap B_t = X_t$, as $v\notin B_t$;
    \item $R_t$ and $S_t$ are disjoint;
    \item $C_{t_1} \cap B_t=C_{t_1}\cap B_{t_1} = R_t \cup S_t$, since $v \notin C_{t_1}$;
    \item Since $v\notin C_{t_1}$, $R_t$ is the union of $C\cap B_{t}$ for all cliques $C$ of $C_{t_1}$ for which $C$ contains a vertex in $V_{t} \setminus B_{t}$.
\end{itemize}

Hence, we are done with this case.

\medskip
\noindent\textbf{Case 2: $N_{Y_t}(v) \subseteq R_t$ and
$T[t_1, X_t, R_t \cup \{v\}, S_t] = 1$.}

Since $T[t_1, X_t, R_t \cup \{v\}, S_t] = 1$, there exists a monopolar partition
$(I_{t_1}, C_{t_1})$ of $G_{t_1}$ such that
\begin{itemize}
    \item $\tilde{I} \cap V_{t_1} \subseteq I_{t_1}$;
    \item $\tilde{C} \cap V_{t_1} \subseteq C_{t_1}$;
    \item $I_{t_1}\cap B_{t_1}=X_t$;
    \item $(R_t\cup\{v\})$ and $S_t$ are disjoint;
    \item $C_{t_1}\cap B_{t_1}=(R_t\cup\{v\})\cup S_t$;
    \item $R_t\cup\{v\}$ is the union of $C\cap B_{t_1}$ for all cliques $C$ of
          $C_{t_1}$ such that $C$ contains a vertex in
          $V_{t_1}\setminus B_{t_1}$.
\end{itemize}
Here we have $v\in C_{t_1}$.
Let $C'$ be the clique of $C_{t_1}$ containing $v$.
By definition of $R_t\cup\{v\}$, this clique $C'$ intersects
$V_{t_1}\setminus B_{t_1}$.

Again $V_t = V_{t_1}$, we have $G_t=G_{t_1}$. Hence $(I_{t_1},C_{t_1})$ is a monopolar partition for $G_t$. 
We also have
\begin{itemize}
    \item $\tilde{I}\cap V_t=\tilde{I}\cap V_{t_1}\subseteq I_{t_1}$;

    \item $\tilde{C}\cap V_t=\tilde{C}\cap V_{t_1}\subseteq C_{t_1}$;
    \item $I_{t_1} \cap B_t=I_{t_1}\cap B_{t_1} = X_t$, as $v\notin I_{t_1}$;
    \item $R_t$ and $S_t$ are disjoint, as $v\notin R_t\cup S_t$, $R_t\cup\{v\}$ and $S_t$ are disjoint;
    \item $C_{t_1} \cap B_t=(C_{t_1}\cap B_{t_1})\setminus \{v\} = R_t \cup S_t$, since $v \notin B_{t}$;
    \item Let $C'$ be the clique of $C_{t_1}$ containing $v$. Then notice that $R_t\setminus C'$ is the union of $C\cap B_{t}$ for all cliques $C$ of $C_{t_1}-C'$ for which $C$ contains a vertex in $V_{t} \setminus B_{t}$. Again, since $C'\cap B_{t_1}=N_{Y_t}[v]$, we have $R_t\setminus C'=R_t\setminus N_{Y_t}(v)$. We also have $C'\cap B_t=N_{Y_t}(v)$, as $v\notin B_t$ and $N_{Y_t}(v)\subseteq R_t$. Thus, ultimately, $(R_t\setminus N_{Y_t}(v))\cup N_{Y_t}(v)=R_t$ is the union of $C\cap B_{t}$ for all cliques $C$ of $C_{t_1}$ for which $C$ contains a vertex in $V_{t} \setminus B_{t}$.
\end{itemize}

Hence, we are done with this case as well.

\medskip
\noindent\textbf{Case 3: $N_{Y_t}(v) \subseteq R_t$ and
$T[t_1, X_t, R_t \setminus N_{Y_t}(v), S_t \cup N_{Y_t}[v]] = 1$.}

Since
$T[t_1, X_t, R_t \setminus N_{Y_t}(v), S_t \cup N_{Y_t}[v]] = 1$,
there exists a monopolar partition $(I_{t_1}, C_{t_1})$ of $G_{t_1}$ such that
\begin{itemize}
    \item $\tilde{I} \cap V_{t_1} \subseteq I_{t_1}$;
    \item $\tilde{C} \cap V_{t_1} \subseteq C_{t_1}$;
    \item $I_{t_1}\cap B_{t_1}=X_t$;
    \item $(R_t\setminus N_{Y_t}(v))$ and $(S_t\cup N_{Y_t}[v])$ are disjoint;
    \item $C_{t_1}\cap B_{t_1}
    =(R_t\setminus N_{Y_t}(v))\cup (S_t\cup N_{Y_t}[v])$;
    \item
    $R_t\setminus N_{Y_t}(v)$ is the union of $C\cap B_{t_1}$ for all cliques $C$
    of $C_{t_1}$ such that $C$ contains a vertex in
    $V_{t_1}\setminus B_{t_1}$.
\end{itemize}

Since $v\in N_{Y_t}[v]\subseteq S_t\cup N_{Y_t}[v]$, we have $v\in C_{t_1}$.
Let $C'$ be the clique of $C_{t_1}$ containing $v$.
By the definition of this case, the clique $C'$ does \emph{not} intersect $V_{t_1}\setminus B_{t_1}$.

Again $V_t = V_{t_1}$, we have $G_t=G_{t_1}$. Hence $(I_{t_1},C_{t_1})$ is a monopolar partition for $G_t$. Moreover, 

\begin{itemize}
    \item $\tilde{I}\cap V_t=\tilde{I}\cap V_{t_1}\subseteq I_{t_1}$;

    \item $\tilde{C}\cap V_t=\tilde{C}\cap V_{t_1}\subseteq C_{t_1}$;
    \item $I_{t_1} \cap B_t=I_{t_1}\cap B_{t_1} = X_t$, as $v\notin I_{t_1}$;
    \item $R_t$ and $S_t$ are disjoint, as $N_{Y_t}(v)\subseteq R_t$, $v\notin R_t\cup S_t$, $(R_t\setminus N_{Y_t}(v))$ and $(S_t\cup N_{Y_t}[v])$ are disjoint;
    \item $C_{t_1} \cap B_t=(C_{t_1}\cap B_{t_1})\setminus \{v\}=(R_t\setminus N_{Y_t}(v))\cup (S_t\cup N_{Y_t}(v))= R_t \cup S_t$, since $v \notin B_{t}$;

    \item
    We now verify the defining property of $R_t$.
    Since $v\notin B_t$ and $V_t=V_{t_1}$, we have $v\in V_t\setminus B_t$.
    Thus, the clique $C'$ containing $v$ now intersects
    $V_t\setminus B_t$.

    By the definition of $R_t$, all bag vertices of this clique must belong to
    $R_t$. These vertices are exactly $N_{Y_t}(v)$.
    Accordingly, the vertices of $N_{Y_t}(v)$ are part of $R_t$.
 All other cliques are unaffected and continue to contribute to $R_t$ exactly as before.
\end{itemize}




Hence, we have completed this case and the forward direction.
\paragraph{Converse Direction.}

Let us assume that there exists a monopolar partition $(I_t, C_t)$ of $G_t$ satisfying
\begin{itemize}
    \item $\tilde{I}\cap V_t\subseteq I_t$;

    \item $\tilde{C}\cap V_t\subseteq C_t$;
    \item $I_t \cap B_t = X_t$;
    \item $R_t$ and $S_t$ are disjoint;
    \item $C_t \cap B_t = R_t \cup S_t$;
    \item $R_t$ is the union of $C\cap B_t$ for all cliques $C$ of $C_t$
          for which $C$ contains a vertex in $V_t \setminus B_t$.
\end{itemize}

Since $t$ is a forget node with child $t_1$, we have
\[
B_{t_1} = B_t \cup \{v\}
\quad\text{and}\quad
V_t = V_{t_1}.
\]

We need to show that $T[t, X_t, R_t, S_t] = 1$.

Note that the graphs $G_t$ and $G_{t_1}$ are the same.
Thus, $(I_t, C_t)$ is a monopolar partition of $G_{t_1}$ as well.
Since $v \in V_{t_1}$, either $v \in I_t$ or $v \in C_t$.

\medskip
\noindent\textbf{Case 1: $v \in I_t$.}

We verify the bag conditions at node $t_1$:
\begin{itemize}
    \item $\tilde{I}\cap V_{t_1}=\tilde{I}\cap V_t\subseteq I_t$;
    \item $\tilde{C}\cap V_{t_1}=\tilde{C}\cap V_t\subseteq C_t$;
    \item $I_{t}\cap B_{t_1}=I_t\cap (B_t\cup \{v\}) = X_t\cup \{v\}$, as $v\in I_t$;
    \item $C_{t}\cap B_{t_1} = R_t \cup S_t$, as $v\notin C_t$;
\item Since $v \notin C_t$ and $V_{t_1}\setminus B_{t_1} = (V_t\setminus B_t)\setminus \{v\}$,
a clique of $C_t$ contains a vertex of $V_{t_1}\setminus B_{t_1}$ if and only
if it contains a vertex of $V_t\setminus B_t$. Hence $R_t$ is also the union of
$C \cap B_{t_1}$ over all cliques $C$ of $C_t$ containing a vertex of
$V_{t_1}\setminus B_{t_1}$, as $v \notin C$ for every such clique.
\end{itemize}

Hence,
\[
T[t_1, X_t \cup \{v\}, R_t, S_t] = 1,
\]
and by the recursion for the forget node, we conclude that
\[
T[t, X_t, R_t, S_t] = 1.
\]

\medskip
\noindent\textbf{Case 2: $v \in C_t$.}
Let $C'$ be the clique in $C_t$ containing $v$. Then $I_t\cap B_{t_1}=I_t\cap B_t=X_t=X_{t_1}$, as $v\notin I_t$. Now let us consider the following two subcases.

\medskip
\noindent\textbf{Subcase 2.1: $C'$ has empty intersection with $V_{t_1} \setminus B_{t_1}$.}

In this case, the clique $C'$ is entirely contained in $B_{t_1}$.

Also, we have
\begin{itemize}
    \item $\tilde{I}\cap V_{t_1}=\tilde{I}\cap V_t\subseteq I_t$;
    \item $\tilde{C}\cap V_{t_1}=\tilde{C}\cap V_t\subseteq C_t$;
    \item $I_{t}\cap B_{t_1} = X_t$, as $v\notin I_t$;
    \item $C_{t}\cap B_{t_1}=C_t\cap (B_t\cup \{v\})  = (R_t\cup \{v\}) \cup S_t=(R_t\setminus N_{Y_t}(v))\cup (S_t\cup N_{Y_t}[v])$, as $v\in C_t$;
\item We know that $R_t$ is the union of $C\cap B_t$ all cliques $C$ in $C_t$ containing a vertex from
$V_t \setminus B_t$. Since, $v\in C' \cap (V_t\setminus B_t)$, we must have that neighbors of $v$ in $C'\cap B_t$ is contained in $R_t$. But the clique $C'$ is entirely contained in $B_{t_1}$. Thus, $C'\cap (V_{t_1}\setminus B_{t_1})=\emptyset$, and $C'$ is only such clique of $C_t$ whose vertex is present in $(V_t\setminus B_t)$, but missing in $(V_{t_1}\setminus B_{t_1})$. Thus, we can say that $R_t\setminus N_{Y_t}(v)$ is the union $C\cap B_{t_1}$ for all cliques $C$ in $C_{t_1}$ containing a vertex from
$V_{t_1} \setminus B_{t_1}$.
\end{itemize}

 Hence, now all conditions for
\[
T[t_1, X_t, R_t \setminus N_{Y_t}(v), S_t \cup N_{Y_t}[v]] = 1,
\] is satisfied,
which implies
\[
T[t, X_t, R_t, S_t] = 1.
\]

\medskip
\noindent\textbf{Subcase 2.2: $C'$ has non-empty intersection with $V_{t_1} \setminus B_{t_1}$.}

Let $C'$ be the clique of $C_t$ containing $v$.
Since $v\in V_t\setminus B_t$, by the definition of $R_t$ we have
\[
C'\cap B_t = N_{Y_t}(v) \subseteq R_t.
\]

We verify the bag conditions at node $t_1$:
\begin{itemize}
    \item $\tilde{I}\cap V_{t_1}=\tilde{I}\cap V_t\subseteq I_t$;
    \item $\tilde{C}\cap V_{t_1}=\tilde{C}\cap V_t\subseteq C_t$;
    \item $I_{t}\cap B_{t_1} = X_t$, as $v\notin I_t$;
    \item $C_{t}\cap B_{t_1}=C_t\cap (B_t\cup \{v\})  = (R_t\cup \{v\}) \cup S_t$, as $v\in C_t$;
\item We now justify the definition of $R_t\cup\{v\}$ at node $t_1$.
Let $C'$ be the clique of $C_t$ containing $v$.
By assumption of this case, $C'$ intersects $V_{t_1}\setminus B_{t_1}$.
Hence, by the defining property of $R_t$ at node $t$,
all vertices of $C'\cap B_t = N_{Y_t}(v)$ belong to $R_t$.

At node $t_1$, the bag is $B_{t_1}=B_t\cup\{v\}$.
Thus, the bag intersection of $C'$ at node $t_1$ is
\[
C'\cap B_{t_1} = (C'\cap B_t)\cup\{v\} = N_{Y_t}(v)\cup\{v\}.
\]

All other cliques of $C_t$ that intersect $V_{t_1}\setminus B_{t_1}$
are unaffected by the presence of $v$ and contribute exactly the same
bag vertices as before.
Therefore, the union of $C\cap B_{t_1}$ over all such cliques is precisely
$R_t\cup\{v\}$.

\end{itemize}
Hence,
\[
T[t_1, X_t, R_t \cup \{v\}, S_t] = 1,
\]
and consequently,
\[
T[t, X_t, R_t, S_t] = 1.
\]

\medskip
Since all possible cases lead to one of the transitions in the recursion,
the converse direction holds.
\end{proof}

\begin{lemma}
The recursion for the join node is correct.
\end{lemma}

\begin{proof}
Let $t$ be a join node with children $t_1$ and $t_2$. Recall that
\[
B_{t_1} = B_{t_2} = B_t,
\qquad
V_t = V_{t_1} \cup V_{t_2},
\qquad
V_{t_1} \cap V_{t_2} = B_t,
\]
and that there are no edges between $V_{t_1}\setminus B_t$ and
$V_{t_2}\setminus B_t$. Let $Q_1,\ldots,Q_p$ be the connected components of
$G[R_t\cup S_t]$ contained in $R_t$. For a state $(X,R,S)$ of a node, we call
$R$ the $R$-part and $S$ the $S$-part of the state. By the definition of the
table, if a monopolar partition realises a state, then the $R$-part consists
of the bag parts of exactly those cliques which contain a vertex outside the
bag. In particular, a clique with non-empty bag part lying in the $S$-part has no
vertex outside the bag.

Fix a tripartition $(X_t,R_t,S_t)$ of $B_t$ such that $(X_t,R_t\cup S_t)$ is
a monopolar partition of $G[B_t]$, no connected component of $G[R_t\cup S_t]$
intersects both $R_t$ and $S_t$, $\tilde{I}\cap B_t\subseteq X_t$ and
$\tilde{C}\cap B_t\subseteq R_t\cup S_t$; otherwise both sides of the
recursion are $0$.

\paragraph{Forward Direction.}
Assume that $T[t,X_t,R_t,S_t]=1$. Then there is a split $(R^1_t,R^2_t)$ of
$R_t$ with
\[
T[t_1,X_t,R^1_t,S_t\cup R^2_t]=1
\quad\text{and}\quad
T[t_2,X_t,R^2_t,S_t\cup R^1_t]=1.
\]
For $i\in\{1,2\}$, let $(I_{t_i},C_{t_i})$ be a monopolar partition of
$G_{t_i}$ realising the corresponding state, that is,
$\tilde{I}\cap V_{t_i}\subseteq I_{t_i}$,
$\tilde{C}\cap V_{t_i}\subseteq C_{t_i}$,
$I_{t_i}\cap B_t=X_t$, $C_{t_i}\cap B_t=R_t\cup S_t$, the $R$-part of the
state of $t_1$ is $R^1_t$ and that of $t_2$ is $R^2_t$.
Let us define
\[
I_t = I_{t_1}\cup I_{t_2},
\qquad
C_t = C_{t_1}\cup C_{t_2}.
\]
Since $V_{t_1}\cap V_{t_2}=B_t$ and the two partitions agree on the bag, the
sets $I_t$ and $C_t$ are disjoint and form a partition of $V_t$. Since
$I_{t_1}$ and $I_{t_2}$ are independent and there are no edges between
$V_{t_1}\setminus B_t$ and $V_{t_2}\setminus B_t$, the set $I_t$ is
independent in $G_t$.

We now show that $C_t$ induces a cluster graph. Let $D$ be a connected
component of $G[C_t]$. If $D$ contains no vertex of $B_t$, then $D$ lies
entirely in $V_{t_1}\setminus B_t$ or entirely in $V_{t_2}\setminus B_t$, and
so $D$ is a clique of $C_{t_1}$ or of $C_{t_2}$.

So let $D$ contain a vertex $q\in B_t$, and let $Q$ be the connected component
of $G[R_t\cup S_t]$ containing $q$. As $(X_t,R_t\cup S_t)$ is a monopolar
partition of $G[B_t]$, the graph $G[R_t\cup S_t]$ is a cluster graph and the
vertices of $Q$ are pairwise adjacent. The vertices of $Q$ lie in $C_{t_1}$
and in $C_{t_2}$, so there is a clique $K_1$ of $C_{t_1}$ and a clique $K_2$
of $C_{t_2}$ containing $Q$. Any vertex of $K_1\cap B_t$ is adjacent to all of
$Q$ and lies in $R_t\cup S_t$, so it belongs to $Q$. Thus $K_1\cap B_t=Q$, and
similarly $K_2\cap B_t=Q$.

Next we show that at most one of $K_1$, $K_2$ contains a vertex outside
$B_t$. By the choice of the tripartition, either $Q\subseteq S_t$ or
$Q\subseteq R_t$. If $Q\subseteq S_t$, then $Q$ is contained in the $S$-parts
$S_t\cup R^2_t$ and $S_t\cup R^1_t$ of both child states, and so neither $K_1$
nor $K_2$ has a vertex outside $B_t$. If $Q\subseteq R_t$, then $Q$ is one of
$Q_1,\ldots,Q_p$, and hence $Q\subseteq R^1_t$ or $Q\subseteq R^2_t$. Without
loss of generality, let $Q\subseteq R^1_t$. Then $Q$ is contained in the
$S$-part $S_t\cup R^1_t$ of the state of $t_2$, so $K_2$ has no vertex outside
$B_t$, while $Q$ is contained in the $R$-part $R^1_t$ of the state of $t_1$,
so $K_1$ has a vertex outside $B_t$.

It follows that $K_1\cup K_2$ is equal to $Q$, or to $K_1$, or to $K_2$, and
in each case it is a clique. We claim that $D=K_1\cup K_2$. Since
$K_1\cup K_2$ is a clique containing $q$, it is contained in $D$. Conversely,
let $x\in K_1\cup K_2$ and let $y\in C_t$ be adjacent to $x$. Then $x$ and $y$
lie in a common $V_{t_i}$, as there are no edges between
$V_{t_1}\setminus B_t$ and $V_{t_2}\setminus B_t$. Now
$(K_1\cup K_2)\cap V_{t_i}=K_i$ because $K_1\cap B_t=K_2\cap B_t=Q$, so
$x\in K_i$, and since $y\in C_{t_i}$ is adjacent to $x$, also $y\in K_i$.
Hence there is no edge from $K_1\cup K_2$ to $C_t\setminus(K_1\cup K_2)$, and
as $D$ is connected, $D=K_1\cup K_2$. Thus $D$ is a clique. Note also that
$D\cap B_t=Q$, and that $D$ has a vertex outside $B_t$ if and only if
$Q\subseteq R_t$.

Therefore $C_t$ is a cluster graph and $(I_t,C_t)$ is a monopolar partition
of $G_t$. Moreover,
\begin{itemize}
    \item $\tilde{I}\cap V_t=(\tilde{I}\cap V_{t_1})\cup(\tilde{I}\cap V_{t_2})
    \subseteq I_{t_1}\cup I_{t_2}=I_t$;
    \item $\tilde{C}\cap V_t=(\tilde{C}\cap V_{t_1})\cup(\tilde{C}\cap V_{t_2})
    \subseteq C_{t_1}\cup C_{t_2}=C_t$;
    \item $I_t\cap B_t=(I_{t_1}\cap B_t)\cup(I_{t_2}\cap B_t)=X_t$;
    \item $C_t\cap B_t=(C_{t_1}\cap B_t)\cup(C_{t_2}\cap B_t)=R_t\cup S_t$;
    \item a clique of $C_t$ disjoint from $B_t$ contributes nothing to the
    $R$-part, and by the above a clique $D$ of $C_t$ meeting $B_t$ has a vertex
    outside $B_t$ exactly when its bag part $Q$ lies in $R_t$. Hence the union
    of $D\cap B_t$ over all cliques $D$ of $C_t$ with a vertex in
    $V_t\setminus B_t$ is exactly $R_t$.
\end{itemize}
Thus $(I_t,C_t)$ is a monopolar partition of $G_t$ satisfying all required
conditions, and we are done with the forward direction.

\paragraph{Converse Direction.}
Assume that there exists a monopolar partition $(I_t,C_t)$ of $G_t$ satisfying
\begin{itemize}
    \item $\tilde{I}\cap V_t\subseteq I_t$;
    \item $\tilde{C}\cap V_t\subseteq C_t$;
    \item $I_t\cap B_t=X_t$;
    \item $C_t\cap B_t=R_t\cup S_t$;
    \item $R_t$ is the union of $C\cap B_t$ for all cliques $C$ of $C_t$
    for which $C$ contains a vertex in $V_t\setminus B_t$.
\end{itemize}
We need to show that $T[t,X_t,R_t,S_t]=1$. For $i\in\{1,2\}$, define
\[
I_{t_i} = I_t\cap V_{t_i},
\qquad
C_{t_i} = C_t\cap V_{t_i}.
\]
Then $I_{t_i}$ is an independent set and $C_{t_i}$ is a cluster in
$G_{t_i}$, and the two sets partition $V_{t_i}$, so $(I_{t_i},C_{t_i})$ is a
monopolar partition of $G_{t_i}$. Moreover
$\tilde{I}\cap V_{t_i}\subseteq I_t\cap V_{t_i}=I_{t_i}$,
$\tilde{C}\cap V_{t_i}\subseteq C_{t_i}$, $I_{t_i}\cap B_t=X_t$ and
$C_{t_i}\cap B_t=R_t\cup S_t$.

We now define the split. Let $j\in\{1,\ldots,p\}$ and let $K$ be the clique of
$C_t$ containing $Q_j$. Two vertices of $C_t\cap B_t$ are in the same clique
of $C_t$ if and only if they are adjacent, so $K\cap B_t=Q_j$. Since
$Q_j\subseteq R_t$, the clique $K$ has a vertex in $V_t\setminus B_t$, and
this vertex lies in $V_{t_1}\setminus B_t$ or in $V_{t_2}\setminus B_t$. It
cannot have vertices in both, since the vertices of $K$ are pairwise adjacent
and there are no edges between $V_{t_1}\setminus B_t$ and
$V_{t_2}\setminus B_t$. So exactly one of the children contains the vertices
of $K$ outside $B_t$, and we assign $Q_j$ to that child. Let $R^1_t$ be the
union of the $Q_j$ assigned to $t_1$ and $R^2_t$ the union of those assigned
to $t_2$. Then $(R^1_t,R^2_t)$ is a split of $R_t$.

Now fix $i\in\{1,2\}$. Every clique of $C_{t_i}$ is of the form
$K\cap V_{t_i}$ for a clique $K$ of $C_t$. A clique $K$ disjoint from $B_t$
contributes nothing to the $R$-part of either child, so we only consider
cliques $K$ with $K\cap B_t\neq\emptyset$. If $K\cap B_t\subseteq S_t$, then
by the last condition on $(I_t,C_t)$ the clique $K$ has no vertex in
$V_t\setminus B_t$, so $K\cap V_{t_i}$ has no vertex in
$V_{t_i}\setminus B_t$. If $K\cap B_t=Q_j$ for some $j$, then by the
assignment, $K\cap V_{t_i}$ has a vertex in $V_{t_i}\setminus B_t$ if and only
if $Q_j\subseteq R^i_t$. Hence the $R$-part realised by $(I_{t_1},C_{t_1})$ is
$R^1_t$ and its $S$-part is $S_t\cup R^2_t$, while the $R$-part realised by
$(I_{t_2},C_{t_2})$ is $R^2_t$ and its $S$-part is $S_t\cup R^1_t$. Therefore
\[
T[t_1,X_t,R^1_t,S_t\cup R^2_t]=1
\quad\text{and}\quad
T[t_2,X_t,R^2_t,S_t\cup R^1_t]=1,
\]
and since $(R^1_t,R^2_t)$ is one of the splits in the recursion, we get
$T[t,X_t,R_t,S_t]=1$. This completes the proof for the join node.
\end{proof}

\section{Missing Proof of \cref{lem:join-time}}\label{appen:join-time}

\begin{proof}
We handle the tripartitions of $B_t$ grouped by their first entry. Fix a set
$X_t \subseteq B_t$ and put $Y = B_t \setminus X_t$. If $X_t$ is not an
independent set, or $G[Y]$ is not a cluster graph, or
$\tilde{I} \cap B_t \nsubseteq X_t$, or $\tilde{C} \cap B_t \nsubseteq Y$, then
every entry $T[t, X_t, R_t, S_t]$ with this first entry is zero, and there is
nothing to compute. So assume none of these happens, and let
$D_1, \dots, D_q$ be the connected components of $G[Y]$, where $q \le |Y|$.

By the condition imposed in the recursion, an entry $T[t, X_t, R_t, S_t]$ can
be one only if each $D_j$ is contained in $R_t$ or in $S_t$. Hence the
tripartitions with first entry $X_t$ that matter are in one-to-one
correspondence with the subsets $A \subseteq \{1, \dots, q\}$: the subset $A$
corresponds to the tripartition with $R_t = \bigcup_{j \in A} D_j$ and
$S_t = Y \setminus R_t$. We write $R(A)$ for this set $R_t$, and for
$i \in \{1,2\}$ we define
\[
  f_i(A) = T\bigl[t_i, X_t, R(A), Y \setminus R(A)\bigr] \in \{0,1\},
  \qquad A \subseteq \{1, \dots, q\}.
\]
Each of these $2 \cdot 2^q$ values is an entry in the table of $t_1$ or of
$t_2$, which has already been computed.

Now consider the recursion for the tripartition corresponding to $A$. The
components of $G[R_t \cup S_t] = G[Y]$ contained in $R_t = R(A)$ are exactly
the sets $D_j$ with $j \in A$, so a split of $R_t$ is the same as a partition
of $A$ into two disjoint sets $A_1$ and $A_2$, with $R^1_t = R(A_1)$ and
$R^2_t = R(A_2)$. For such a split we have
$S_t \cup R^2_t = Y \setminus R(A_1)$ and
$S_t \cup R^1_t = Y \setminus R(A_2)$, so the two child entries in the
recursion are $f_1(A_1)$ and $f_2(A_2)$. Hence
\[
  T\bigl[t, X_t, R(A), Y \setminus R(A)\bigr] = 1 \text{ if and only if }
  h(A) \;:=\; \sum_{A_1 \subseteq A} f_1(A_1) \cdot f_2(A \setminus A_1)
  \;\ge\; 1 ,
\]
where the sum is over the integers: each term is one exactly for a split
making both child entries one, and zero otherwise.

The function $h$ is the subset convolution of $f_1$ and $f_2$ over the ground
set $\{1, \dots, q\}$. Björklund, Husfeldt, Kaski and
Koivisto~\cite{DBLP:conf/stoc/BjorklundHKK07} showed that, over any
commutative ring, the subset convolution of two functions on the subsets of a
$q$-element set can be computed for all $2^q$ arguments simultaneously using
$2^q \cdot q^{\mathcal{O}(1)}$ ring operations. We apply their algorithm in
the ring of integers modulo $2^{q+1}$. Every value $h(A)$ is a sum of at most
$2^{|A|} \le 2^q$ terms, each $0$ or $1$, so $0 \le h(A) < 2^{q+1}$, and
therefore $h(A)$ is determined by its residue modulo $2^{q+1}$. In this ring
every number has at most $q+1$ bits, so each ring operation takes polynomial
time. Thus all
entries with first entry $X_t$ are computed in time
$2^q \cdot n^{\mathcal{O}(1)} \le 2^{|B_t \setminus X_t|} \cdot n^{\mathcal{O}(1)}$.

Summing over all choices of $X_t$ gives a total of
\[
  \sum_{X_t \subseteq B_t} 2^{|B_t \setminus X_t|} \cdot n^{\mathcal{O}(1)}
  \;=\; (1+2)^{|B_t|} \cdot n^{\mathcal{O}(1)}
  \;=\; 3^{|B_t|} \cdot n^{\mathcal{O}(1)} ,
\]
which completes the proof.\qed
\end{proof}

\section{Missing Proof of \Cref{th:mrtw}}\label{appenpf:mrtw}

\begin{proof}
Given a tree decomposition $\mathcal{T} = (T, \{B_t\}_{t \in V(T)})$ of width $\tw$
for a graph $G$, it is well known that one can compute, in polynomial time, a nice tree
decomposition of $G$ with the same width $\tw$~\cite{DBLP:books/sp/CyganFKLMPPS15} (see \Cref{lm:twnice}).
Hence, throughout we assume that
$\mathcal{T} = (T, \{B_t\}_{t \in V(T)})$ is a nice tree decomposition of $G$
of width $\tw$.

Using the dynamic programming recursions described above, we compute the value
$T[t,X_t,R_t,S_t]$ for every node $t \in V(T)$ and every tripartition
$(X_t,R_t,S_t)$ of the bag $B_t$.
The value $T[t,X_t,R_t,S_t]=1$ indicates the existence of a monopolar partition
$(I_t,C_t)$ of the subgraph $G_t$ such that $\tilde{I}\cap V_t\subseteq I_t$; $\tilde{C}\cap V_t\subseteq C_t$; $I_t \cap B_t = X_t$;
$C_t \cap B_t = R_t \cup S_t$; and $R_t$ consists precisely of the vertices in $B_t$
that belong to cliques of $C_t$ containing at least one vertex from
$V_t \setminus B_t$.

The computation proceeds bottom-up along the tree $T$, starting from the leaf nodes.
Let $r$ be the root of $T$.
Since $B_r=\emptyset$ and $G_r = G$, we conclude that the input graph $G$ admits a
monopolar partition extending $(\tilde{I},\tilde{C})$ if and only if
$T[r,\emptyset,\emptyset,\emptyset]=1$.

For each bag $B_t$, we evaluate $T[t,X_t,R_t,S_t]$ for all possible tripartitions
$(X_t,R_t,S_t)$.
As each vertex of $B_t$ can belong to exactly one of the three sets,
there are at most $3^{\tw+1}$ such tripartitions.
Computing a single table entry at an introduce node or at a forget node takes
only polynomial time, since each such entry depends on a constant number of
entries of the child, and hence the total time required at such a node is
$3^{\tw} \cdot n^{\Oh(1)}$. At a join node, the same bound
follows from \Cref{lem:join-time}.
Since the number of nodes in $T$ is polynomial in $n$,
the overall running time of the algorithm is
$3^{\tw}\cdot n^{\Oh(1)}$.\qed
\end{proof}

\section{Missing Proofs of \Cref{sec:fptmvd}}\label{appen:fptmvd}

\begin{lemma}
The recurrence for the introduce node is correct.
\end{lemma}

\begin{proof}
Let $t$ be an introduce node with child $t_1$, and let $v$ be the vertex introduced at $t$.
Then $B_t = B_{t_1}\cup\{v\}$ and $V_t = V_{t_1}\cup\{v\}$.
Since $v$ appears for the first time at $t$, all its neighbors in $V_t$ lie in $B_t$,
and in particular
\[
N(v)\cap (V_t\setminus B_t)=\emptyset.
\]

Fix a partition $(A_t,X_t,R_t,S_t)$ of $B_t$ such that $(X_t,R_t\cup S_t)$ is
a monopolar partition of $G[B_t\setminus A_t]$ and no connected component of
$G[R_t \cup S_t]$ intersects both $R_t$ and $S_t$; otherwise both sides of the
recurrence are $\infty$ and there is nothing to prove. We show that the value given by the recurrence equals $W[t,A_t,X_t,R_t,S_t]$.

\paragraph{($\le$) Direction.}
We show that $W[t,A_t,X_t,R_t,S_t]$ is at most the value returned by the
recurrence. If that value is $\infty$, there is nothing to show, so in each
case below we assume that the child entry on the right-hand side is finite.

We consider the cases of the recurrence.

\textbf{Case 1: $v\in A_t$.} Let $M_{t_1}$ be an optimal solution for $W[t_1,A_t\setminus\{v\},X_t,R_t,S_t]$ and $(I_{t_1},C_{t_1})$ be a monopolar partition for $G_{t_1}-M_{t_1}$ satisfying $M_{t_1} \cap B_{t_1} = A_t\setminus \{v\}$; $I_{t_1} \cap B_{t_1} = X_t$; $R_{t}$ and $S_t$ are disjoint; $C_{t_1} \cap B_{t_1} = R_t \cup S_t$; $R_t$ is the union of $C\cap B_{t_1}$ for all cliques $C$ of $C_{t_1}$ for which $C$ contains a vertex in $V_{t_1} \setminus B_{t_1}$.
Let us define
\[
M_t = M_{t_1}\cup\{v\},\quad
I_t = I_{t_1},\quad
C_t = C_{t_1}.
\]
Then $(I_t,C_t)=(I_{t_1},C_{t_1})$ is a valid monopolar partition of $G_t-M_t$ as $G_{t_1}-M_{t_1}$ and $G_t-M_t$ are same graph and thus $M_t$ is a monopolar vertex deletion set for $G_t$.
Also
\begin{itemize}
    \item $M_t\cap B_t=A_t$ as $M_t\cap B_t=(M_{t_1}\cap B_{t_1})\cup \{v\}$ and $M_{t_1}\cap B_{t_1}=A_t\setminus \{v\}$;

    \item $I_t \cap B_t=I_{t_1}\cap (B_{t_1}\cup \{v\}) =X_t$ as $v\notin I_{t_1}$;

    \item $R_t$ and $S_t$ are disjoint;

    \item $C_t \cap B_t=C_{t_1}\cap (B_{t_1}\cup \{v\})=C_{t_1}\cap B_{t_1} = R_t \cup S_t$; 

    \item For any clique $C$ in $C_t$, we have $C\cap B_t=C\cap (B_{t_1}\cup \{v\})=C\cap B_{t_1}$, as $v\notin C_t$. Thus, $R_t$ is the union of $C\cap B_t$ for all cliques $C$ of $C_t=C_{t_1}$ for which $C$ contains a vertex in $V_t \setminus B_t=V_{t_1}\setminus B_{t_1}$.
\end{itemize}
 
Hence,
\[
W[t,A_t,X_t,R_t,S_t] \le w(M_t)
= W[t_1,A_t\setminus\{v\},X_t,R_t,S_t] + w(v).
\]

\textbf{Case 2: $v\in X_t$.} Let $M_{t_1}$ be an optimal solution for
$W[t_1,A_t,X_t\setminus \{v\},R_t,S_t]$ and $(I_{t_1},C_{t_1})$ be a monopolar partition for $G_{t_1}-M_{t_1}$ satisfying $M_{t_1} \cap B_{t_1} = A_t$; $I_{t_1} \cap B_{t_1} = X_t\setminus \{v\}$; $R_{t}$ and $S_t$ are disjoint; $C_{t_1} \cap B_{t_1} = R_t \cup S_t$; $R_t$ is the union of $C\cap B_{t_1}$ for all cliques $C$ of $C_{t_1}$ for which $C$ contains a vertex in $V_{t_1} \setminus B_{t_1}$.
Let us define
\[
M_t=M_{t_1},\quad
I_t=I_{t_1}\cup\{v\},\quad
C_t=C_{t_1}.
\]
Clearly, $(I_t,C_t)$ is a partition for $V_t\setminus M_t$. Also since $(X_t, R_t \cup S_t)$ is a monopolar partition of $G[B_t\setminus A_t]$ and $v$ has no neighbors in
$V_t \setminus B_t = V_{t_1} \setminus B_{t_1}$, the set $I_t$ is independent in $G_t-M_t$.
Moreover, since $C_t = C_{t_1}$ and $C_{t_1}$ is a cluster graph, $C_t$ is also a cluster graph.
Thus, $(I_t, C_t)$ is a monopolar partition of $G_t-M_t$ and thus $M_t$ is a monopolar vertex deletion set for $G_t$.

Furthermore,
\begin{itemize}
    \item $M_t\cap B_t=M_{t_1}\cap (B_{t_1}\cup \{v\})=M_{t_1}\cap B_{t_1}=A_t$, as $v\notin M_{t_1}$;

    \item $I_t \cap B_t
= (I_{t_1} \cup \{v\}) \cap (B_{t_1} \cup \{v\})
= (I_{t_1}\cap B_{t_1})\cup \{v\}=X_t$;

\item $R_t$ and $S_t$ are disjoint;

\item $C_t \cap B_t = C_{t_1} \cap (B_{t_1}\cup \{v\})=C_{t_1}\cap B_{t_1} = R_t \cup S_t$, as $v\notin C_{t_1}$;

\item Since $V_t \setminus B_t = V_{t_1} \setminus B_{t_1}$ and $v\notin C_t$, the definition of $R_t$ is preserved.
\end{itemize}

Hence,
\[
W[t,A_t,X_t,R_t,S_t] \le W[t_1,A_t,X_t\setminus\{v\},R_t,S_t].
\]

\textbf{Case 3: $v\in R_t\cup S_t$.}

Let us first assume that $v\in R_t$. By definition, $R_t$ consists of all vertices in $B_t$ that belong to a clique
$C$ of $C_t$ such that
\[
C \cap (V_t \setminus B_t) \neq \emptyset.
\]
Since $v$ is introduced at node $t$, all neighbors of $v$ in $G_t$ lie in
the bag $B_t$, and therefore
\[
N(v) \cap (V_t \setminus B_t) = \emptyset.
\]
Hence, the clique of $C_t$ containing $v$ does not intersect
$V_t \setminus B_t$. Thus, we cannot have a monopolar vertex deletion set $M_t$ such that $(I_t, C_t)$ is a monopolar partition of $G_t-M_t$ satisfying:

\begin{itemize}
    \item $M_t \cap B_t = A_t$;
    \item $I_t \cap B_t = X_t$;
    \item $R_t$ and $S_t$ are disjoint;
    \item $C_t \cap B_t = R_t \cup S_t$;
    \item $R_t$ is the union of $C\cap B_t$ for all cliques $C$ of $C_t$ for which 
    $C$ contains a vertex in $V_t \setminus B_t$.
\end{itemize}

Hence, $W[t,A_t,X_t,R_t,S_t]=\infty.$

Now, let $v\in S_t$. Also let $M_{t_1}$ be an optimal solution for
$W[t_1,A_t,X_t,R_t,S_t\setminus \{v\}]$ and $(I_{t_1},C_{t_1})$ be a monopolar partition for $G_{t_1}-M_{t_1}$ satisfying $M_{t_1} \cap B_{t_1} = A_t$; $I_{t_1} \cap B_{t_1} = X_t$; $R_{t}$ and $S_t\setminus \{v\}$ are disjoint; $C_{t_1} \cap B_{t_1} = R_t \cup (S_t\setminus \{v\})$; $R_t$ is the union of $C\cap B_{t_1}$ for all cliques $C$ of $C_{t_1}$ for which $C$ contains a vertex in $V_{t_1} \setminus B_{t_1}$.
Let us define
\[
M_t = M_{t_1}, \qquad I_t = I_{t_1}, \qquad C_t = C_{t_1} \cup \{v\}.
\]
By the assumption on the partition, the connected component of $v$ in
$G[R_t \cup S_t]$ is contained in $S_t$, so $v$ has no neighbor in $R_t$.
Since $v$ has no neighbor in $V_t \setminus B_t$ and the vertices of $A_t$
are deleted, the neighbors of $v$ in $C_{t_1}$ are exactly the vertices of
$N_{S_t}(v)$.

If $N_{S_t}(v) = \emptyset$, then $v$ has no neighbor in $C_{t_1}$, so $C_t$
is a cluster graph in which $v$ forms a clique of its own, and this clique
contains no vertex of $V_t \setminus B_t$.

Otherwise, fix $u \in N_{S_t}(v)$ and let $K$ be the clique of $C_{t_1}$
containing $u$. The vertex $u$ lies in $S_t \setminus \{v\}$, so $K$ is not
one of the cliques contributing to $R_t$. By the definition of $R_t$ at node
$t_1$, such a clique contains no vertex of $V_{t_1} \setminus B_{t_1}$. Hence
$K \subseteq B_{t_1}$, and so $K \cup \{v\} \subseteq R_t \cup S_t$. Now
$G[R_t \cup S_t]$ is a cluster graph, since $(X_t, R_t \cup S_t)$ is a
monopolar partition of $G[B_t \setminus A_t]$. As $v$ is adjacent to $u$, the
vertex $v$ lies in the same clique of $G[R_t \cup S_t]$ as $u$, and so does
every vertex of $K$, because $K$ is a clique containing $u$. Therefore $v$ is
adjacent to every vertex of $K$, that is, $N_{S_t}(v) = K$. Consequently
$C_t$ is a cluster graph, obtained from $C_{t_1}$ by replacing $K$ with the
clique $K \cup \{v\}$, and this clique still contains no vertex of
$V_t \setminus B_t$.

The set $I_t = I_{t_1}$ is independent in $G_t - M_t$, and $(I_t, C_t)$ is a
partition of $V_t \setminus M_t$, so $(I_t, C_t)$ is a monopolar partition of
$G_t - M_t$.

Moreover,

\begin{itemize}
    \item $M_t\cap B_t=M_{t_1}\cap (B_{t_1}\cup \{v\})=M_{t_1}\cap B_{t_1}=A_t$, as $v\notin M_{t_1}$;

    \item $I_t \cap B_t
= I_{t_1} \cap (B_{t_1} \cup \{v\})
= I_{t_1}\cap B_{t_1}=X_t$, as $v\notin I_{t_1}$;

\item $R_t$ and $S_t$ are disjoint as $R_t$ and $S_t\setminus \{v\}$ are disjoint and $v\notin R_t$;

\item $C_t \cap B_t = (C_{t_1}\cap B_{t_1})\cup \{v\}= R_t \cup S_t, \text{ as } v\in C_t\cap B_t$;

\item Since $V_t \setminus B_t = V_{t_1} \setminus B_{t_1}$ and $v\notin R_t$, the definition of $R_t$ is preserved.
\end{itemize}

Thus, $M_t$ is a feasible solution for $W[t, A_t, X_t, R_t, S_t]$ with weight $w(M_{t_1})$, and hence
\[
W[t, A_t, X_t, R_t, S_t]
\le
W[t_1, A_t, X_t, R_t, S_t \setminus \{v\}].
\]

\paragraph{($\ge$) Direction.}
If $W[t,A_t,X_t,R_t,S_t]=\infty$, the inequality holds trivially. Otherwise,
let $M_t$ be a minimum-weight deletion set corresponding to
$W[t,A_t,X_t,R_t,S_t]$, and let $(I_t,C_t)$ be the corresponding
monopolar partition of $G_t-M_t$. Then we have 

\begin{itemize}
    \item $M_t \cap B_t = A_t$;
    \item $I_t \cap B_t = X_t$;
    \item $R_t$ and $S_t$ are disjoint;
    \item $C_t \cap B_t = R_t \cup S_t$;
    \item $R_t$ is the union of $C\cap B_t$ for all cliques $C$ of $C_t$ for which 
    $C$ contains a vertex in $V_t \setminus B_t$.
\end{itemize}
We distinguish cases depending on the position of $v$.

\textbf{Case 1: $v\in M_t$.} Then $v\in A_t$.
Let us consider $M_{t_1}=M_t\setminus\{v\}$; $I_{t_1}=I_{t}$; $C_{t_1}=C_{t}$.
Since the graphs $G_t-M_t$ and $G_{t_1}-M_{t_1}$ are same, $M_{t_1}$ is a monopolar vertex deletion set for $G_{t_1}$ with the monopolar partition $(I_t,C_t)=(I_{t_1},C_{t_1})$ as one monopolar partition for $G_{t_1}-M_{t_1}$ which satisfies: 

\begin{itemize}
    \item $M_{t_1} \cap B_{t_1}=(M_t\setminus \{v\})\cap (B_t\setminus \{v\})=(M_t\cap B_t)\setminus \{v\} = A_t\setminus \{v\}$;
    \item $I_{t_1} \cap B_{t_1} = I_t\cap (B_{t_1}\cup \{v\})=I_t\cap B_t = X_t$, as $v\notin I_t$;
    \item $R_t$ and $S_t$ are disjoint;
    \item $C_{t_1} \cap B_{t_1}=C_t\cap (B_{t_1}\cup \{v\})=C_t\cap B_t = R_t \cup S_t$, as $v\notin C_t$;
    \item Note that for any clique $C$ in $C_t=C_{t_1}$, $C\cap B_{t_1}=C\cap B_t$, as $v\notin C_t$ and $V_t\setminus B_t=V_{t_1}\setminus B_{t_1}$. Thus, $R_t$ is the union of $C\cap B_{t_1}$ for all cliques $C$ of $C_t=C_{t_1}$ for which 
    $C$ contains a vertex in $V_{t_1} \setminus B_{t_1}$.
\end{itemize}

This shows that $M_{t_1}$ is a valid solution for
$W[t_1,A_t\setminus\{v\},X_t,R_t,S_t]$.
Hence,
\[
W[t,A_t,X_t,R_t,S_t]
= w(M_{t_1}) + w(v)
\ge W[t_1,A_t\setminus\{v\},X_t,R_t,S_t] + w(v).
\]

\textbf{Case 2: $v\in I_t$.} Then $v\in X_t$. Let us consider $M_{t_1}=M_t$; $I_{t_1}=I_t\setminus \{v\}$; $C_{t_1}=C_t$. Since $G_{t_1}-M_{t_1}$ is a subgraph of $G_{t}-M_t$, $M_{t_1}$ is a monopolar vertex deletion set for $G_{t_1}$ and $(I_t\cap V_{t_1}, C_t\cap V_{t_1})=(I_{t_1},C_{t_1})$ is one monopolar partition for $G_{t_1}-M_{t_1}$. Moreover,

\begin{itemize}
    \item $M_{t_1} \cap B_{t_1}=M_t\cap (B_t\setminus \{v\})=M_t\cap B_t = A_t$, as $v\notin M_t$;
    \item $I_{t_1} \cap B_{t_1} = (I_t\setminus \{v\})\cap (B_{t}\setminus \{v\})=(I_t\cap B_t)\setminus \{v\} = X_t\setminus \{v\}$;
    \item $R_t$ and $S_t$ are disjoint;
    \item $C_{t_1} \cap B_{t_1}=C_t\cap (B_{t_1}\cup \{v\})=C_t\cap B_t = R_t \cup S_t$, as $v\notin C_t$;
    \item Note that for any clique $C$ in $C_t=C_{t_1}$, $C\cap B_{t_1}=C\cap B_t$, as $v\notin C_t$ and $V_t\setminus B_t=V_{t_1}\setminus B_{t_1}$. Thus, $R_t$ is the union of $C\cap B_{t_1}$ for all cliques $C$ of $C_t=C_{t_1}$ for which 
    $C$ contains a vertex in $V_{t_1} \setminus B_{t_1}$.
\end{itemize}

This shows that $M_{t_1}$ is a valid solution for
$W[t_1,A_t,X_t\setminus\{v\},R_t,S_t]$,
and therefore
\[
W[t,A_t,X_t,R_t,S_t]
\ge W[t_1,A_t,X_t\setminus\{v\},R_t,S_t].
\]

\textbf{Case 3: $v\in C_t$.} Then $v\in R_t\cup S_t$. Let us first assume that $v\in R_t$. By definition, $R_t$ consists of all vertices in $B_t$ that belong to a clique
$C$ of $C_t$ such that
\[
C \cap (V_t \setminus B_t) \neq \emptyset.
\]
Since $v$ is introduced at node $t$, all neighbors of $v$ in $G_t$ lie in
the bag $B_t$, and therefore
\[
N(v) \cap (V_t \setminus B_t) = \emptyset.
\]
Hence, the clique of $C_t$ containing $v$ does not intersect
$V_t \setminus B_t$. Thus, we can not have a monopolar vertex deletion set $M_t$ such that $(I_t, C_t)$ is a monopolar partition of $G_t-M_t$ satisfying:

\begin{itemize}
    \item $M_t \cap B_t = A_t$;
    \item $I_t \cap B_t = X_t$;
    \item $R_t$ and $S_t$ are disjoint;
    \item $C_t \cap B_t = R_t \cup S_t$;
    \item $R_t$ is the union of $C\cap B_t$ for all cliques $C$ of $C_t$ for which 
    $C$ contains a vertex in $V_t \setminus B_t$.
\end{itemize}

Hence, $W[t,A_t,X_t,R_t,S_t]=\infty.$

Now let $v\in S_t$. Let us consider $M_{t_1}=M_t$; $I_{t_1}=I_t$; $C_{t_1}=C_t\setminus \{v\}$. Since $G_{t_1}-M_{t_1}$ is a subgraph of $G_{t}-M_t$, $M_{t_1}$ is a monopolar vertex deletion set for $G_{t_1}$ and $(I_t\cap V_{t_1}, C_t\cap V_{t_1})=(I_{t_1},C_{t_1})$ is one monopolar partition for $G_{t_1}-M_{t_1}$. Moreover,

\begin{itemize}
    \item $M_{t_1} \cap B_{t_1}=M_t\cap (B_t\setminus \{v\})=M_t\cap B_t = A_t$, as $v\notin M_t$;
    \item $I_{t_1} \cap B_{t_1} = I_t\cap (B_{t_1}\cup \{v\})=I_t\cap B_t = X_t$, as $v\notin I_t$;
    \item $R_t$ and $S_t\setminus \{v\}$ are disjoint as $R_t$ and $S_t$ are disjoint;
    \item $C_{t_1} \cap B_{t_1}=(C_t\setminus \{v\})\cap (B_{t}\setminus \{v\})=(C_t\cap B_t)\setminus \{v\} = R_t \cup (S_t\setminus \{v\})$;
    \item Note that $v\notin R_t$ and $V_t\setminus B_t=V_{t_1}\setminus B_{t_1}$. Thus, $R_t$ is the union of $C\cap B_{t_1}$ for all cliques $C$ of $C_{t_1}=C_t\setminus \{v\}$ for which $C$ contains a vertex in $V_{t_1} \setminus B_{t_1}$.
\end{itemize}

This shows that $M_{t_1}$ is a valid solution for $W[t_1,A_t,X_t,R_t,S_t\setminus\{v\}]$ and hence,
\[
W[t,A_t,X_t,R_t,S_t]=w(M_{t})=w(M_{t_1})
\ge W[t_1,A_t,X_t,R_t,S_t\setminus\{v\}].
\]

This completes the proof of the lemma.
\end{proof}

\begin{lemma}
The recurrence for the forget node is correct.
\end{lemma}
\begin{proof}
Let $t$ be a forget node with child $t_1$, and let $v$ be the vertex forgotten
at $t$. Then
\[
B_{t_1} = B_t \cup \{v\}
\quad\text{and}\quad
V_t = V_{t_1}.
\]
Throughout, we use that $C_t$ is a cluster graph, so for $v \in C_t$ the
vertices of $B_{t_1}$ lying in the clique of $v$ are exactly $N_{Y_t}[v]$,
where $Y_t = R_t \cup S_t$.

Fix a partition $(A_t,X_t,R_t,S_t)$ of $B_t$ such that $(X_t,R_t\cup S_t)$ is
a monopolar partition of $G[B_t\setminus A_t]$ and no connected component of
$G[R_t \cup S_t]$ intersects both $R_t$ and $S_t$; otherwise both sides of the
recurrence are $\infty$ and there is nothing to prove.
We show that the value given by the recurrence equals
$W[t,A_t,X_t,R_t,S_t]$.

\paragraph{($\le$) Direction.}
We show that $W[t,A_t,X_t,R_t,S_t]$ is at most the value returned by the recurrence.
We consider the four terms in the minimum separately. If a term is $\infty$, it does not affect the minimum, so in each case below we assume the child entry is finite.

\medskip
\noindent\textbf{Case 1: $W[t_1,A_t\cup\{v\},X_t,R_t,S_t]$.}

Let $M_{t_1}$ be an optimal solution for $W[t_1,A_t\cup\{v\},X_t,R_t,S_t]$ and $(I_{t_1},C_{t_1})$ be one monopolar partition for $G_{t_1}-M_{t_1}$ satisfying the following.

\begin{itemize}
    \item $M_{t_1} \cap B_{t_1}=A_t\cup \{v\}$;
    \item $I_{t_1} \cap B_{t_1} = X_t$;
    \item $R_t$ and $S_t$ are disjoint;
    \item $C_{t_1} \cap B_{t_1}=R_t\cup S_t$;
    \item $R_t$ is the union of $C\cap B_{t_1}$ for all cliques $C$ of $C_{t_1}$ for which $C$ contains a vertex in $V_{t_1} \setminus B_{t_1}$.
\end{itemize}
From here, we can say that $v\in M_{t_1}$.

Since $V_t = V_{t_1}$, we have $G_t=G_{t_1}$ and $G_t-M_{t_1}=G_{t_1}-M_{t_1}$. Hence $M_{t_1}$ is a monopolar vertex deletion set for $G_t$ and $(I_{t_1},C_{t_1})$ is a monopolar partition for $G_t-M_{t_1}$. 

Moreover,
\begin{itemize}
    \item $M_{t_1} \cap B_{t}=A_t$, as $v\notin B_t$;
    \item $I_{t_1} \cap B_{t}=I_{t_1}\cap B_{t_1} = X_t$, as $v\notin I_{t_1}$;
    \item $R_t$ and $S_t$ are disjoint;
    \item $C_{t_1}\cap B_t=C_{t_1} \cap B_{t_1}=R_t\cup S_t$, as $v\notin C_{t_1}$;
    \item Since $v\notin C_{t_1}$, $R_t$ is the union of $C\cap B_{t}$ for all cliques $C$ of $C_{t_1}$ for which $C$ contains a vertex in $V_{t} \setminus B_{t}$.
\end{itemize}

Thus,
\[
W[t,A_t,X_t,R_t,S_t]
\le
W[t_1,A_t\cup\{v\},X_t,R_t,S_t].
\]

\medskip
\noindent\textbf{Case 2: $W[t_1, A_t, X_t \cup \{v\}, R_t, S_t]$.}

Let $M_{t_1}$ be an optimal solution for
$W[t_1, A_t, X_t \cup \{v\}, R_t, S_t]$,
and let $(I_{t_1}, C_{t_1})$ be a monopolar partition of
$G_{t_1} - M_{t_1}$ satisfying:

\begin{itemize}
    \item $M_{t_1} \cap B_{t_1} = A_t$;
    \item $I_{t_1} \cap B_{t_1} = X_t \cup \{v\}$;
    \item $R_t$ and $S_t$ are disjoint;
    \item $C_{t_1} \cap B_{t_1} = R_t \cup S_t$;
    \item $R_t$ is the union of $C \cap B_{t_1}$ for all cliques $C$ of $C_{t_1}$
          such that $C$ contains a vertex in $V_{t_1} \setminus B_{t_1}$.
\end{itemize}

Clearly, $v \in I_{t_1}$.
Since $V_t = V_{t_1}$, we have $G_t=G_{t_1}$ and $G_t-M_{t_1}=G_{t_1}-M_{t_1}$. Hence $M_{t_1}$ is a monopolar vertex deletion set for $G_t$ and $(I_{t_1},C_{t_1})$ is a monopolar partition for $G_t-M_{t_1}$. 

We now verify the required conditions at node $t$:

\begin{itemize}
    \item $M_{t_1} \cap B_t =M_{t_1}\cap B_{t_1}= A_t$, since $v \notin M_{t_1}$;
    \item $I_{t_1} \cap B_t = X_t$, as $v\notin B_t$;
    \item $R_t$ and $S_t$ are disjoint;
    \item $C_{t_1} \cap B_t=C_{t_1}\cap B_{t_1} = R_t \cup S_t$, since $v \notin C_{t_1}$;
    \item Since $v\notin C_{t_1}$, $R_t$ is the union of $C\cap B_{t}$ for all cliques $C$ of $C_{t_1}$ for which $C$ contains a vertex in $V_{t} \setminus B_{t}$.
\end{itemize}

Therefore, $M_{t_1}$ is a feasible solution for node $t$ with the same weight, and
\[
W[t, A_t, X_t, R_t, S_t]
\le
W[t_1, A_t, X_t \cup \{v\}, R_t, S_t].
\]

\medskip
\noindent\textbf{Case 3:
$N_{Y_t}(v)\subseteq R_t$ and
$W[t_1,A_t,X_t,R_t\cup\{v\},S_t]$.}

Let $M_{t_1}$ be an optimal solution for
$W[t_1,A_t,X_t,R_t\cup\{v\},S_t]$,
and let $(I_{t_1},C_{t_1})$ be a monopolar partition of
$G_{t_1}-M_{t_1}$ satisfying:
\begin{itemize}
    \item $M_{t_1}\cap B_{t_1}=A_t$;
    \item $I_{t_1}\cap B_{t_1}=X_t$;
    \item $(R_t\cup\{v\})$ and $S_t$ are disjoint;
    \item $C_{t_1}\cap B_{t_1}=(R_t\cup\{v\})\cup S_t$;
    \item $R_t\cup\{v\}$ is the union of $C\cap B_{t_1}$ for all cliques $C$ of
          $C_{t_1}$ such that $C$ contains a vertex in
          $V_{t_1}\setminus B_{t_1}$.
\end{itemize}

Here we have $v\in C_{t_1}$.
Let $C'$ be the clique of $C_{t_1}$ containing $v$.
By definition of $R_t\cup\{v\}$, this clique $C'$ intersects
$V_{t_1}\setminus B_{t_1}$.

Again $V_t = V_{t_1}$, we have $G_t=G_{t_1}$ and $G_t-M_{t_1}=G_{t_1}-M_{t_1}$. Hence $M_{t_1}$ is a monopolar vertex deletion set for $G_t$ and $(I_{t_1},C_{t_1})$ is a monopolar partition for $G_t-M_{t_1}$. 

We now verify the required conditions at node $t$:

\begin{itemize}
    \item $M_{t_1} \cap B_t =M_{t_1}\cap B_{t_1}= A_t$, since $v \notin M_{t_1}$;
    \item $I_{t_1} \cap B_t=I_{t_1}\cap B_{t_1} = X_t$, as $v\notin I_{t_1}$;
    \item $R_t$ and $S_t$ are disjoint, as $v\notin R_t\cup S_t$, $R_t\cup\{v\}$ and $S_t$ are disjoint;
    \item $C_{t_1} \cap B_t=(C_{t_1}\cap B_{t_1})\setminus \{v\} = R_t \cup S_t$, since $v \notin B_{t}$;
    \item Let $C'$ be the clique of $C_{t_1}$ containing $v$. Then notice that $R_t\setminus C'$ is the union of $C\cap B_{t}$ for all cliques $C$ of $C_{t_1}-C'$ for which $C$ contains a vertex in $V_{t} \setminus B_{t}$. Again, since $C'\cap B_{t_1}=N_{Y_t}[v]$, we have $R_t\setminus C'=R_t\setminus N_{Y_t}(v)$. We also have $C'\cap B_t=N_{Y_t}(v)$, as $v\notin B_t$ and $N_{Y_t}(v)\subseteq R_t$. Thus, ultimately, $(R_t\setminus N_{Y_t}(v))\cup N_{Y_t}(v)=R_t$ is the union of $C\cap B_{t}$ for all cliques $C$ of $C_{t_1}$ for which $C$ contains a vertex in $V_{t} \setminus B_{t}$.
\end{itemize}

Hence, $M_{t_1}$ is a feasible solution for node $t$ with the same weight, and
\[
W[t,A_t,X_t,R_t,S_t]
\le
W[t_1,A_t,X_t,R_t\cup\{v\},S_t].
\]

\medskip
\noindent\textbf{Case 4:
$N_{Y_t}(v)\subseteq R_t$ and
$W[t_1,A_t,X_t,R_t\setminus N_{Y_t}(v),S_t\cup N_{Y_t}[v]]$.}

Let $M_{t_1}$ be an optimal solution for
$W[t_1,A_t,X_t,R_t\setminus N_{Y_t}(v),S_t\cup N_{Y_t}[v]]$,
and let $(I_{t_1},C_{t_1})$ be a monopolar partition of
$G_{t_1}-M_{t_1}$ satisfying:
\begin{itemize}
    \item $M_{t_1}\cap B_{t_1}=A_t$;
    \item $I_{t_1}\cap B_{t_1}=X_t$;
    \item $(R_t\setminus N_{Y_t}(v))$ and $(S_t\cup N_{Y_t}[v])$ are disjoint;
    \item $C_{t_1}\cap B_{t_1}
    =(R_t\setminus N_{Y_t}(v))\cup (S_t\cup N_{Y_t}[v])$;
    \item
    $R_t\setminus N_{Y_t}(v)$ is the union of $C\cap B_{t_1}$ for all cliques $C$
    of $C_{t_1}$ such that $C$ contains a vertex in
    $V_{t_1}\setminus B_{t_1}$.
\end{itemize}

Since $v\in N_{Y_t}[v]\subseteq S_t\cup N_{Y_t}[v]$, we have $v\in C_{t_1}$.
Let $C'$ be the clique of $C_{t_1}$ containing $v$.
By the definition of this case, the clique $C'$ does \emph{not} intersect $V_{t_1}\setminus B_{t_1}$.

Again $V_t = V_{t_1}$, we have $G_t=G_{t_1}$ and $G_t-M_{t_1}=G_{t_1}-M_{t_1}$. Hence $M_{t_1}$ is a monopolar vertex deletion set for $G_t$ and $(I_{t_1},C_{t_1})$ is a monopolar partition for $G_t-M_{t_1}$. 

We now verify the required conditions at node $t$:

\begin{itemize}
    \item $M_{t_1} \cap B_t =M_{t_1}\cap B_{t_1}= A_t$, since $v \notin M_{t_1}$;
    \item $I_{t_1} \cap B_t=I_{t_1}\cap B_{t_1} = X_t$, as $v\notin I_{t_1}$;
    \item $R_t$ and $S_t$ are disjoint, as $N_{Y_t}(v)\subseteq R_t$, $v\notin R_t\cup S_t$, $(R_t\setminus N_{Y_t}(v))$ and $(S_t\cup N_{Y_t}[v])$ are disjoint;
    \item $C_{t_1} \cap B_t=(C_{t_1}\cap B_{t_1})\setminus \{v\}=(R_t\setminus N_{Y_t}(v))\cup (S_t\cup N_{Y_t}(v))= R_t \cup S_t$, since $v \notin B_{t}$;

    \item
    We now verify the defining property of $R_t$.
    Since $v\notin B_t$ and $V_t=V_{t_1}$, we have $v\in V_t\setminus B_t$.
    Thus, the clique $C'$ containing $v$ now intersects
    $V_t\setminus B_t$.

    By the definition of $R_t$, all bag vertices of this clique must belong to
    $R_t$. These vertices are exactly $N_{Y_t}(v)$.
    Accordingly, the vertices of $N_{Y_t}(v)$ are part of $R_t$.
 All other cliques are unaffected and continue to contribute to $R_t$ exactly as before.
\end{itemize}

Thus, $M_{t_1}$ is a feasible solution for node $t$ with the same weight, and
\[
W[t,A_t,X_t,R_t,S_t]
\le
W[t_1,A_t,X_t,R_t\setminus N_{Y_t}(v),S_t\cup N_{Y_t}[v]].
\]

Finally, we get   \[
    \begin{aligned}
    W[t,A_t,X_t,R_t,S_t]
    \leq \min \Big\{&
        W[t_1, A_t \cup \{v\}, X_t, R_t, S_t], \\[3pt]
        & W[t_1, A_t, X_t \cup \{v\}, R_t, S_t], \\[3pt]
        & [N_{Y_t}(v) \subseteq R_t]\; W[t_1, A_t, X_t, R_t \cup \{v\}, S_t], \\[3pt]
        & [N_{Y_t}(v) \subseteq R_t]\; W[t_1, A_t, X_t, R_t \setminus N_{Y_t}(v), S_t \cup N_{Y_t}[v]]
    \Big\},
    \end{aligned}
    \]
    where $Y_t = R_t \cup S_t$ and by $[N_{Y_t}(v) \subseteq R_t]$ we mean we would consider the term only if $N_{Y_t}(v) \subseteq R_t$.

\paragraph{$(\ge)$ Direction.}

If $W[t,A_t,X_t,R_t,S_t]=\infty$, the inequality holds trivially. Otherwise,
let $M_t$ be a minimum-weight monopolar vertex-deletion set witnessing
$W[t,A_t,X_t,R_t,S_t]$, and let $(I_t,C_t)$ be a monopolar partition of
$G_t-M_t$ satisfying:
\begin{enumerate}
    \item $M_t \cap B_t = A_t$;
    \item $I_t \cap B_t = X_t$;
    \item $R_t$ and $S_t$ are disjoint;
    \item $C_t \cap B_t = R_t \cup S_t$;
    \item $R_t$ is the union of $C\cap B_t$ for all cliques $C$ of $C_t$
          for which $C$ contains a vertex in $V_t \setminus B_t$.
\end{enumerate}
We show that the value returned by the recurrence is at most $w(M_t)$.

Since $t$ is a forget node with child $t_1$, we have
\[
B_{t_1} = B_t \cup \{v\}
\quad\text{and}\quad
V_t = V_{t_1}.
\]
The graphs $G_t$ and $G_{t_1}$ are identical. Since $G_t = G_{t_1}$, we may set
\[
M_{t_1} = M_t, \qquad I_{t_1} = I_t, \qquad C_{t_1} = C_t,
\]
so that $M_{t_1}$ is a monopolar vertex deletion set for $G_{t_1}$ and
$(I_{t_1}, C_{t_1})$ is a monopolar partition of $G_{t_1} - M_{t_1}$. We
distinguish cases according to the position of $v$ in $(M_t, I_t, C_t)$.

\medskip
\noindent\textbf{Case 1: $v \in M_t$.}

Since, $V_t = V_{t_1}$, we have $G_t=G_{t_1}$ and $G_t-M_{t}=G_{t_1}-M_{t}$. Hence $M_{t}$ is a monopolar vertex deletion set for $G_t$ and $(I_{t},C_{t})$ is a monopolar partition for $G_t-M_{t}$.

We verify the bag conditions at node $t_1$:
\begin{itemize}
    \item $M_{t}\cap B_{t_1}=M_t\cap (B_t\cup\{v\}) = A_t \cup \{v\}$, as $v\in M_t$;
    \item $I_{t}\cap B_{t_1} = X_t$, as $v\notin I_t$;
    \item $C_{t}\cap B_{t_1} = R_t \cup S_t$, as $v\notin C_t$;
\item Since $v \notin C_t$ and $V_{t_1}\setminus B_{t_1} = (V_t\setminus B_t)\setminus \{v\}$,
a clique of $C_t$ contains a vertex of $V_{t_1}\setminus B_{t_1}$ if and only
if it contains a vertex of $V_t\setminus B_t$. Hence $R_t$ is also the union of
$C \cap B_{t_1}$ over all cliques $C$ of $C_t$ containing a vertex of
$V_{t_1}\setminus B_{t_1}$, as $v \notin C$ for every such clique.
\end{itemize}

Thus,
\[
W[t_1,A_t\cup\{v\},X_t,R_t,S_t]
\le w(M_{t_1}) = w(M_t),
\]
and therefore the recurrence value is at most $w(M_t)$.

\medskip
\noindent\textbf{Case 2: $v \in I_t$.}

Since, $V_t = V_{t_1}$, we have $G_t=G_{t_1}$ and $G_t-M_{t}=G_{t_1}-M_{t}$. Hence $M_{t}$ is a monopolar vertex deletion set for $G_t$ and $(I_{t},C_{t})$ is a monopolar partition for $G_t-M_{t}$.

We verify the bag conditions at node $t_1$:
\begin{itemize}
        \item $M_{t}\cap B_{t_1}=M_t\cap (B_t\cup\{v\}) = A_t$, since $v\notin M_t$;
    \item $I_{t}\cap B_{t_1}=I_t\cap (B_t\cup \{v\}) = X_t\cup \{v\}$, as $v\in I_t$;
    \item $C_{t}\cap B_{t_1} = R_t \cup S_t$, as $v\notin C_t$;
\item Since $v \notin C_t$ and $V_{t_1}\setminus B_{t_1} = (V_t\setminus B_t)\setminus \{v\}$,
a clique of $C_t$ contains a vertex of $V_{t_1}\setminus B_{t_1}$ if and only
if it contains a vertex of $V_t\setminus B_t$. Hence $R_t$ is also the union of
$C \cap B_{t_1}$ over all cliques $C$ of $C_t$ containing a vertex of
$V_{t_1}\setminus B_{t_1}$, as $v \notin C$ for every such clique.
\end{itemize}

Thus,
\[
W[t_1,A_t,X_t\cup\{v\},R_t,S_t]
\le w(M_t),
\]
and again the recurrence value is at most $w(M_t)$.

\medskip
\noindent\textbf{Case 3: $v \in C_t$ and the clique containing $v$
intersects $V_{t_1}\setminus B_{t_1}$.}

Let $C'$ be the clique of $C_t$ containing $v$.
Since $v\in V_t\setminus B_t$, by the definition of $R_t$ we have
\[
C'\cap B_t = N_{Y_t}(v) \subseteq R_t.
\]

Again $V_t = V_{t_1}$, we have $G_t=G_{t_1}$ and $G_t-M_{t}=G_{t_1}-M_{t}$. Hence $M_{t}$ is a monopolar vertex deletion set for $G_t$ and $(I_{t},C_{t})$ is a monopolar partition for $G_t-M_{t}$.

We verify the bag conditions at node $t_1$:
\begin{itemize}
    \item $M_{t}\cap B_{t_1}=M_t\cap (B_t\cup\{v\}) = A_t$, as $v\notin M_t$;
    \item $I_{t}\cap B_{t_1} = X_t$, as $v\notin I_t$;
    \item $C_{t}\cap B_{t_1}=C_t\cap (B_t\cup \{v\})  = (R_t\cup \{v\}) \cup S_t$, as $v\in C_t$;
\item We now justify the definition of $R_t\cup\{v\}$ at node $t_1$.
Let $C'$ be the clique of $C_t$ containing $v$.
By assumption of this case, $C'$ intersects $V_{t_1}\setminus B_{t_1}$.
Hence, by the defining property of $R_t$ at node $t$,
all vertices of $C'\cap B_t = N_{Y_t}(v)$ belong to $R_t$.

At node $t_1$, the bag is $B_{t_1}=B_t\cup\{v\}$.
Thus, the bag intersection of $C'$ at node $t_1$ is
\[
C'\cap B_{t_1} = (C'\cap B_t)\cup\{v\} = N_{Y_t}(v)\cup\{v\}.
\]

All other cliques of $C_t$ that intersect $V_{t_1}\setminus B_{t_1}$
are unaffected by the presence of $v$ and contribute exactly the same
bag vertices as before.
Therefore, the union of $C\cap B_{t_1}$ over all such cliques is precisely
$R_t\cup\{v\}$.

\end{itemize}

Thus,
\[
W[t_1,A_t,X_t,R_t\cup\{v\},S_t]
\le w(M_t),
\]
which is one of the recurrence terms.

\medskip
\noindent\textbf{Case 4: $v \in C_t$ and the clique containing $v$
does not intersect $V_{t_1}\setminus B_{t_1}$.}

Let $C'$ be the clique of $C_t$ containing $v$.
Then $C'$ is fully contained in $B_{t_1}$ and does not intersect
$V_{t_1}\setminus B_{t_1}$.
However, at node $t$, the vertex $v$ lies in $V_t\setminus B_t$, and hence
$C'$ now intersects $V_t\setminus B_t$ via $v$.

By the definition of $R_t$, all vertices of $C'\cap B_t = N_{Y_t}(v)$ must
belong to $R_t$.

Again $V_t = V_{t_1}$, we have $G_t=G_{t_1}$ and $G_t-M_{t}=G_{t_1}-M_{t}$. Hence $M_{t}$ is a monopolar vertex deletion set for $G_t$ and $(I_{t},C_{t})$ is a monopolar partition for $G_t-M_{t}$.

We verify the bag conditions at node $t_1$:
\begin{itemize}
    \item $M_{t}\cap B_{t_1}=M_t\cap (B_t\cup\{v\}) = A_t$, as $v\notin M_t$;
    \item $I_{t}\cap B_{t_1} = X_t$, as $v\notin I_t$;
    \item $C_{t}\cap B_{t_1}=C_t\cap (B_t\cup \{v\})  = (R_t\cup \{v\}) \cup S_t=(R_t\setminus N_{Y_t}(v))\cup (S_t\cup N_{Y_t}[v])$, as $v\in C_t$;
\item We now justify the definition of $R_t\setminus N_{Y_t}(v)$ at node $t_1$.
Let $C'$ be the clique of $C_t$ containing $v$.
By assumption of this case, $C'$ does not intersect
$V_{t_1}\setminus B_{t_1}$, and hence it does not contribute any bag
vertices to the $R$-set at node $t_1$.

At node $t$, the vertex $v$ lies in $V_t\setminus B_t$.
Therefore, the clique $C'$ intersects $V_t\setminus B_t$ via $v$,
and by the definition of $R_t$, all vertices of
\[
C'\cap B_t = N_{Y_t}(v)
\]
must belong to $R_t$.

For all other cliques $C''$ of $C_t$ that intersect
$V_{t_1}\setminus B_{t_1}$, their intersections with $B_t$ are unchanged
when passing from node $t_1$ to node $t$.
Hence, these cliques contribute exactly the same bag vertices to $R_t$
as they did at node $t_1$.

Consequently, the union of $C\cap B_{t_1}$ over all cliques $C$ of $C_{t_1}$
intersecting $V_{t_1}\setminus B_{t_1}$ is precisely $R_t \setminus N_{Y_t}(v),$ as claimed.

\end{itemize}

Therefore,
\[
W[t_1,A_t,X_t,R_t\setminus N_{Y_t}(v),S_t\cup N_{Y_t}[v]]
\le w(M_t),
\]
which again matches one of the recurrence terms.

\medskip
Since in all possible cases the recurrence returns a value at most $w(M_t)$,
we conclude that
\[
    \begin{aligned}
    W[t,A_t,X_t,R_t,S_t]
    \geq \min \Big\{&
        W[t_1, A_t \cup \{v\}, X_t, R_t, S_t], \\[3pt]
        & W[t_1, A_t, X_t \cup \{v\}, R_t, S_t], \\[3pt]
        & [N_{Y_t}(v) \subseteq R_t]\; W[t_1, A_t, X_t, R_t \cup \{v\}, S_t], \\[3pt]
        & [N_{Y_t}(v) \subseteq R_t]\; W[t_1, A_t, X_t, R_t \setminus N_{Y_t}(v), S_t \cup N_{Y_t}[v]]
    \Big\},
    \end{aligned}
    \]
    where $Y_t = R_t \cup S_t$ and by $[N_{Y_t}(v) \subseteq R_t]$ we mean we would consider the term only if $N_{Y_t}(v) \subseteq R_t$.
This completes the proof of the $(\ge)$ direction.

Combining $(\leq)$ and $(\geq)$ direction, we get \[
    \begin{aligned}
    W[t,A_t,X_t,R_t,S_t]
    = \min \Big\{&
        W[t_1, A_t \cup \{v\}, X_t, R_t, S_t], \\[3pt]
        & W[t_1, A_t, X_t \cup \{v\}, R_t, S_t], \\[3pt]
        & [N_{Y_t}(v) \subseteq R_t]\; W[t_1, A_t, X_t, R_t \cup \{v\}, S_t], \\[3pt]
        & [N_{Y_t}(v) \subseteq R_t]\; W[t_1, A_t, X_t, R_t \setminus N_{Y_t}(v), S_t \cup N_{Y_t}[v]]
    \Big\},
    \end{aligned}
    \]
    where $Y_t = R_t \cup S_t$ and by $[N_{Y_t}(v) \subseteq R_t]$ we mean we would consider the term only if $N_{Y_t}(v) \subseteq R_t$.

    Hence, the proof for the forget node.
\end{proof}

\begin{lemma}
The recurrence for the join node is correct.
\end{lemma}

\begin{proof}
Let $t$ be a join node with children $t_1$ and $t_2$. Recall that
\[
B_{t_1} = B_{t_2} = B_t,
\qquad
V_t = V_{t_1} \cup V_{t_2},
\qquad
V_{t_1} \cap V_{t_2} = B_t,
\]
and that there are no edges between $V_{t_1}\setminus B_t$ and
$V_{t_2}\setminus B_t$. Let $Q_1,\ldots,Q_p$ be the connected components of
$G[R_t\cup S_t]$ contained in $R_t$. For a state $(A,X,R,S)$ of a node, we
call $R$ the $R$-part and $S$ the $S$-part of the state. By the definition of
the table, if a solution realises a state, then the $R$-part consists of the
bag parts of exactly those cliques which contain a vertex outside the bag. In particular, a clique with non-empty bag part lying in the $S$-part has no
vertex outside the bag.

Fix a partition $(A_t,X_t,R_t,S_t)$ of $B_t$ such that $(X_t,R_t\cup S_t)$ is
a monopolar partition of $G[B_t\setminus A_t]$ and no connected component of
$G[R_t\cup S_t]$ intersects both $R_t$ and $S_t$; otherwise both sides of the
recurrence are $\infty$.

\paragraph{$(\le)$ Direction.}
Let $(R^1_t,R^2_t)$ be a split of $R_t$ at which the minimum in the recurrence
is attained. We may assume that both child entries are finite. For
$i\in\{1,2\}$, let $M_{t_i}$ be an optimal solution for the entry of $t_i$,
and let $(I_{t_i},C_{t_i})$ be a monopolar partition of $G_{t_i}-M_{t_i}$
realising the corresponding state, that is, $M_{t_i}\cap B_t=A_t$,
$I_{t_i}\cap B_t=X_t$, $C_{t_i}\cap B_t=R_t\cup S_t$, the $R$-part of the
state of $t_1$ is $R^1_t$ and that of $t_2$ is $R^2_t$.
Let us define
\[
M_t = M_{t_1}\cup M_{t_2},
\qquad
I_t = I_{t_1}\cup I_{t_2},
\qquad
C_t = C_{t_1}\cup C_{t_2}.
\]
Since $V_{t_1}\cap V_{t_2}=B_t$ and the two solutions agree on the bag, the
sets $M_t$, $I_t$, $C_t$ are pairwise disjoint and form a partition of $V_t$.
Since $I_{t_1}$ and $I_{t_2}$ are independent and there are no edges between
$V_{t_1}\setminus B_t$ and $V_{t_2}\setminus B_t$, the set $I_t$ is
independent in $G_t-M_t$.

We now show that $C_t$ induces a cluster graph. Let $D$ be a connected
component of $G[C_t]$. If $D$ contains no vertex of $B_t$, then $D$ lies
entirely in $V_{t_1}\setminus B_t$ or entirely in $V_{t_2}\setminus B_t$, and
so $D$ is a clique of $C_{t_1}$ or of $C_{t_2}$.

So let $D$ contain a vertex $q\in B_t$, and let $Q$ be the connected component
of $G[R_t\cup S_t]$ containing $q$. As $(X_t,R_t\cup S_t)$ is a monopolar
partition of $G[B_t\setminus A_t]$, the graph $G[R_t\cup S_t]$ is a cluster
graph and the vertices of $Q$ are pairwise adjacent. The vertices of $Q$ lie
in $C_{t_1}$ and in $C_{t_2}$, so there is a clique $K_1$ of $C_{t_1}$ and a
clique $K_2$ of $C_{t_2}$ containing $Q$. Any vertex of $K_1\cap B_t$ is
adjacent to all of $Q$ and lies in $R_t\cup S_t$, so it belongs to $Q$. Thus
$K_1\cap B_t=Q$, and similarly $K_2\cap B_t=Q$.

Next we show that at most one of $K_1$, $K_2$ contains a vertex outside
$B_t$. By the choice of the partition, either $Q\subseteq S_t$ or
$Q\subseteq R_t$. If $Q\subseteq S_t$, then $Q$ is contained in the $S$-parts
$S_t\cup R^2_t$ and $S_t\cup R^1_t$ of both child states, and so neither $K_1$
nor $K_2$ has a vertex outside $B_t$. If $Q\subseteq R_t$, then $Q$ is one of
$Q_1,\ldots,Q_p$, and hence $Q\subseteq R^1_t$ or $Q\subseteq R^2_t$. Without
loss of generality, let $Q\subseteq R^1_t$. Then $Q$ is contained in the
$S$-part $S_t\cup R^1_t$ of the state of $t_2$, so $K_2$ has no vertex outside
$B_t$, while $Q$ is contained in the $R$-part $R^1_t$ of the state of $t_1$,
so $K_1$ has a vertex outside $B_t$.

It follows that $K_1\cup K_2$ is equal to $Q$, or to $K_1$, or to $K_2$, and
in each case it is a clique. We claim that $D=K_1\cup K_2$. Since
$K_1\cup K_2$ is a clique containing $q$, it is contained in $D$. Conversely,
let $x\in K_1\cup K_2$ and let $y\in C_t$ be adjacent to $x$. Then $x$ and $y$
lie in a common $V_{t_i}$, as there are no edges between
$V_{t_1}\setminus B_t$ and $V_{t_2}\setminus B_t$. Now
$(K_1\cup K_2)\cap V_{t_i}=K_i$ because $K_1\cap B_t=K_2\cap B_t=Q$, so
$x\in K_i$, and since $y\in C_{t_i}$ is adjacent to $x$, also $y\in K_i$.
Hence there is no edge from $K_1\cup K_2$ to $C_t\setminus(K_1\cup K_2)$, and
as $D$ is connected, $D=K_1\cup K_2$. Thus $D$ is a clique. Note also that
$D\cap B_t=Q$, and that $D$ has a vertex outside $B_t$ if and only if
$Q\subseteq R_t$.

Therefore $C_t$ is a cluster graph, $(I_t,C_t)$ is a monopolar partition of
$G_t-M_t$, and $M_t$ is a monopolar vertex deletion set of $G_t$. Moreover,
\begin{itemize}
    \item $M_t\cap B_t=(M_{t_1}\cap B_t)\cup(M_{t_2}\cap B_t)=A_t$;
    \item $I_t\cap B_t=(I_{t_1}\cap B_t)\cup(I_{t_2}\cap B_t)=X_t$;
    \item $C_t\cap B_t=(C_{t_1}\cap B_t)\cup(C_{t_2}\cap B_t)=R_t\cup S_t$;
    \item a clique of $C_t$ disjoint from $B_t$ contributes nothing to the
    $R$-part, and by the above a clique $D$ of $C_t$ meeting $B_t$ has a vertex
    outside $B_t$ exactly when its bag part $Q$ lies in $R_t$. Hence the union
    of $D\cap B_t$ over all cliques $D$ of $C_t$ with a vertex in
    $V_t\setminus B_t$ is exactly $R_t$.
\end{itemize}
Thus $M_t$ is a feasible solution for $W[t,A_t,X_t,R_t,S_t]$. Finally, since
$M_{t_1}\cap M_{t_2}=A_t$, we have $w(M_t)=w(M_{t_1})+w(M_{t_2})-w(A_t)$,
and hence
\[
W[t,A_t,X_t,R_t,S_t]
\le
W[t_1,A_t,X_t,R^1_t,S_t\cup R^2_t]
+
W[t_2,A_t,X_t,R^2_t,S_t\cup R^1_t]
-
w(A_t).
\]

\paragraph{$(\ge)$ Direction.}
Let $M_t$ be a minimum-weight monopolar vertex-deletion set witnessing
$W[t,A_t,X_t,R_t,S_t]$, and let $(I_t,C_t)$ be a monopolar partition of
$G_t-M_t$ satisfying the four conditions. If no such $M_t$ exists, the left
side is $\infty$ and we are done. For $i\in\{1,2\}$, define
\[
M_{t_i} = M_t\cap V_{t_i},
\qquad
I_{t_i} = I_t\cap V_{t_i},
\qquad
C_{t_i} = C_t\cap V_{t_i}.
\]
Then $I_{t_i}$ is an independent set and $C_{t_i}$ is a cluster in
$G_{t_i}-M_{t_i}$, and the three sets partition $V_{t_i}$. So $M_{t_i}$ is a
monopolar vertex deletion set for $G_{t_i}$ with $(I_{t_i},C_{t_i})$ as a
monopolar partition of $G_{t_i}-M_{t_i}$, and
$M_{t_i}\cap B_t=A_t$, $I_{t_i}\cap B_t=X_t$, $C_{t_i}\cap B_t=R_t\cup S_t$.

We now define the split. Let $j\in\{1,\ldots,p\}$ and let $K$ be the clique of
$C_t$ containing $Q_j$. Two vertices of $C_t\cap B_t$ are in the same clique
of $C_t$ if and only if they are adjacent, so $K\cap B_t=Q_j$. Since
$Q_j\subseteq R_t$, the clique $K$ has a vertex in $V_t\setminus B_t$, and
this vertex lies in $V_{t_1}\setminus B_t$ or in $V_{t_2}\setminus B_t$. It
cannot have vertices in both, since the vertices of $K$ are pairwise adjacent
and there are no edges between $V_{t_1}\setminus B_t$ and
$V_{t_2}\setminus B_t$. So exactly one of the children contains the vertices
of $K$ outside $B_t$, and we assign $Q_j$ to that child. Let $R^1_t$ be the
union of the $Q_j$ assigned to $t_1$ and $R^2_t$ the union of those assigned
to $t_2$. Then $(R^1_t,R^2_t)$ is a split of $R_t$.

Now fix $i\in\{1,2\}$. Every clique of $C_{t_i}$ is of the form
$K\cap V_{t_i}$ for a clique $K$ of $C_t$. A clique $K$ disjoint from $B_t$
contributes nothing to the $R$-part of either child, so we only consider
cliques $K$ with $K\cap B_t\neq\emptyset$. If $K\cap B_t\subseteq S_t$, then
by the last condition on $(I_t,C_t)$ the clique $K$ has no vertex in
$V_t\setminus B_t$, so $K\cap V_{t_i}$ has no vertex in
$V_{t_i}\setminus B_t$. If $K\cap B_t=Q_j$ for some $j$, then by the
assignment, $K\cap V_{t_i}$ has a vertex in $V_{t_i}\setminus B_t$ if and only
if $Q_j\subseteq R^i_t$. Hence the $R$-part realised by $(I_{t_1},C_{t_1})$ is
$R^1_t$ and its $S$-part is $S_t\cup R^2_t$, while the $R$-part realised by
$(I_{t_2},C_{t_2})$ is $R^2_t$ and its $S$-part is $S_t\cup R^1_t$. Therefore
\[
W[t_1,A_t,X_t,R^1_t,S_t\cup R^2_t]\le w(M_{t_1})
\quad\text{and}\quad
W[t_2,A_t,X_t,R^2_t,S_t\cup R^1_t]\le w(M_{t_2}).
\]
Since $M_{t_1}\cup M_{t_2}=M_t$ and $M_{t_1}\cap M_{t_2}=A_t$, we have
$w(M_{t_1})+w(M_{t_2})=w(M_t)+w(A_t)$. Adding the two inequalities and
subtracting $w(A_t)$,
\[
\begin{aligned}
W[t_1,A_t,X_t,R^1_t,S_t\cup R^2_t]
&+ W[t_2,A_t,X_t,R^2_t,S_t\cup R^1_t] - w(A_t)\\
&\le w(M_t) = W[t,A_t,X_t,R_t,S_t].
\end{aligned}
\]
As $(R^1_t,R^2_t)$ is one of the splits in the recurrence, the value of the
recurrence is at most $W[t,A_t,X_t,R_t,S_t]$.

From both directions, the recurrence computes $W[t,A_t,X_t,R_t,S_t]$.
\end{proof}

\section{Missing Proof of \cref{lem:join-time-mvd}}\label{appen:join-time-mvd}

\begin{proof}
We first prove the bound $5^{|B_t|}\cdot n^{\Oh(1)}$. For a partition
$(A_t,X_t,R_t,S_t)$ of $B_t$ and a split $(R^1_t,R^2_t)$ of $R_t$, the term
inside the minimum is the sum of two entries of the tables of $t_1$ and $t_2$,
which have already been computed, so it is evaluated in polynomial time. Every such pair assigns each vertex of $B_t$ to exactly one of $A_t$, $X_t$,
$S_t$, $R^1_t$, $R^2_t$, and distinct pairs give distinct assignments. Hence
there are at most $5^{|B_t|}$ pairs of a partition and a split. Enumerating
all $5^{|B_t|}$ assignments and discarding those that do not arise from a
partition and a split, we compute all entries at $t$ in time
$5^{|B_t|}\cdot n^{\Oh(1)}$.

Now let the weights be integers bounded by a polynomial in $n$, and let
$\Omega = w(V(G))$; then $\Omega$ is bounded by a polynomial in $n$, and every
finite table entry lies between $0$ and $\Omega$. We handle the partitions of
$B_t$ grouped by their first two entries. Fix disjoint sets
$A_t, X_t\subseteq B_t$ and put $Y = B_t\setminus(A_t\cup X_t)$. If
$(X_t,Y)$ is not a monopolar partition of $G[B_t\setminus A_t]$, then every
entry with these first two entries is $\infty$, and we set them accordingly.
So assume it is, and let $D_1,\ldots,D_q$ be the connected components of
$G[Y]$, where $q\le |Y|$.

An entry $W[t,A_t,X_t,R_t,S_t]$ can be finite only if each $D_j$ is contained
in $R_t$ or in $S_t$. Hence the partitions with first two entries $A_t, X_t$
that can have a finite value are exactly those of the form
$(A_t,X_t,R(A),Y\setminus R(A))$ for a subset $A\subseteq\{1,\ldots,q\}$,
where $R(A)=\bigcup_{j\in A}D_j$. For $i\in\{1,2\}$ and
$A\subseteq\{1,\ldots,q\}$, let
\[
g_i(A) = W\bigl[t_i,A_t,X_t,R(A),Y\setminus R(A)\bigr] \in
\{0,1,\ldots,\Omega\}\cup\{\infty\},
\]
which is an entry of the table of $t_i$. As in the proof of
\cref{lem:join-time}, a split of $R(A)$ is the same as a partition of $A$ into
two sets $A_1$ and $A_2$, and the two child entries in the recurrence are
$g_1(A_1)$ and $g_2(A_2)$. So the recurrence reads
\[
W\bigl[t,A_t,X_t,R(A),Y\setminus R(A)\bigr]
= \min_{A_1\subseteq A}\bigl(g_1(A_1)+g_2(A\setminus A_1)\bigr) - w(A_t).
\]

To compute this minimum for all $A$ at once, we encode values as exponents.
Let $\mathcal{R}$ be the ring of polynomials in one variable $x$ with
coefficients in the integers modulo $2^{q+1}$. For $i\in\{1,2\}$ define
$P_i:2^{\{1,\ldots,q\}}\to\mathcal{R}$ by $P_i(A)=x^{g_i(A)}$ if $g_i(A)$ is
finite and $P_i(A)=0$ otherwise. Consider the subset convolution
\[
P(A) = \sum_{A_1\subseteq A} P_1(A_1)\cdot P_2(A\setminus A_1)
= \sum_{\substack{A_1\subseteq A\\ g_1(A_1),\,g_2(A\setminus A_1)<\infty}}
  x^{\,g_1(A_1)+g_2(A\setminus A_1)} .
\]
The coefficient of $x^a$ in $P(A)$ counts the sets $A_1\subseteq A$ with
$g_1(A_1)+g_2(A\setminus A_1)=a$. This count is at most $2^{|A|}\le 2^q$, so it
is determined by its residue modulo $2^{q+1}$. Hence
$\min_{A_1\subseteq A}\bigl(g_1(A_1)+g_2(A\setminus A_1)\bigr)$ is the smallest
$a$ such that the coefficient of $x^a$ in $P(A)$ is non-zero, and it is
$\infty$ if $P(A)=0$.

By the result of Björklund, Husfeldt, Kaski and
Koivisto~\cite{DBLP:conf/stoc/BjorklundHKK07}, which holds over any
commutative ring, all values $P(A)$ can be computed using
$2^q\cdot q^{\Oh(1)}$ additions and multiplications in $\mathcal{R}$. The
polynomials $P_1(A)$ and $P_2(A)$ have degree at most $\Omega$, and their
algorithm multiplies two such polynomials at most once on any path of the
computation and otherwise only adds and subtracts, so every polynomial
occurring has degree at most $2\Omega$. Each coefficient has at most $q+1$
bits. Thus one ring operation takes time polynomial in $\Omega$ and $q$, hence
polynomial in $n$, and all entries with first two entries $A_t, X_t$ are
computed in time $2^q\cdot n^{\Oh(1)}\le 2^{|Y|}\cdot n^{\Oh(1)}$.

Summing over all choices of the disjoint sets $A_t, X_t$, each vertex of
$B_t$ is in $A_t$, in $X_t$, or in $Y$, and the last option is weighted by
$2$. So the total time is
\[
\sum_{A_t,\,X_t} 2^{|B_t\setminus(A_t\cup X_t)|}\cdot n^{\Oh(1)}
= (1+1+2)^{|B_t|}\cdot n^{\Oh(1)}
= 4^{|B_t|}\cdot n^{\Oh(1)} .
\]
\end{proof}

\section{Proof of \cref{th:mvdtw}}\label{appenpf:mvdtw}
\begin{proof}
    Given a tree decomposition $\mathcal{T} = (T, \{\mathcal{B}_t\}_{t \in V(T)})$ of width $\tw$
for a graph $G$, it is known that one can compute, in polynomial time, a nice tree
decomposition of $G$ with the same width $\tw$~\cite{DBLP:books/sp/CyganFKLMPPS15}.
Hence, we may assume without loss of generality that
$\mathcal{T} = (T, \{\mathcal{B}_t\}_{t \in V(T)})$ is a nice tree decomposition of $G$
of width $\tw$.

Using the dynamic programming recurrences described above, we correctly compute the
value $W[t,A_t,X_t,R_t,S_t]$ for every node $t \in V(T)$ and every partition
$(A_t,X_t,R_t,S_t)$ of the bag $B_t$.
The value $W[t,A_t,X_t,R_t,S_t]$ denotes the minimum weight of a monopolar
vertex-deletion set $M_t$ for the subgraph $G_t$ such that $G_t - M_t$ admits a
monopolar partition $(I_t,C_t)$ satisfying: $M_t \cap B_t = A_t$; $I_t \cap B_t = X_t$; $C_t \cap B_t = R_t \cup S_t$; and $R_t$ is exactly the union of $C \cap B_t$ over all cliques $C$ of $C_t$ that contain at least one vertex from $V_t \setminus B_t$.

The computation proceeds bottom-up along the tree $T$, starting from the leaf nodes.
Let $r$ be the root of $T$.
Since $B_r = \emptyset$ and $G_r = G$, the value
$W[r,\emptyset,\emptyset,\emptyset,\emptyset]$ gives the minimum weight of a monopolar
vertex-deletion set for the entire graph $G$.

For each bag $B_t$, we compute $W[t,A_t,X_t,R_t,S_t]$ for all possible
partitions $(A_t,X_t,R_t,S_t)$. As each vertex of $B_t$ can belong to exactly
one of the four sets, there are at most $4^{|B_t|}\le 4^{\tw+1}$ such
partitions. At a leaf, introduce or forget node, each entry is computed in
polynomial time from a constant number of entries of the child, so the time
spent at such a node is $4^{\tw}\cdot n^{\Oh(1)}$. At a join node, by
\cref{lem:join-time-mvd}, the time spent is $5^{\tw}\cdot n^{\Oh(1)}$ in
general, and $4^{\tw}\cdot n^{\Oh(1)}$ if the weights are integers bounded by
a polynomial in $n$. Since the number of nodes in $T$ is polynomial in $n$, the
overall running time is $5^{\tw}\cdot n^{\Oh(1)}$ in general and
$4^{\tw}\cdot n^{\Oh(1)}$ in the bounded integer case.\qed
\end{proof}
\end{document}